\documentclass[12pt]{iopart}

\usepackage{graphicx}        

\newcommand{\be}{\begin{equation}}
\newcommand{\ee}{\end{equation}}
\newcommand{\bea}{\begin{eqnarray}}
\newcommand{\eea}{\end{eqnarray}}
\newcommand{\ba}{\begin{array}}
\newcommand{\ea}{\end{array}}
\newcommand{\bc}{\begin{center}}
\newcommand{\ec}{\end{center}}
\newcommand{\Dslash}{D\hspace*{-0.23cm}/\,}
\newcommand{\bm}[1]{\mbox{\boldmath${#1}$}}
\def\lsim{\mathrel {\vcenter {\baselineskip 0pt \kern 0pt
    \hbox{$<$} \kern 0pt \hbox{$\sim$} }}}
\def\gsim{\mathrel {\vcenter {\baselineskip 0pt \kern 0pt
    \hbox{$>$} \kern 0pt \hbox{$\sim$} }}}

\def\N{{\mathcal N}}

\def\x{{\bf x}}

\def\Rads{L}
\def\llangle{\left\langle}
\def\rrangle{\right\rangle}
\def\lsim{\mbox{~{\protect\raisebox{0.4ex}{$<$}}\hspace{-1.1em}
	{\protect\raisebox{-0.6ex}{$\sim$}}~}}

\begin{document}

\title[Nearly Perfect Fluidity]{Nearly Perfect Fluidity: From 
Cold Atomic Gases to Hot Quark Gluon Plasmas}

\author{Thomas Sch\"afer$^1$ and Derek Teaney$^{2,3}$}

\address{$^1$Department of Physics,
     North Carolina State University,
     Raleigh, NC 27695}

\address{$^2$Department of Physics,
     State University of New York at Stony Brook,
     Stony Brook, NY 11794}

\address{$^3$ RIKEN-BNL Research Center, 
     Brookhaven National Laboratory, Upton, NY 11973}

\begin{abstract}
 Shear viscosity is a measure of the amount of dissipation 
in a simple fluid. In kinetic theory shear viscosity is related 
to the rate of momentum transport by quasi-particles, and the 
uncertainty relation suggests that the ratio of shear viscosity 
$\eta$ to entropy density $s$ in units of $\hbar/k_B$ is bounded 
by a constant. Here, $\hbar$ is Planck's constant and $k_B$ is 
Boltzmann's constant. A specific bound has been proposed on the 
basis of string theory where, for a large class of theories, one 
can show that $\eta/s\geq \hbar/(4\pi k_B)$. We will refer to a 
fluid that saturates the string theory bound as a perfect fluid.
In this review we summarize 
theoretical and experimental information on the properties of 
the three main classes of quantum fluids that are known to have 
values of $\eta/s$ that are smaller than $\hbar/k_B$.
These fluids are strongly coupled Bose fluids, in particular 
liquid helium, strongly correlated ultracold Fermi gases, and 
the quark gluon plasma. We discuss the main theoretical approaches 
to transport properties of these fluids: kinetic theory, 
numerical simulations based on linear response theory, and 
holographic dualities. We also summarize the experimental 
situation, in particular with regard to the observation of
hydrodynamic behavior in ultracold Fermi gases and the quark 
gluon plasma. 

\end{abstract}

\maketitle

\tableofcontents

\section{Introduction}
\label{sec_intro}

 A fluid is a material that can be described by the laws of
fluid dynamics. These laws imply that the response of a 
fluid to slowly varying external perturbations is completely 
governed by conservation laws. In the case of simple fluids,
such as water, the conserved quantities are mass, energy, 
and momentum. 

 The study of fluids is one of the oldest problems in physics
\cite{Darrigol:2005}. Understanding why certain materials make 
good fluids, and others do not, has nevertheless remained a
very difficult question. The quality of a fluid can be 
characterized by its shear viscosity $\eta$. Shear viscosity 
is defined in terms of the friction force $F$ per unit area 
$A$ created by a shear flow with transverse flow gradient 
$\nabla_{\! y} v_x$, 
\be 
\label{eta_fric}
 \frac{F}{A}=\eta\, \nabla_{\! y} v_x\, . 
\ee
Viscosity causes dissipation which converts part of the 
kinetic energy of the flow to heat. A good fluid is therefore 
characterized by a small shear viscosity. Indeed, the inverse of 
shear viscosity, $\varphi=1/\eta$, is sometimes called fluidity. 

 The molecular theory of transport phenomena in dilute gases 
goes back to Maxwell. Maxwell realized that shear viscosity is 
related to momentum transport by individual molecules. A simple
estimate of the shear viscosity of a dilute gas is 
\be 
\label{eta_mfp}
\eta = \frac{1}{3}\,n p l_{\it mfp}\, , 
\ee
where $n$ is the density, $p$ is the average momentum of the 
molecules, and $l_{\it mfp}$ is the mean free path. The mean free
path can be written as $l_{\it mfp}=1/(n\sigma)$ where $\sigma$ 
is a suitable transport cross section. This implies that the
shear viscosity of a dilute gas grows with temperature (as
$p\sim T^{1/2}$) but is approximately independent of density. 
This counterintuitive result is confirmed by experiment, going
back to experiments carried out by Maxwell himself \cite{Brush:1986}. 
Equ.~(\ref{eta_mfp}) also shows that the viscosity of 
an ideal gas is infinite, not zero. In order to achieve thermal 
equilibrium we have to view the ideal gas as the limit of an 
interacting system in which the scattering cross section $\sigma$ 
is taken to zero. In this limit the mean free path, and with it
the viscosity, goes to infinity.

 At low temperature gases condense into the liquid (or solid)
state. In a liquid transport is no longer governed by the motion 
of individual molecules. A simple picture, due to Frenkel, Eyring,
and others, is that momentum transport is due to processes
that involve the motion of vacancies \cite{Frenkel:1955}. These 
processes can be viewed as thermally activated transitions in 
which a molecule or a cluster moves from one local energy minimum 
to another. The viscosity scales as \cite{Eyring:1936,Tabor:1979}
\be 
\label{eta_eyring}
\eta \simeq  h n e^{E/(k_BT)}\, ,
\ee 
where $E$ is an activation energy and $h$ is Planck's constant. 
We note that the viscosity of a liquid has a very strong 
dependence on temperature. We also observe that the overall scale
involves Planck's constant. The appearance of $h$ is related to 
Eyring's assumption that the collision time of the molecules 
is $h/(k_BT)$, the shortest time scale in a liquid. We will 
come back to this assumption below.
Equ.~(\ref{eta_eyring}) shows that the viscosity of a liquid 
grows as the temperature is lowered. Together with equ.~(\ref{eta_mfp}) 
this result implies that the viscosity of a typical fluid has 
a minimum as a function of temperature, and that the minimum is likely 
to occur in the vicinity of the liquid-gas phase transition.

 Experimental results show that the minimum value of the viscosity 
of good fluids, like water, liquid helium and liquid nitrogen, differs
by many orders of magnitude, see the data in Table \ref{tab_eta}. The 
SI unit for viscosity is Pascal second (Pa $\cdot$ s), the CGS unit is 
Poise (P). Note that 1 Pa $\cdot$ s = 10 Poise. Clearly, it is desirable 
to normalize the viscosity to a suitable thermodynamic quantity in 
order to make more useful comparisons. Equations (\ref{eta_mfp}) and 
(\ref{eta_eyring}) indicate that a suitable ratio is provided
by $\eta/n$. We note that the ratio of viscosity over mass
density $\rho =mn$ is known as the kinematic viscosity $\nu=\eta/\rho$. 
The behavior of solutions of the Navier-Stokes equation is governed
by the Reynolds number   
\be 
\label{Reynolds}
  {\it Re} = \left(\frac{\rho}{\eta}\right)  v L\, ,
\ee 
which is the ratio of a property of the flow, its characteristic 
velocity $v$ multiplied by its characteristic length scale $L$,
over a property of the fluid, its kinematic viscosity. Good fluids 
attain larger Reynolds numbers, and are more likely to exhibit 
turbulent flow. Data for the ratio $\eta/n$ are tabulated in 
Table \ref{tab_eta}. We observe that the ratios $\eta/n$ for 
good fluids are indeed similar in magnitude.

\begin{table}[t]
\bc\begin{tabular}{|c|c|c|c|c|c|} 
\hline
fluid & $P$ [Pa] & $T$ [K] & $\eta$ [Pa$\cdot$s] & 
   $\eta/n$   $[\hbar]$ & 
   $\eta/s$  $[\hbar/k_B]$ \\ \hline
H$_2$O   &  0.1$\cdot 10^6$   & 370    & $2.9\cdot 10^{-4}$ 
         &   85               & 8.2     \\
$^4$He   &  0.1$\cdot 10^6$   & 2.0    & $1.2\cdot 10^{-6}$ 
         &  0.5               & 1.9   \\
H$_2$O   & 22.6$\cdot 10^6$   & 650    & $6.0\cdot 10^{-5}$ 
         & 32                 & 2.0  \\
$^4$He   & 0.22$\cdot 10^6$   & 5.1    & $1.7\cdot 10^{-6}$ 
         & 1.7                & 0.7  \\
$^6$Li ($a=\infty$)   
         &  12$\cdot 10^{-9}$ & 23$\cdot 10^{-6}$ & $\leq 1.7\cdot 10^{-15}$  
         &  $\leq 1$          & $\leq 0.5$  \\
QGP      &  88$\cdot 10^{33}$ &  2$\cdot 10^{12}$ & $\leq 5 \cdot 10^{11}$  
         &                    & $\leq 0.4$ \\ \hline
\hline\end{tabular}\ec
\caption{\label{tab_eta}
Viscosity $\eta$, viscosity over density, and viscosity over
entropy density ratio for several fluids. Data for water and 
helium taken from \cite{nist,codata} and \cite{Wilks:1966}, data 
for Li and the quark gluon plasma (QGP) will be explained in 
Sect.~\ref{sec_exp}. For water and helium we show data at 
atmospheric pressure and temperatures just below the boiling
point and the $\lambda$ transition, respectively. These data points 
roughly correspond to the minimum of $\eta/n$ at atmospheric pressure.
We also show and data near the tri-critical point which roughly
corresponds to the global minimum of $\eta/s$. Note that the 
quark gluon plasma does not have a well defined density.  }
\end{table}

 A disadvantage of considering the ratio $\eta/n$ is that it is 
not possible to include relativistic fluids in the comparison. 
In the case of a relativistic fluid the number of particles 
is not conserved. As a consequence the quantity $n$ is not be well 
defined in an interacting system. In a quark gluon plasma, 
for example, only the net number of quarks (the number of quarks 
minus the number of anti-quarks) is well defined, but the number 
of quarks or the number of gluons is not. In Sect.~\ref{sec_hydro} 
we will show that the Reynolds number of a relativistic fluid is 
defined in terms of the ratio $\eta/(sT)$, where $s$ is the entropy 
density and $T$ is the temperature. This indicates that we should 
consider the ratio $\eta/s$ instead of $\eta/n$. We note that this 
ratio is well defined in both the relativistic and non-relativistic 
limit, and that $s \sim n k_B$ for many fluids. For example, in a 
non-interacting relativistic Bose gas $s/n\simeq 3.6\,k_B$, and 
for non-interacting fermions $s/n\simeq 4.2\, k_B$. In a weakly 
interacting non-relativistic gas $s/n\simeq k_B$ up to logarithms 
of $gn/(mT)^{3/2}$, where $g$ is the degeneracy factor.
Data for $\eta/s$ in units of $\hbar/k_B$ are also given in Table 
\ref{tab_eta}. 

 We observe that good fluids are characterized by $\eta/s\sim \hbar/k_B$.
This value is consistent with simple theoretical estimates. 
Consider the kinetic theory estimate in equ.~(\ref{eta_mfp}). In the 
strong coupling limit the mean free path becomes very small, but 
the uncertainty relation suggests that $pl_{\it mfp}\gsim \hbar$ 
\cite{Danielewicz:1984ww}. For a rough estimate we may use the
relation $s\simeq 3.6k_B n$ and conclude that $\eta/s\gsim \hbar
/(10.8\, k_b)$. A bound on $\eta/s$ is also indicated by the 
high temperature limit of Eyring's formula equ.~(\ref{eta_eyring}).
We note, however, that kinetic theory is not reliable in the 
regime $\eta/s\sim  \hbar/k_B$, and other methods are needed
to determine the minimum value of $\eta/s$.

  A precise value of the viscosity bound was proposed
based on results from string theory. Kovtun et al.~conjectured 
that  \cite{Kovtun:2004de}
\be 
\label{ks_bound}
 \frac{\eta}{s} \geq \frac{\hbar}{4\pi k_B} \, , 
\ee
for all fluids. We will call a fluid that saturates the bound
(\ref{ks_bound}) a ``perfect fluid''. A perfect fluid dissipates
the smallest possible amount of energy, and satisfies the laws
of fluid dynamics in the largest possible domain. In a typical 
fluid, hydrodynamics is an effective description of long 
wavelength fluctuations, but in a perfect fluid hydrodynamics 
is reliable at distances as short as the inter-particle spacing.

 The viscosity bound conjecture raises a number of interesting 
questions:

\begin{itemize} 
\item Is the conjecture in equ.~(\ref{ks_bound}) correct? Does this
bound, or some other bound on $\eta/s$, follow from the general 
principles of quantum mechanics and statistical mechanics?

\item Is there a ``perfect fluid'' in nature, i.e. can we observe
a fluid that attains the value $\eta/s=\hbar/(4\pi k_B)$? If yes, 
what are the characteristics of such a fluid? Is it possible to 
describe the fluid in terms of quasi-particles? 

\item How is $\eta/s$ correlated with other transport properties,
like bulk viscosity, diffusion constants, conductivities? Are
there bounds on other transport properties? 
\end{itemize}

 We will not be able to provide definitive answers to these 
questions in this review. There are, however, a number of 
recent results, from both theory and experiment, that shed 
light on these issues:

\begin{itemize}
\item The experimental realization of new classes of quantum 
fluids. Prior to 1995 the only bulk quantum fluids that could
be studied in the laboratory were the two isotopes of liquid
helium, $^4$He and $^3$He. In 1995 several groups achieved 
quantum degeneracy is dilute atomic Bose gases. In 1999 
experimentalists also succeeded in producing degenerate
atomic Fermi gases \cite{DeMarco:1999,OHara:1999}. Using Feshbach 
resonances it is possible to experimentally control the interaction 
between the atoms, and to study equilibrium and transport properties 
as a function of the interaction strength. 

\item The experimental discovery of almost ideal hydrodynamic
flow in a completely different physical system, the quark
gluon plasma created in heavy ion collisions at the relativistic
heavy ion collider (RHIC) at Brookhaven National Laboratory
\cite{Adler:2003kt,Back:2004mh,Adams:2004bi}. The quark gluon
plasma also exhibits a large energy loss for high energy 
colored particles, and a very small heavy quark diffusion 
constant. 

\item Progress in non-equilibrium field theory culminated  
in the calculation of transport coefficients of weakly coupled 
gauge theory plasmas \cite{Baym:1990uj,Arnold:2000dr,Arnold:2003zc}. 
These results complete the program of using kinetic theory to calculate 
the transport properties of the three main classes of quantum liquids: 
Bose gases, Fermi gases, and gauge theory plasmas. 

\item The theoretical discovery of a completely new method for 
computing the transport properties of very strongly coupled
fluids \cite{Policastro:2001yc}. This method is based on the 
holographic duality between certain strongly coupled field 
theories in $d=4$ space-time dimensions and weakly coupled 
string theory in $d=10$ \cite{Maldacena:1997re}. For gauge 
theories that have a weakly coupled string dual the shear 
viscosity to entropy density ratio at infinite coupling is 
$\eta/s=\hbar/(4\pi k_B)$. It was also shown that the first 
correction to this result at finite but large coupling increases 
$\eta/s$ \cite{Buchel:2004di}, and it was conjectured that $\eta/s
=\hbar/(4\pi k_B)$ is a universal lower bound \cite{Kovtun:2004de}.

\end{itemize}

 It is the goal of this review to summarize these recent 
developments. For this purpose we shall concentrate on three 
representative fluids: $^4$He, a strongly coupled Bose fluid; 
atomic Fermi gases near a Feshbach resonance, which are the 
most strongly coupled Fermi liquids; and the quark gluon plasma 
near the critical temperature for condensation into hadron gas, 
which is a very strongly coupled plasma. The review is structured 
as follows: In Sect.~\ref{sec_qfl} we discuss the thermodynamics of 
these quantum fluids. In Sect.~\ref{sec_transp} we review theoretical 
approaches to transport properties. We briefly summarize the 
hydrodynamic description of relativistic and non-relativistic fluids, 
as well as superfluids in Sects.~\ref{sec_hydro}-\ref{sec_dyn}. 
A general connection between transport coefficients and the 
underlying field theory is provided by Kubo relations, which
we introduce in Sect.~\ref{sec_kubo}. In Sect.~\ref{sec_kin} we 
concentrate on fluids that can be described in terms of weakly 
coupled quasi-particles. In this case transport properties can be 
computed using kinetic theory. In Sect.~\ref{sec_ads} we summarize 
results for transport coefficients that have been obtained using 
holographic dualities. Finally, in Sect.~\ref{sec_exp} we discuss 
experimental results for the viscosity and other transport properties 
of strongly coupled quantum fluids. 

 Needless to say, a review of this size cannot adequately summarize 
all the work that has been done on the transport properties of quantum 
fluids. A standard reference on the properties of liquid helium is 
\cite{Wilks:1966}, recent reviews on strongly coupled Fermi gases 
are \cite{Bloch:2007,Giorgini:2008}, and the physics of the strongly 
coupled quark gluon plasma is discussed in \cite{Shuryak:2008eq}. The 
kinetic theory of dilute Bose and Fermi gases is covered in textbooks, see 
\cite{Khalatnikov:1965,Baym:1991}, and the kinetic theory of gauge 
fields was reviewed in \cite{Arnold:2007pg}. Reviews of the AdS/CFT 
correspondence with an emphasis on transport theory are 
\cite{Son:2007vk,Rangamani:2009xk}, and reviews of relativistic 
hydrodynamics can be found in 
\cite{Heinz:2009xj,Romatschke:2009im,Teaney:2009qa}.

\section{Strongly coupled quantum fluids}
\label{sec_qfl}

 In this section we will discuss equilibrium properties of strongly 
interacting quantum fluids. We will specify the effective action 
for bosonic, fermionic, and gauge theory fluids, identify the relevant 
physical scales, and discuss the nature of low energy excitations.

\subsection{Bose fluids: Dilute Bose gases}
\label{sec_bl}

 A gas of atoms satisfying Bose statistics can be described in 
terms of a scalar field $\psi({\bf x},t)$ governed by the 
Hamiltonian 
\bea 
\label{H_bose}
 H &=& \int d^3x\; 
  \psi^*({\bf x},t)\left(
        -\frac{\hbar^2{\bm \nabla}^2}{2m}\right)\psi({\bf x},t) 
   \\
 & & \mbox{} +\frac{1}{2}\int d^3x_1 \int d^3x_2\; 
  \psi^*({\bf x}_1,t)\psi({\bf x}_1,t) V({\bf x}_1-{\bf x}_2) 
  \psi^*({\bf x}_2,t)\psi({\bf x}_2,t) \, , 
   \nonumber 
\eea
where $m$ is the mass of the boson, and $V({\bf x})$ is a potential. 
The Hamiltonian is invariant with respect to translations and rotations, 
as well as under the $U(1)$ symmetry $\psi\to\exp(i\alpha)\psi$. 
Symmetries correspond to conservation laws. Translations and rotations
correspond to the conservation of linear and angular momentum, and 
the $U(1)$ symmetry is associated with the conservation of the number 
of atoms. The Schr\"odinger equation is also invariant under Galilean 
transformations ${\bf x}\to {\bf x}-{\bf v}t$ which act on the field 
as $\psi({\bf x},t) \to \exp(im{\bf v}\cdot{\bf x}-\frac{i}{2}m{\bf v}^2t)
\psi({\bf x}- {\bf v}t,t)$. This symmetry will play a role when we 
consider the motion of fluids.

 If the typical momenta are small compared to $1/r_0$, where $r_0$ 
is the range of the potential, we can approximate the interaction 
by a contact term. For very small momenta the leading contribution 
is an $s$-wave interaction 
\be 
\label{V_sr}
 V({\bf x}_1-{\bf x}_2) = C_0 \delta({\bf x}_1-{\bf x}_2) \, ,
\ee
where $C_0$ can be related to the scattering length $a$, $C_0 = 
(4\pi\hbar^2 a)/m$. 
In order to make connections with the physics of 
relativistic fluids it is also useful to introduce the lagrangian
${\cal L}=i\hbar \psi\partial_0\psi -{\cal H}$, where ${\cal H}$ is 
the Hamiltonian density. The lagrangian is 
\be 
\label{l_bose}
{\cal L} = \psi^\dagger \left( i\hbar \partial_0 +
 \frac{\hbar^2{\bm \nabla}^2}{2m} \right) \psi 
 - \frac{C_0}{2} \left(\psi^\dagger \psi\right)^2 .
\ee
In the following we shall consider many-body systems described by 
this lagrangian. We first study the relevant scales in a weakly 
interacting Bose gas governed by the $s$-wave scattering length. 
At high temperature the Bose gas is a classical Boltzmann gas. 
The average energy of the atoms is $\frac{3}{2}k_BT$, and the 
average momenta are of order $(mk_BT)^{1/2}$. The importance of 
quantum statistics is governed by the parameter $nv_Q$, where $n$ 
is the density, $v_Q=\lambda^3$ is the quantum volume and 
\be
\label{v_Q}
\lambda = \frac{2\pi\hbar}{\sqrt{2\pi mk_BT}}
\ee 
is the thermal wave length. Quantum statistics becomes important 
for $nv_Q\sim 1$, and Bose condensation in an ideal gas occurs 
at $nv_Q=2.61$, corresponding to a critical temperature
\be
\label{T_c_BEC}
T_c = \frac{2\pi\hbar^2}{mk_B}
  \left(\frac{n}{\zeta(3/2)}\right)^{2/3}\, . 
\ee
The effects of a non-zero scattering length can be taken into 
account order by order in an expansion in $an^{1/3}$. At high
temperature this is the standard virial expansion 
\be 
\label{P_vir}
P= nk_BT\left\{ 1 + b_2 n + O(n^2) \right\}\, , 
\ee
where $P$ is the pressure and $b_2$ is the second virial coefficient. 
In the limit of small $a$ the second virial coefficient of 
a single component Bose gas is given by \cite{Huang:1987}
\be 
\label{b_2}
b_2 = -\frac{1}{4\sqrt{2}}\lambda^3 + 2a\lambda^2\, ,
\ee
where the first term is due to quantum statistics, and the second
term is related to the interaction. The second virial coefficient
is finite in the limit of a large scattering length. As $a\to\infty$
the interaction part approaches $-\sqrt{2}\lambda^3$, and the role 
of interactions is governed by the same parameter that controls 
the effects of quantum statistics. 

 The interaction also shifts the critical temperature for Bose
condensation. The calculation of this shift is a non-perturbative 
problem, even if the scattering length is small. This is related
to the fact that fluctuations become large in the vicinity of a
second order phase transition, and perturbation theory breaks down. 
One can show that $\delta T_c \sim an^{1/3}T_c$, and 
there is a physical argument that a repulsive scattering length 
($a>0$) increases the value of $T_c$ \cite{Baym:1999}.
A numerical calculation using the Landau-Ginzburg
effective lagrangian gives \cite{Arnold:2001mu}
\be 
\label{Del_T_c}
\Delta T_c = (1.32\pm 0.02)  an^{1/3}T_c .
\ee
Dilute Bose gases in which the scattering length is attractive 
are not stable, but it is possible to create metastable systems
confined by external fields. Weak repulsive interactions increase 
the transition temperature but suppress the condensate fraction.
If $an^{1/3}$ is large then the Bose fluid will typically 
solidify, but the phase structure and critical density depend on 
details of the interaction. A hard sphere gas freezes at $an^{1/3}
\simeq 0.24$ \cite{Kalos:1974}.

 An issue which is very important for transport properties
is the nature of the quasi-particle excitations. 
At high temperature the cross section for binary
scattering between the atoms decreases with the thermal wavelength,
$\sigma \sim \lambda^2 \sim 1/(mT)$, and the mean free path 
$l_{\it mfp} \sim 1/(n\sigma)$ is large. As a consequence the 
the atoms are good quasi-particles. At very low temperature the 
system is superfluid and there is a Goldstone boson, the phonon, 
related to the breaking of the $U(1)$ phase symmetry. Phonons are 
derivatively coupled and the interaction at low energy is weak. 
This implies that the mean free path at low temperature, $T\ll T_c$, 
is also large. The phonon dispersion relation in a weakly non-ideal 
($na^3<1$) Bose gas was first computed by Bogoliubov. The result is 
\be 
\epsilon_p = \frac{1}{2m}\sqrt{ 
 \left( p^2 + 8\pi a n\right)^2 - \left( 8\pi a n\right)^2 } \, .
\ee
For small momenta the dispersion relation is linear, 
$\epsilon_p\simeq c_s p$, and the phonon velocity is given 
by $c_{s}= \sqrt{4\pi an}/m$.

\subsection{Bose fluids: Liquid helium}
\label{sec_he}

  A simple $s$-wave interaction is sufficient for understanding 
the properties of trapped atomic Bose gases, but more accurate 
potentials are required for even a qualitative description of 
liquid $^4$He. Accurate $^4$He potentials can be written as the
sum of a short range term and a long range van der Waals potential, 
\be 
\label{V_vdw}
V(r)=V_{sr}(r) - \frac{C_6}{r^6} \, .
\ee
The coefficient $C_6$ defines the van der Waals length scale 
$l_{VdW}= (mC_6/\hbar^2)^{1/4}$. Accurate parametrizations of 
$V_{sr}$ can be found \cite{Aziz:1991,Tang:1995}. These
potentials have a van der Waals length $l_{VdW}\simeq 10.2\,
a_0$, an effective range $r\simeq 14\, a_0$, and a very large 
scattering length $a\simeq 189\, a_0$, where $a_0=0.529\,\AA$ 
is the Bohr radius. The large $s$-wave scattering length is 
related to the existence of a very weakly bound $^4$He dimer.
The binding energy of the dimer is $B=-1.1\cdot 10^{-7}$ eV.
There are many interesting universal effects governed by 
the large scattering length \cite{Braaten:2006vd}. The density 
of liquid $^4$He is too large for these phenomena to be
important, but universal effects have been observed in 
dilute atomic gases in which the scattering length is large. 

 In the case of $^4$He the interaction between the atoms is
not weak, and it cannot be characterized in terms of the 
scattering length only. In the high temperature limit 
$^4$He is a classical gas, and corrections to the ideal 
gas behavior are described by the virial expansion. The 
virial expansion provides a very accurate description of 
the equation of state at normal pressure for temperatures 
above 10 K. At temperatures below 10 K one has to rely on 
quantum Monte Carlo methods or variational many-body wave 
functions \cite{Ceperley:1995}. At atmospheric pressure 
$^4$He liquefies at 4.22 K, and it becomes superfluid at 
$T_c=2.17$ K. This temperature can be compared to the critical 
temperature for Bose condensation of an ideal gas with the 
density of liquid helium, $n=1/(3.6\,\AA)^3$, which is 
$T_c^0=3.1$ K. 
The dependence of $T_c$ on the density in the case 
of a hard sphere gas was studied by Gr\"uter et al.~\cite{Gruter:1997}.
Helium is well described by an effective hard sphere parameter 
$a=2.20\,\AA$. Gr\"uter et al.~show that for $na^3\lsim 0.1$
the critical temperature is larger than that of a non-interacting
gas, in agreement with equ.~(\ref{Del_T_c}). The increase in 
$T_c$ is small, reaching about 6\%. For larger values of $na^3$ 
the critical temperature drops rapidly, until freezing occurs 
at $na^3\sim 0.25$. 

 The presence of strong interactions also manifests itself
in a small condensate fraction. Glyde et al.~measured the number 
of condensed atoms $N_0(T)$ using neutron scattering on liquid
$^4$He at saturated vapor pressure \cite{Glyde:2000}. They find
$N_0(T)/N= f(1-(T/T_c)^\gamma$ with $f\simeq (7.25\pm 0.75)\cdot 
10^{-2}$ and $\gamma=5.5\pm 1$.
The superfluid transition is in 
the universality class of the three dimensional $O(2)$ model. 
Renormalization group arguments predict a mild non-analyticity 
in the specific heat, $c_v\sim t^{-\alpha}$ with $t=(T-T_c)/T_c$ 
and $\alpha = -0.0151(3)$ \cite{Campostrini:2006}. This 
prediction agrees reasonably well with micro gravity experiments 
which find $\alpha = -0.01285(4)$ \cite{Lipa:1996}.

 The excitation spectrum of superfluid $^4$He shows important
differences as compared to the spectrum of a dilute Bose
condensed gas. 
As expected, at small momenta the excitations
are phonons with a linear dispersion relation $\epsilon(p)=c_sp$, 
where the speed of sound at normal pressure is $c_s=238$ m/sec. At 
larger momenta the dispersion relation has a second minimum, called 
the roton minimum. The dispersion relation in the vicinity of 
the minimum is
\be 
\label{roton}
\epsilon(p)= \Delta + \frac{(p-p_0)^2}{2m^*}\, , 
\ee
where $\Delta/k_B=8.7$ K, $m^*=0.14\, m$ and $p_0/\hbar=1.9\,\AA^{-1}$. 
The roton plays a significant role in determining the specific
heat and transport properties near the critical temperature. 
The existence of the roton is closely related with strong short 
range correlations in liquid helium. These correlations can be
quantified in terms of the static structure factor $S(q)$, which
is the Fourier transform of the density correlation function
\be 
\label{S(q)}
S(q) = \frac{1}{\rho}\int d^3x\, e^{-i{\bf q}\cdot{\bf x}}
 \left[ \langle \rho(0) \rho({\bf x}) \rangle - \langle \rho(0)\rangle^2
 \right]\, . 
\ee 
The static structure factor vanishes linearly in $q$ for small 
momenta, and approaches a constant value for large $q$. $S(q)$ has 
a sharp maximum at intermediate values of $q$, which reflects the 
presence of correlations at the scale of the typical inter-atomic
distance. Feynman proposed a variational wave function for excitations
in liquid helium which gives $\epsilon(q)= q^2/ (2mS(q))$ 
\cite{Feynman:1954}. This relation can also be derived from an 
effective hydrodynamic Hamiltonian, see \cite{Khalatnikov:1965}.
Feynman's result reproduces both the phonon dispersion relation
at low momentum, and the roton minimum at larger momentum.
In order to study the kinetics of liquid helium one has to 
understand the scattering of phonons and rotons. The phonon-phonon 
and phonon-roton interaction is determined by the equation of state
and by constraints from Galilean invariance and $U(1)$ symmetry 
\cite{Khalatnikov:1965}. We will discuss these constraints in 
more detail in the next section, in connection with superfluid 
Fermi gases.

\subsection{Fermi liquids: The dilute Fermi gas at unitarity}
\label{sec_fl}

 In this section we consider non-relativistic Fermi liquids. 
Fermionic systems are interesting because it is possible 
to make strongly correlated liquids with only zero
range interactions. 
The Fermi liquid is described by the same lagrangian 
as in equ.~(\ref{l_bose})
\be 
\label{l_4f}
{\cal L} = \psi^\dagger \left( i\partial_0 +
 \frac{{\bm \nabla}^2}{2m} \right) \psi 
 - \frac{C_0}{2} \left(\psi^\dagger \psi\right)^2 ,
\ee
were $\psi$ is now a two-component fermion field with mass $m$.
The coupling constants $C_0$ is related to the scattering 
length by the same relation as in the bosonic case, $C_0 = 
(4\pi\hbar^2 a)/m$. 

 Over the last ten years there has been truly remarkable progress
in the study of cold, dilute gases of fermionic atoms in which the
scattering length $a$ of the atoms can be controlled experimentally.
These systems can be realized in the laboratory using Feshbach
resonances, see \cite{Regal:2005} for a review. A small negative
scattering length corresponds to a weak attractive interaction between
the atoms. This regime is known as the BCS (Bardeen-Cooper-Schrieffer)
limit. As the strength of the attractive interaction increases 
the scattering length becomes larger. It diverges at the point where a 
two-body bound state is formed. The point $a=\infty$ is called the 
unitarity limit, because the scattering cross section saturates the 
$s$-wave unitarity bound $\sigma=4\pi/k^2$. On the other side of the 
resonance the scattering length is positive. For large positive 
values of $a$ the two-body binding energy is related to the scattering
length by $B=\hbar^2/(ma^2)$. The regime in which the binding energy becomes
large is called the BEC (Bose-Einstein condensation) limit.
 

 We now consider properties of the the many-body system as a 
function of the $s$-wave scattering lengths. In the high temperature 
limit the equation of state is again that of an ideal gas, and the 
leading correction is described by the virial expansion. For
small $a$ the second virial coefficient is 
\be 
\label{b_2_f}
b_2 = \frac{1}{8\sqrt{2}}\lambda^3 + \frac{1}{2}a\lambda^2\, . 
\ee
In the limit $a\to\infty$ the interaction term goes to 
$-\lambda^3/(2\sqrt{2})$. The Fermi gas becomes degenerate 
as $n\lambda^3 \sim 1$. In the limit in which the scattering 
length is large the Fermi gas becomes strongly interacting 
at the same temperature at which quantum effects become 
important.

 At low temperature and in the BCS limit, $a<0$ and 
$n^{1/3}|a|<1$, the Fermi gas can be described as a Landau Fermi 
liquid. The excitations are weakly interacting particles and 
holes which carry the quantum numbers of the elementary fermions. 
At very low temperature the particle-particle interaction 
near the Fermi surface becomes large, and the Fermi 
liquid undergoes a phase transition to a BCS superfluid. The 
transition temperature is \cite{Gorkov:1961}
\be
\label{gap_bcs}
T_c  = \frac{8e^\gamma E_F}{(4e)^{1/3}e^2\pi}
   \exp\left(-\frac{\pi}{2k_F|a|}\right), 
\ee
where $\gamma$ is the Euler constant. The Fermi momentum $k_F$ is 
defined by the relation
\be
 n= \frac{k_F^3}{3\pi^2}\, ,
\ee
and $E_F=k_F^2/(2m)$ is the Fermi energy. This relation defines a
``Fermi momentum'' even in the case that no sharp Fermi surface 
exists. Note that $T_F\equiv E_F$ is the degeneracy temperature 
(we have set $k_B=1$). Also note that $n^{1/3}|a|<1$ implies
$T_c\ll T_F$. 

 In the Bose-Einstein limit the fermions form tightly bound molecules. 
The residual interaction between the molecules is repulsive, and
the many-body system behaves as a weakly non-ideal Bose gas. The Bose 
gas condenses at the critical temperature given in equ.~(\ref{T_c_BEC}).
Using the fact that the mass of molecules is $2m$, and that their
density is $n/2$, we get 
\be
 T_c = 0.21 E_F\, . 
\ee
Variational calculations suggest that at zero temperature the 
evolution from weak to strong coupling is smooth \cite{Nozieres:1984}.
The system is a pair condensate for all values of the coupling, but
the size of the pairs evolves from being much smaller than the 
inter-particle spacing in the BEC limit to being much larger in 
the BCS limit. This idea is confirmed by quantum Monte Carlo 
calculations \cite{Carlson:2004} and experimental observations 
\cite{Bartenstein:2004}. 

 Of particular interest is the crossover (``unitarity'') regime 
where $a\to\infty$. The Fermi gas at unitarity possesses a number of 
interesting properties. First of all, the system is scale invariant 
\cite{Mehen:1999nd,Son:2005rv}. This implies, for example, that all 
energy scales in the many body system, like the critical 
temperature, the gap, and the chemical potential, are 
proportional to the Fermi energy
\be 
\label{xi}
 T_c   = \alpha E_F\, , \hspace{0.7cm}
 \Delta= \beta E_F\,  , \hspace{0.7cm}
 \mu   = \xi E_F\, .
\ee
Similarly, all length scales are given by numerical constants
times the inverse Fermi momentum. The values of the universal 
constants $\alpha,\beta,\xi,\ldots$ can be determined using Quantum 
Monte Carlo (QMC) calculations, or from experiments on harmonically
trapped Fermions. QMC calculations performed by Burovski et al.~give 
$T_c=0.152(7)E_F$ \cite{Burovski:2006}, and Carlson et al.~obtained 
$\Delta=0.50(3)E_F$ \cite{Carlson:2005} and
$\mu=0.44(1)E_F$ \cite{Carlson:2003}. A summary of experimental 
results was recently given by Luo and Thomas \cite{Luo:2008}. 

 Second, the unitarity regime is the most strongly correlated simple
many body system. The crossover regime is continuously connected to 
both the non-interacting Fermi gas and the non-interacting Bose gas, 
but neither limit provides a quantitatively accurate description. 
Very important for the purpose of this review is the observation
hydrodynamic behavior and low viscosity in very dilute Fermi gases
in the unitarity limit. 

 In order to study the kinetic description of a dilute Fermi
gas at unitarity we have to determine the nature of the 
quasi-particles and their interaction.  
In the high temperature limit the excitations are elementary 
fermions, even in the limit $a\to\infty$. This is related to 
the fact that the average cross section is of order $\lambda^2$, 
where $\lambda \sim T^{-1/2}$ is the thermal wave length. In the 
low temperature superfluid phase the dominant excitation is the 
phonon. The dispersion relation is 
\be 
\label{disp_ph}
 \epsilon_p = c_s p\, , \hspace{1cm} 
  c_s = \sqrt{\frac{\xi}{3}}\,  v_F\, , 
\ee
where $v_F=k_F/m$ is the Fermi velocity, and $\xi$ is the 
universal parameter defined in equ.~(\ref{xi}).
Corrections to equ.~(\ref{disp_ph}) are of the order 
$p^2/(m\mu)$ \cite{Rupak:2008xq}, and become large when $p\sim k_F$. 
The dispersion relation for momenta near $k_F$ is not well constrained.  
The static structure factor has been measured in quantum Monte 
Carlo simulations, and it does not show a liquid-like peak
\cite{Astrakharchik:2004}. This suggest that the dispersion
relation does not have a roton minimum.

 The three and four-phonon interaction is completely 
fixed by the equation of state and symmetry constraints. These
constraints are most easily derived from an effective lagrangian
for the phonon field. The phonon field is defined as the phase 
of the order parameter
\be
\label{phi_gb}
  \langle\psi\psi\rangle  = |\langle\psi\psi\rangle| e^{2i\varphi}.
\ee
We now construct the most general lagrangian for the field $\varphi$
which is consistent with Galilei invariance and $U(1)$ symmetry. A
$U(1)$ transformation changes the phase of the wave function and
acts as a shift on phonon field, $\varphi\to\varphi+\alpha$. 
Invariance under the $U(1)$ symmetry requires that the lagrangian 
only contains derivatives of $\varphi$. The phase symmetry can be
extended to time-dependent transformations $\psi\to\exp(i\alpha(t))
\psi$ if the chemical potential transform as $\mu\to \mu+\hbar\partial_0
\alpha$. This is a symmetry of the effective lagrangian if the 
chemical potential always appears in the combination $\mu+\hbar
\partial_0\varphi$. Finally, under a Galilei transformation with 
velocity ${\bf v}$ the phonon transforms as $\varphi({\bf x},t) 
\to \varphi({\bf x}-{\bf v}t)-m{\bf v}\cdot{\bf x}+O({\bf v}^2)$. 
This implies that time derivatives of $\varphi$ have to be accompanied 
by spatial derivatives of $\varphi$.
At leading order in derivatives of $\varphi$ we can incorporate 
these symmetries by constructing a lagrangian that only depends 
on the variable
\be
\label{def_X}
X = \mu - \hbar\partial_0\varphi 
      - \frac{(\hbar{\bm \nabla}\varphi)^2}{2m}\, .
\ee
The functional form of the lagrangian ${\cal L}(X)$ is fixed 
by the observation that for a constant phonon field the lagrangian 
reduces to a function of $\mu$. Since differentiating the lagrangian 
with respect to the chemical potential gives the density this function 
must be the pressure $P(\mu)$. We conclude that
\be
\label{l_sfl_pert}
  {\cal L} = P(X) = \frac{2^{5/2} m^{3/2}}{15\pi^2\xi^{3/2}} 
   \left(\mu -\hbar\partial_0\varphi
      -\frac{(\hbar{\bm \nabla}\varphi)^2}{2m} \right)^{5/2}\, ,
\ee
where we have used the fact that, up to a numerical factor, the
pressure of the interacting system is equal to that of a free gas. 
We have also used that this factor can be related to the ratio
$\xi=\mu/E_F$. Phonons are low energy excitations and we can 
expand equ.~(\ref{l_sfl_pert}) in powers of $\partial_0\varphi$ 
and $\nabla_i\varphi$. We find
\bea
\label{ph_vert}
{\cal L} &=& \frac{1}{2}(\partial_0\phi)^2 -\frac{1}{2}c_s^2
  \left({\bm \nabla}\phi\right)^2
  -\alpha\left[ \left(\partial_0\phi\right)^3 
     - 9c_s^2\partial_0\phi\left({\bm \nabla}\phi\right)^2 \right]
  \\ 
& & \mbox{}
   -\frac{3}{2}\alpha^2 \left[\left(\partial_0\phi\right)^4
   + 18 c_s^2\left(\partial_0\phi\right)^2\left({\bm \nabla}\phi\right)^2
   - 27 c_s^4 \left({\bm \nabla}\phi\right)^4\right]+\cdots ,\nonumber
\eea
where we have rescaled the field $\varphi={\it const}\times\phi$ to 
make it canonically normalized. We have also defined $\alpha = \pi 
c_s^{3/2}\xi^{3/4}/ (3^{1/4} 8\mu^2)$. We observe that the three 
and four phonon vertices are completely fixed by the speed of 
sound $c_s$. 
This implies that there are no free parameters that enter into
the kinetic theory of phonons. We also note that equ.~(\ref{l_sfl_pert}) 
generates phonons self interactions to arbitrary order in the phonon
field, but to leading order in the number of derivatives. Terms 
involving higher derivatives of $\varphi$ were constructed in 
\cite{Son:2005rv}. These terms involve non-trivial constraints 
from not just scale invariance, but from the full conformal 
symmetry of the Fermi gas at unitarity.

\vspace*{0.3cm}
\noindent
{\it About Units:} Up to this point, we have explicitly displayed
factors of $\hbar$, $c$ and $k_B$. From now on we will work in
natural units and set $\hbar=k_B=c=1$.

\subsection{Gauge theories: QCD}
\label{sec_qcd}

 Quantumchromodynamics (QCD) is governed by the lagrangian 
\be
\label{l_qcd}
 {\cal L } =  - \frac{1}{4} G_{\mu\nu}^a G_{\mu\nu}^a
  + \sum_f^{N_f} \bar{\psi}_f ( i\Dslash - m_f) \psi_f\, ,
\ee
where $\psi_f$ is a Dirac fermion with flavor index $f$ and $m_f$
is the quark mass. We have suppressed the color ($A=1,\ldots,N_c$)
and spinor ($\alpha=1,\ldots,4$) indices of the fermion fields. 
The covariant derivative acting on the quark fields is 
\be
 i\Dslash \psi = \gamma^\mu \left(
 i\partial_\mu + g A_\mu^a \frac{\lambda^a}{2}\right) \psi\, , 
\ee
where $A_\mu^a$ is a gauge potential and  $\lambda^a$ ($a=1,\ldots,
N_c^2-1$) are the Gell-Mann matrices. The field strength tensor is 
defined by 
\be
 G_{\mu\nu}^a = \partial_\mu A_\nu^a - \partial_\nu A_\mu^a
  + gf^{abc} A_\mu^b A_\nu^c\, ,
\ee
where $f^{abc}$ are the $SU(N_c)$ structure constants, and $g$ is a
coupling constant. In the standard model $N_c=3$ and $N_f=6$, but 
three out of the six flavors are too heavy too play much of a role 
in the dynamics of QCD, and we shall mostly concentrate on $N_f=3$
flavors. The total quark density
\be 
\rho_q = \sum_f\psi^\dagger_f\psi_f
\ee
is conserved and we can introduce a chemical potential $\mu$ coupled
to $\rho_q$. The phase structure and transport properties of QCD at 
finite $\mu$ are an interesting subject \cite{Alford:2007xm}, but in 
this review we will concentrate on QCD at non-zero temperature and 
zero or very small small chemical potential. It is interesting to
note that at low quark density the relevant degrees of freedom 
are protons and neutrons. In the low energy limit the interaction
between neutrons and protons is governed by an effective lagrangian
of the type given in equ.~(\ref{l_4f}). 
The scattering length is a function of the quark masses,
and it is theoretically possible to tune the light quark masses 
to a point where the neutron-neutron scattering length diverges.
The real world is close to this point, as the experimental value 
of the scattering is $a_{nn}\simeq-17$ fm is much larger than 
typical QCD scales. This implies that there is a point in the QCD 
phase diagram where the long distance physics is equivalent to 
that of a dilute atomic Fermi gas at unitarity.

 For many purposes we can consider the first three flavors (up, down,
and strange) to be approximately massless. In this limit the QCD 
lagrangian contains a single dimensionless parameter, the coupling
constant $g$. If quantum effects are taken into account the coupling
becomes scale dependent. At leading order the running coupling 
constant is 
\be
\label{g_1l}
 g^2(q) = \frac{16\pi^2}
  {b_0\log(q^2/\Lambda_{\it QCD}^2)}\, , \hspace{1cm}
 b_0=\frac{11}{3}N_c-\frac{2}{3}N_f\, .
\ee
This result implies that as a quantum theory, QCD is not characterized
by a dimensionless coupling, but by a dimensionful scale, the QCD 
scale parameter $\Lambda_{\it QCD}$. This phenomenon is called 
dimensional transmutation. We also observe that the coupling decreases
with increasing momentum. This is the phenomenon of asymptotic 
freedom. 

 At high temperature the dominant momenta are on the order of $T$, and
for $T\gg\Lambda_{QCD}$ asymptotic freedom implies that bulk thermodynamics
is governed by weak coupling. The weak coupling expansion of the equation
of state is 
\be 
\label{P_pqcd}
 P = T^4 \left\{ c_0 + c_2 g^2 + c_3 g^3 + (c_4'\log(g)+c_4)g^4 
  + \ldots \right\}\, , 
\ee
where the first term is the Stefan-Boltzmann law and
\be
\label{P_SB}
c_0 = \frac{\pi^2}{90}\left( 
  2\left( N_c^2-1 \right) + 4N_cN_f\frac{7}{8} \right) 
\ee
depends on the number of degrees of freedom ($2(N_c^2-1)$ gluons 
and $4N_cN_f$ quarks). We note that in a theory of massless
particles the equation of state is always sensitive to quantum 
statistics, even if the temperature is high. The first correction 
is \cite{Shuryak:1977ut}
\be
c_2 = -\frac{N_c^2-1}{144} \left( N_c +\frac{5}{4} N_f \right)\, . 
\ee
The perturbative expansion in equ.~(\ref{P_pqcd}) is evaluated with 
$g$ taken to be the running coupling constant evaluated at a scale
$q\sim T$. The precise scale is not uniquely determined -- changing 
the scale corresponds to reshuffling higher order corrections in the 
perturbative expansion. The scale is usually chosen to improve the 
apparent rate of convergence. This criterion gives a value close 
to $2\pi T$. 

 We note that the perturbative expansion is not a power series in the 
fine structure constant $\alpha_s=g^2/(4\pi)$. The expansion contains
square roots and logarithms of $\alpha_s$. Non-analytic terms in the 
expansion are related to infrared sensitive diagrams. For example, the 
$g^3$ term in equ.~(\ref{P_pqcd}) is due to ring diagrams (also called
the plasmon term). Ring diagrams are one-loop gluon diagrams in which 
the leading order gluon self energy has been summed to all orders. 
We also note that the weak coupling expansion cannot be extended to 
arbitrarily high powers in $g$. At $O(g^6)$ one encounters infrared 
divergent diagrams which can only be summed non-perturbatively, by 
computing the partition function of three-dimensional QCD at zero
temperature. 

 In order to analyze the relevant scales in high temperature QCD
in more detail we consider the current-current interaction
\be 
 {\cal M} = j_\mu^a \Pi_{\mu\nu}^{ab}  j_\nu^b\, ,
\ee
where $j_\mu^a$ is a color current and $\Pi_{\mu_\nu}^{ab}$ is 
the gluon polarization function. The tensor structure of the 
gluon polarization function can be decomposed into a 
transverse and a longitudinal part
\bea 
\label{proj}
 \Pi_{\mu\nu}(q)&=& \Pi^T(q)P^T_{\mu\nu} +\Pi^L(q)P^L_{\mu\nu}\\
 P_{ij}^T &=& \delta_{ij}-\hat{q}_i\hat{q}_j \, , \hspace{1cm}
 P_{00}^T = P_{0i}^T = 0\, ,    \nonumber \\
 P_{\mu\nu}^L &=& -g_{\mu\nu}+\frac{q_\mu q_\nu}{q^2}
   -P_{\mu\nu}^T \, . \nonumber 
\eea
We will consider the polarization function in the limit 
of weak coupling ($g<1$), and for $\omega\ll q\ll T$, where 
$\omega,q$ are the energy and momentum transfer. We find
\bea
\label{Pi_L}
\Pi^{ab\, L}(q) &=& \frac{\delta^{ab}}{{\bf q}^2+m_D^2}\, , \\
\label{Pi_T}
\Pi^{ab\, T}(q) &=& \frac{\delta^{ab}}{{\bf q}^2
         -i\frac{\pi}{4}m_D^2\frac{\omega}{|{\bf q}|}} \, , 
\eea
where 
\be
\label{m_D}
m_D^2 = g^2T^2 \left(1+\frac{N_f}{6}\right)
\ee
is called the Debye mass. The longitudinal term governs the 
color-Coulomb interaction between static charges. We observe 
that the Coulomb interaction is screened at distances $r\sim m_D^{-1}
\sim 1/(gT)$. In perturbation theory the static magnetic interaction 
is unscreened \cite{Baym:1990uj}, but non-static magnetic interactions 
are dynamically screened at a distance $r\sim (m_D^2\omega)^{-1/3}$. 
This phenomenon, known as Landau damping, is due to 
the coupling of gluons to particle-hole (or particle-anti-particle)
pairs, and also play a role in electro-magnetic plasmas. 
Unlike classical plasmas the QCD plasma has a non-perturbative 
static magnetic screening mass $m_M\sim g^2 T$. This is 
the scale that determines the non-perturbative $g^6$ term
in the pressure. Modes below the magnetic screening scale 
contribute
\be 
\label{P_mag}
 P \sim T \int^{m_M} d^3k \sim g^6T^4\, .
\ee
The gluon polarization tensor also determines the propagation
of gluonic modes. For this purpose we need the full energy and 
momentum dependence of $\Pi^{T,L}$, see \cite{LeBellac:1996}.
For momenta $q\gg gT$ there are two transverse modes with dispersion
relation $\omega\simeq q$. For momenta $q< gT$ there are two 
transverse and one longitudinal mode. The longitudinal mode 
is sometimes called the plasmon. The energy of all three
modes approaches $\omega=\omega_p=m_D/\sqrt{3}$ as $q\to 0$. 
The quantity $\omega_p$ is known as the plasma frequency. 
The gluon (and plasmon) decay constant in the limit $q\to 0$
is \cite{Braaten:1990it}
\be
\label{gam_plas}
\gamma = 6.64 \frac{g^2N_c T}{24\pi} \, . 
\ee
An important issue is how small the coupling has to be in order 
for the perturbative estimates to be applicable. The convergence 
properties of the weak coupling expansion for the pressure are 
extremely poor. The series shows no signs of converging unless
the coupling is taken to be much smaller than one, $g\ll 1$,
corresponding to completely unrealistic temperatures on the order 
of 1 TeV. The problem is mostly due to the non-analytic terms in
the expansion, and convergence can be improved significantly
by considering resummation schemes or self-consistent quasi-particle
expansions \cite{Blaizot:2003tw}. Convergence can also be 
improved by using a hierarchy of effective field theories
for the hard ($p\sim T$), electric $(p\sim gT$), and magnetic
($p\sim g^2 T$) sectors of the QCD plasma \cite{Braaten:1995cm}.
Ordinary perturbation theory corresponds to treating the hard
and the electric sector perturbatively, but convergence can be 
improved by treating both the electric and the magnetic 
sector non-perturbatively \cite{Hietanen:2008tv}.

 Despite these advances accurate results at temperatures that 
can be reached in heavy ion collisions at RHIC have to rely on 
numerical simulations of the QCD partition function on a space-time 
lattice, see \cite{Laermann:2003cv} for a review. Lattice simulations 
with realistic quark masses find a phase transition at the critical 
temperature $T_c=192(8)$ MeV \cite{Karsch:2007vw}. The transition is 
a rapid (but smooth) crossover from a low temperature phase that 
exhibits chiral symmetry breaking and confinement to a chirally
restored and deconfined high temperature phase\footnote{This issue
is not completely settled. Aoki et al.~find distinct crossover
transitions at significantly lower temperatures, $T_{\chi}=151$ MeV 
for chiral symmetry restoration, and $T_{dec}=175$ MeV for deconfinement
\cite{Aoki:2006br}.}. The energy density reaches about 85\% of the 
ideal gas value at $T\simeq 2T_c$ and then evolves very slowly towards 
the non-interacting limit. 

 Below the critical temperature the degrees of freedom are hadrons. 
The lightest hadrons are pions, which are the Goldstone bosons
associated with the spontaneous breaking of the chiral 
symmetry of the QCD lagrangian. 
We can view pions as a spin-isospin sound wave that 
propagates in the QCD vacuum. Because quarks are not massless
the chiral symmetry is not exact, and pions have non-zero masses.
The masses of the charged and neutral pions are $m_{\pi^\pm}=139$ 
MeV and $m_{\pi^0}=135$ MeV. The lightest particle which 
is not a type of sound wave is the rho meson, with a mass of 
770 MeV. Chiral symmetry constrains the pion scattering amplitudes.
As in the case of phonons, these constraints are obtained 
most easily from the low energy effective chiral lagrangian. 
At leading order we have
\be
\label{l_chpt}
{\cal L} = \frac{f_\pi^2}{4} {\rm Tr}\left[
 \partial_\mu\Sigma\partial^\mu\Sigma^\dagger\right] 
 +\Big[ B {\rm Tr}(M\Sigma^\dagger) + h.c. \Big]
+ \ldots \, , 
\ee
where $\Sigma=\exp(i\phi^a\lambda^a/f_\pi)$ ($a=1,\ldots,8$) is the 
chiral field, $f_\pi=93$ MeV is the pion decay constant, $B$ is 
proportional to the quark condensate, and $M={\rm diag}(m_u,m_d,m_s)$ 
is the mass matrix.
We note that $f_\pi$ can be viewed as the stiffness of 
the QCD vacuum, 
\be 
 f_\pi^2 = \frac{2m_q}{m_\pi^2}
   \frac{\partial P_{\it vac}}{\partial m_q}\, ,
\ee
where $P_{\it vac}\simeq 0.5\,{\rm GeV}/{\rm fm}^3$ is the vacuum
pressure, and $m_q=(m_u+m_d)/2$. This result follows from the 
Gell-Mann-Oakes-Renner relation $m_\pi^2f_\pi^2=(m_u+m_d)\langle
\bar\psi\psi\rangle$ together with $(\partial P_{\it vac})/(\partial m)=
\langle\bar\psi\psi\rangle$.
An expansion of $\Sigma$ in powers of the field $\phi^a$ 
determines the interaction between pions. Restricting ourselves 
to the $SU(2)$ flavor sector (pions only) we get 
\be 
\label{l_chpt_comp}
{\cal L} =\frac{1}{2}(\partial_\mu\phi^a)^2
-\frac{1}{2}m_\pi^2 (\phi^a)^2
+\frac{1}{6f_\pi^2}\left[ (\phi^a\partial_\mu \phi^a)^2
  -(\phi^a)^2(\partial_\mu\phi^b)^2 \right] + \ldots \, ,
\ee
where $\phi^a$ ($a=1,2,3$) is the pion field. 
This result is clearly analogous to the phonon interaction 
in equ.~(\ref{ph_vert}). There are, however, some minor differences.
Because of parity and isospin symmetry there are no vertices with 
an odd number of pions. We also note that the leading four-pion
interaction has two derivatives, while the four-phonon term
involves four derivatives.

\begin{figure}
\begin{center}
\includegraphics[width=6.5cm]{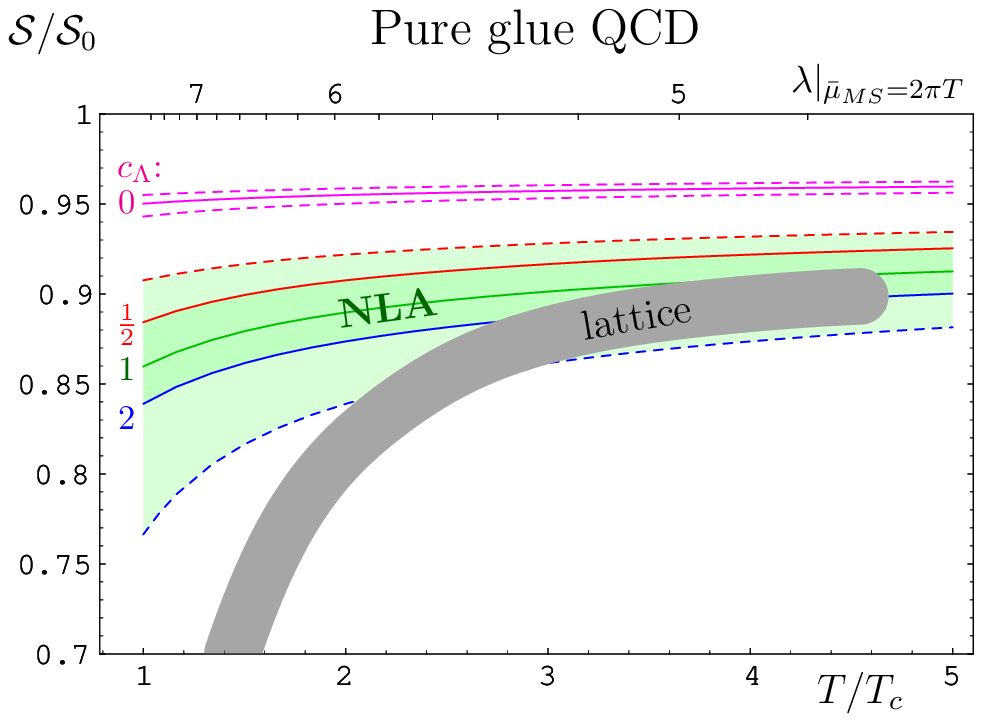}
\includegraphics[width=6.2cm]{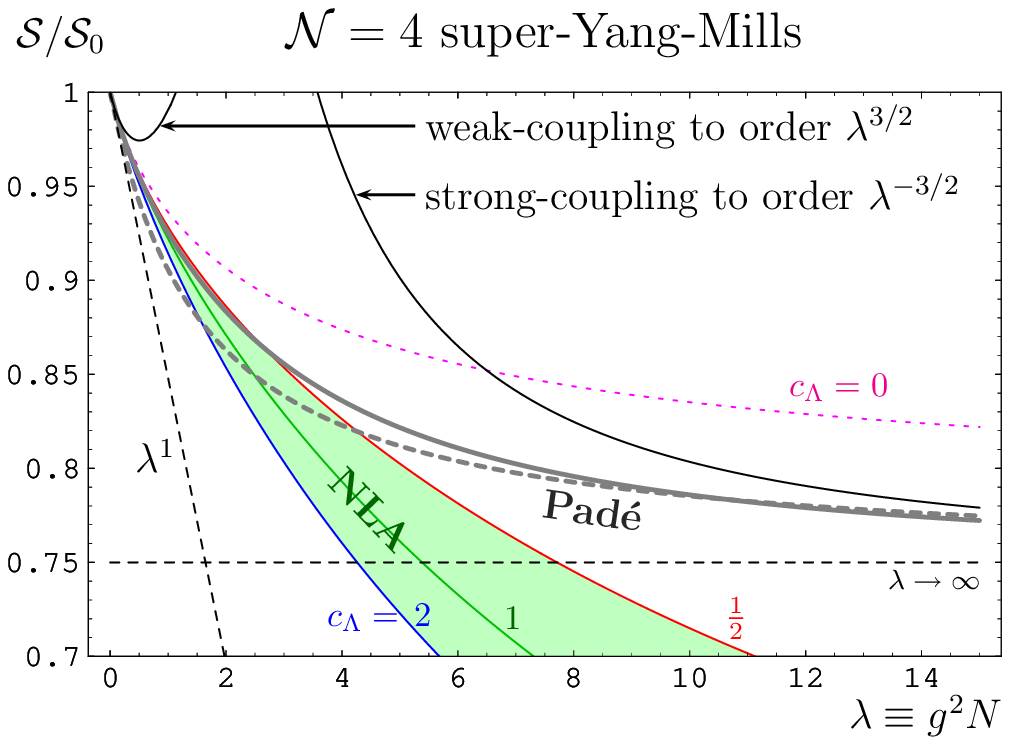}
\end{center}
\caption{\label{fig_eos_qcd}
Entropy density in units of the Stefan-Boltzmann value for pure 
gauge QCD and ${\cal N}=4$ supersymmetric QCD. The left panel
shows the entropy density of pure gauge QCD as a function of 
$T/T_c$. The grey band is the lattice result. The solid lines
show a resummed QCD calculation \cite{Blaizot:2003tw}. The different 
lines correspond to different choices for a non-perturbative 
parameter $c_\Lambda$. The dashed lines mark an error band
determined by variations in the QCD renormalization scale. The 
right panel shows the entropy density of SUSY QCD as a function
of the 't Hooft coupling $\lambda$. The curves are labeled as
in the left  panel.   }
\end{figure}

\subsection{Gauge theories: Superconformal QCD}
\label{sec_susy}

 QCD is a complicated theory, and a significant amount of
effort has been devoted to the study of generalizations of 
QCD that possess a larger amount of symmetry, in particular
supersymmetry. 
Supersymmetry is a symmetry that relates bosonic
and fermionic fields. The simplest supersymmetric cousin
of QCD is SUSY gluodynamics, a theory of gluons and massless 
fermions in the adjoint representation of the color group called 
gluinos. Theories with more supersymmetry involve extra
fermions and colored scalar fields. The most supersymmetric 
extension of QCD is a theory with four supersymmetries, 
called ${\cal N}=4$ SUSY QCD. Theories with even more 
supersymmetry contain fields with spin 3/2 and 2, and 
therefore involve gravitational interactions. These
theories are known as supergravity.

 The lagrangian of ${\cal N}=4$ SUSY QCD is
\be
\label{l_sqcd}
 {\cal L } = - \frac{1}{4} G_{\mu\nu}^a G_{\mu\nu}^a
 -i\bar\lambda^a_i\sigma^\mu D_\mu \lambda^a_i
 + D^\mu\phi^{\dagger\, a}_{ij}D_\mu\phi^a_{ij}
 + {\cal L}_{\lambda\lambda\phi} + {\cal L}_{\phi^4} \, ,
\ee
where $G_{\mu\nu}^a$ is the usual field strength tensor, 
$\lambda^{a}_i$ is the gluino field, and $\phi^a_{ij}$ is a colored
Higgs field. The gluino is a two-component (Weyl) fermion in 
the adjoint representation of the color group. The index $i$
($i=1,\ldots,4$) transforms in the fundamental representation
of a global $SU(4)_R$ ``R-symmetry''. 
This symmetry interchanges the bosons and fermions
that are related by the four supersymmetries, and is analogous
to the flavor symmetry of QCD.
The Higgs is a scalar 
field in the adjoint representation of color, and in an 
anti-symmetric tensor (six dimensional) of $SU(4)_R$. Note 
that the total number of fermionic fields, $8(N_c^2-1)$, is 
indeed equal to the number of bosonic fields. We have not 
explicitly displayed the Yukawa couplings ${\cal L}_{\lambda
\lambda\phi}$ and Higgs self couplings ${\cal L}_{\phi^4}$, see
\cite{Terning:2006}. Both interaction terms only involve the
dimensionless gauge coupling $g$. 

 ${\cal N}=4$ SUSY QCD has a vanishing beta function and is 
believed to be a conformal field theory (CFT). As a consequence
there is no dimensional transmutation, no confinement or 
spontaneous symmetry breaking, and no phase transition. The
theory is an Coulomb phase for all values of the coupling 
$g$ and the temperature $T$. However, if $g$ is not small
then there is no obvious way to compute thermodynamic or 
transport properties of the plasma.

  An interesting new approach is provided by the duality 
between strongly coupled large $N_c$ gauge theory and weakly 
coupled string theory on ${\rm AdS}_5\times {\rm S}_5$ discovered 
by Maldacena \cite{Maldacena:1997re}. We will have more to say 
about this approach in Section \ref{sec_ads}. For now we observe 
that the correspondence can be extended to finite temperature. 
In this case the relevant configurations is an ${\rm AdS}_5\times
{\rm S}_5$ black hole. The temperature of the gauge theory is given 
by the Hawking temperature of the black hole, and the entropy 
is given by the Hawking-Beckenstein formula $S=A/(4G)$, where 
$A$ is the surface area of the event horizon and $G$ is Newton's
constant.

 The AdS/CFT correspondence makes predictions for the 
thermodynamics of the gauge theory in the limit of a large 
number of colors, $N_c\to\infty$. The perturbative expansion
of a $SU(N_c)$ Yang-Mills theory involves the 't Hooft coupling
$\lambda=g^2N_c$. In the weak coupling limit we take $N_c\to
\infty$ with $\lambda={\it const}$ and $\lambda\ll 1$. Using 
the AdS/CFT correspondence we can also study the strong 
coupling limit $N_c\to\infty$ with $\lambda\gg 1$. ${\cal N}=4$
SUSY QCD is a conformal theory and scale invariance implies 
that $\epsilon=3P$ as well as $s=4P/T$, where $\epsilon$ is the 
energy density and $s$ is the entropy density. By dimensional 
analysis the entropy density of the interacting system must be 
proportional to the entropy density $s_0$ of the free 
system. The weak and strong coupling expansions for $s/s_0$ are
\cite{Gubser:1998nz,Fotopoulos:1998es,Nieto:1999kc}
\be 
\label{s_susy_ym}
\frac{s}{s_0} = \left\{
\begin{array}{ll}
\frac{3}{4} +  \frac{45\zeta(3)}{32}\lambda^{-3/2} + \ldots 
              & \lambda\gg 1\, ,  \\
 1 -\frac{3}{2\pi^2}\lambda + \frac{\sqrt{2}+3}{\pi^3}\lambda^{3/2}           
   + \ldots   & \lambda\ll 1\, . 
\end{array}\right.  
\ee
This result has a number of remarkable features. First we observe
that the entropy density at infinite coupling only differs by a
factor 3/4 from the result in the free theory. We also note that 
the first non-trivial corrections in the strong and weak coupling 
limit are consistent with the idea that the evolution from weak
to strong coupling is smooth. Equation (\ref{s_susy_ym}) was
compared with resummed perturbation theory and Pade approximants
in \cite{Blaizot:2006tk}, see Fig.~\ref{fig_eos_qcd}. The authors 
argue that at the ``QCD-like'' point $s/s_0=0.85$ neither the strong 
nor the weak coupling expansion are quantitatively reliable, but 
that resummed perturbation theory is useful in this regime.

\section{Transport theory}
\label{sec_transp}

 In this section we summarize theoretical approaches to transport 
phenomena in strongly coupled quantum fluids. The most general of
these approaches is hydrodynamics. Hydrodynamics is based on the 
observation that correlation functions at low energy and small
momentum are governed by the evolution of conserved charges.
Conservation laws imply that the densities of conserved charges
cannot relax locally, but have to propagate or diffuse out to
large distance. This corresponds to hydrodynamic excitations with 
dispersion laws of the form $\omega\sim q$ (sound) or $\omega\sim
iq^2$ (diffusion).

 Hydrodynamics can be developed as an expansion in 
derivatives of the fluid velocity and the thermodynamic variables. 
The leading order theory, called ideal hydrodynamics, only depends 
on the equation of state, and is exactly time-reversible. The next 
order theory, (first order) viscous hydrodynamics, involves a new 
set of parameters called transport coefficients, and describes 
dissipative, time-irreversible phenomena. The validity of 
hydrodynamics is controlled by the relative size of the 
next-to-leading order terms. If dissipation is dominated by 
shear viscosity\footnote{If dissipation is dominated by 
heat transport, then the expansion parameter is $1/({\it Re}
\cdot{\it Pr})$, where ${\it Pr}$ is the Prandtl number
defined in equ.~(\ref{Prandtl}).}
then the expansion parameter is $1/{\it Re}$, where ${\it Re}$
is the Reynolds number defined in equ.~(\ref{Reynolds}).

 The values of the transport coefficients can be extracted 
from experiment, or computed from an underlying field theory. 
The connection between transport coefficients and correlation
functions in a (quantum) field theory is provided by linear
response theory. Using linear response theory one can relate
transport coefficients to the zero energy and zero momentum
limit of a retarded correlation function. These relations are
known as Kubo formulas, see Sect.~\ref{sec_kubo}.

 Calculations based on the Kubo formula are difficult, 
in particular if the interaction is not weak.
The situation simplifies if the system allows a microscopic description 
in terms of quasi-particles. In that case we can use an intermediate 
effective theory, known as kinetic theory, to relate the microscopic 
lagrangian to the hydrodynamic description. 
Kinetic theory also provides a more microscopic criterion
for the applicability of hydrodynamics. Using the kinetic estimate
for the shear viscosity in equ.~(\ref{eta_mfp}) we get $1/{\it Re}
\sim (v/c_s)(l_{\it mfp}/L)$ where $c_s$ is the speed of sound
and the ratio
\be 
\label{Knudsen}
{\it Kn} = \frac{l_{\it mfp}}{L} 
\ee
is called the Knudsen number. Hydrodynamics is valid if the 
mean free path is much smaller than the characteristic size, 
and the Knudsen number is small. 

The calculation of 
transport coefficients in kinetic theory is reviewed in 
Sect.~\ref{sec_kin}. If the interaction between quasi-particles
is strong then the kinetic description breaks down. A new approach 
to extracting transport properties from a strongly coupled field 
theory is the holographic method which we will discuss in the next 
section. Using holography the calculation of the retarded correlator
can be reduced to a classical computation in a suitable dual 
theory.

\subsection{Hydrodynamics}
\label{sec_hydro}

\subsubsection{Non-relativistic fluids}
\label{sec_nr_fluid}

 The hydrodynamics of a one-component non-relativistic 
fluid is governed by the conservation laws of energy, mass (particle 
number), and momentum, 
\bea
\label{e_cons}
\frac{\partial \epsilon}{\partial t} 
 + {\bm \nabla}\cdot{\bf j}^{\; \epsilon} &=& 0 \, ,\\
\label{rho_cons}
\frac{\partial \rho}{\partial t} 
 + {\bm \nabla} \cdot{\bf g} &=& 0 \, , \\
\label{g_cons}
\frac{\partial g_i}{\partial t } 
 + \nabla_j \Pi_{ij} &=& 0  \, . 
\eea
Here, $\epsilon$ is the energy density, $\rho$ is the mass 
density, ${\bf g}$ is the momentum density, and $\Pi_{ij}$ is 
the stress tensor. The relations between the conserved currents 
and the hydrodynamic variables are called constitutive relations. 
These relations can be determined order by order in an expansion 
in derivatives of the flow velocity and the thermodynamic 
variables. The leading order result is called ``ideal 
hydrodynamics''. At this order the constitutive relations are 
completely fixed by Galilean invariance, rotational invariance, 
and conservation of entropy. The result is 
\bea
\label{jeps_nrfl}
 {\bf j}^{\;\epsilon} &=& {\bf v}(\epsilon+P)\, , \\
\label{g_nrfl} 
 {\bf g}  &=& \rho{\bf v} \, ,  \\
\label{pij_nrfl}
 \Pi_{ij} &=& P\delta_{ij} +\rho v_i v_j\, ,  
\eea
where $\epsilon=\epsilon_0+\frac{1}{2}\rho v^2$ and $\epsilon_0$ is 
the energy density in the rest frame of the fluid. There are six 
hydrodynamic variables, ${\bf v}$, $\rho$, $\epsilon$ 
and $P$, which are determined by the five conservation laws
(\ref{e_cons}-\ref{g_cons}). In order for the equations to 
close we need to supply an equation of state $P=P(\epsilon,\rho)$.
Since hydrodynamic variables evolve slowly, the equation of 
state is the one in thermal equilibrium. 

In ideal hydrodynamics the equations of continuity and momentum 
conservation are 
\bea 
\label{cont_nrfl}
 \frac{\partial\rho}{\partial t}  
 + {\bm \nabla}\cdot\left(\rho{\bf v}\right) &=& 0 \, , \\
\label{euler}
 \frac{\partial{\bf v}}{\partial t} + 
  \left({\bf v}\cdot{\bm \nabla}\right) {\bf v} 
 &=& -\frac{1}{\rho}{\bm \nabla}P \, .
\eea
The equation of momentum conservation is known as the Euler equation.
In the case of ideal hydrodynamics the equation of energy conservation
can be rewritten as conservation of entropy,
\be 
 \frac{\partial s}{\partial t} 
  + {\bm \nabla}\cdot \left({\bf v} s\right) =0 \, . 
\ee
At next order in the derivative expansion dissipative terms appear. The
size of these terms is controlled by new parameters called transport 
coefficients. The relation ${\bf g}=\rho{\bf v}$ is not modified 
(it follows from Galilean invariance), but two new coefficients 
appear in the stress tensor. We can write $\Pi_{ij}=P\delta_{ij}
+\rho v_iv_j+\delta\Pi_{ij}$ with  
\be
\label{del_pij_nrfl}
\delta\Pi_{ij} = -\eta\left(\nabla_iv_j-\nabla_jv_i-
 \frac{2}{3}\delta_{ij}{\bm \nabla}\cdot{\bf v} \right)
 - \zeta \delta_{ij}\left({\bm \nabla}\cdot{\bf v}\right)\, . 
\ee
Here, $\eta$ is the shear viscosity and $\zeta$ is the bulk viscosity. 
The correction to the energy current has the form $j_i^{\epsilon}= 
v_i(\epsilon+P)+v_j\delta\Pi_{ij}+Q_i$ with 
\be
\label{del_q_nrfl}
 {\bf Q} = -\kappa {\bm \nabla}T ,  
\ee
where $T$ is the temperature and $\kappa$ is the thermal conductivity.
The second law of thermodynamics implies that $\eta,\zeta,\kappa\geq 0$. 
The equation of momentum conservation with the viscous stresses 
(\ref{del_pij_nrfl}) included is known as the Navier-Stokes equation. 

 The linearized hydrodynamic equations describe the propagation of 
sound and diffusive modes. In the case of a non-relativistic fluid
there is a pair of sound modes that couple to the pressure/density 
and the longitudinal velocity, a pair of diffusive shear modes that 
couple to the transverse velocity, and a diffusive heat mode. The
longitudinal and transverse components of the velocity are defined
by ${\bf v}={\bf v}^{\,L}+{\bf v}^{\,T}$ with ${\bm \nabla}\times 
{\bf v}^{\,L}=0$ and ${\bm \nabla}\cdot{\bf v}^{\,T}=0$. The 
hydrodynamic modes govern the hydrodynamic correlation functions. 
The transverse velocity correlation function is defined by
\be
\label{S_vv_def}
 S^{vv}_{ij}(\omega,{\bf k}) = 
 \langle \delta v^T_i \delta v^T_j\rangle_{\omega,{\bf k}}
 = \int d^3x\,dt\; e^{i(\omega t-{\bf k}\cdot{\bf x})}
 \langle \delta v^T_i({\bf x},t) \delta v^T_j(0,0) \rangle \, , 
\ee 
where $\delta v^T_i({\bf x},t)=v^T_i({\bf x},t)-\langle v^T_i({\bf x},t)
\rangle$ is a fluctuation of the velocity. Linearized hydrodynamics 
gives
\be 
\label{S_vv}
 S^{vv}_{ij}(\omega,{\bf k}) = 
  \frac{2T}{\rho} \left( \delta_{ij} -\hat{k}_i\hat{k}_j \right)
 \frac{\nu {\bf k}^2}{\omega^2+\nu^2{\bf k}^4}\, ,
\ee
where $\nu=\eta/\rho$ is the kinetic viscosity. The dependence of 
the correlation function on $\omega$ and ${\bf k}$ is determined
by the laws of hydrodynamics, equ.~(\ref{e_cons}-\ref{g_cons}), 
and the overall normalization is fixed by the thermodynamic
relation
\be 
\langle \delta v_i({\bf x},t) \delta v_j({\bf x}',t)\rangle = 
 \frac{T}{\rho}\delta_{ij} \delta({\bf x}-{\bf x}')\, . 
\ee
We observe that the transverse velocity correlation function
has a diffusive pole, where the diffusion constant is given 
by the kinematic viscosity. The entropy correlation function 
has a diffusive pole governed by the thermal diffusion 
constant $\chi=\kappa/(c_p\rho)$, where $c_p$ is the specific 
heat at constant pressure. The correlation function is  
\be 
\label{S_ss}
S^{ss}(\omega,{\bf k}) = 
 \langle \delta s \delta s\rangle_{\omega,{\bf k}}
 =  \frac{2c_p}{\rho} 
    \frac{\chi {\bf k}^2}{\omega^2+\chi^2 {\bf k}^4} \, .
\ee
The pressure correlation function contains the sound pole and 
is given by 
\be 
\label{S_pp}
S^{pp}(\omega,{\bf k}) = 
 \langle \delta p \delta p\rangle_{\omega,{\bf k}}
 = 4\rho T c_s^3 \; 
 \frac{\gamma c_s^2 {\bf k}^2 +\gamma_T (\omega^2-c_s^2 {\bf k}^2)}
      {(\omega^2-c_s^2{\bf k}^2)^2+4\gamma^2c_s^2\omega^2}\, , 
\ee
where $c_s=[(\partial P)/(\partial \rho)|_s]^{1/2}$ is the speed of 
sound and $\gamma= \gamma_{\eta,\zeta} + \gamma_T$ is the coefficient 
of sound absorption (the inverse sound attenuation length). The 
contributions to $\gamma$ from viscosity and thermal conductivity
are given by
\be
 \gamma_{\eta,\zeta} = \frac{{\bf k}^2}{2\rho c_s} 
   \left( \zeta + \frac{4}{3}\eta \right)\, , \hspace{0.5cm}
 \gamma_T = \frac{{\bf k}^2 c_s\rho}{2T} \, \chi 
    \left( \frac{\partial T}{\partial P}\right)_s\, .
\ee
These results illustrates the criterion for the validity 
of hydrodynamics given above. Hydrodynamics is based on a small 
momentum expansion. Applied to equ.~(\ref{S_pp}) this implies that 
$\omega\sim c_sk\gg k^2\eta/\rho$. Taking the characteristic 
size to be $L\sim 1/k$ this is equivalent to $\eta/(c_s\rho L)\ll 1$. 
A more microscopic criterion follows from the kinetic estimate of 
the shear viscosity given in equ.~(\ref{eta_mfp}): Hydrodynamics 
describes sound waves with a wave length that is large compared 
to the mean free path: $L\gg l_{mfp}$.

\subsubsection{Superfluid hydrodynamics}
\label{sec_sfl_hydro}

 Superfluidity is characterized by the spontaneous breakdown of
the $U(1)$ symmetry associated with the conserved particle number.
By Goldstone's theorem the spontaneous breaking of a continuous
symmetry leads to the appearance of a gapless mode. This mode
has to be included in the hydrodynamic description of the 
system. We introduced the Goldstone boson field $\varphi$ in 
equ.~(\ref{phi_gb}). The quantity ${\bf v}_s={\bm \nabla}\varphi/m$
can be interpreted as the superfluid velocity. Since ${\bf v}_s$
is the gradient of a phase the superfluid velocity is irrotational, 
${\bm \nabla}\times{\bf v}_s=0$.

 We have to generalize the constitutive equations to include both
the normal fluid velocity ${\bf v}_n$ and the superfluid velocity 
${\bf v}_s$. In the ideal fluid case (no dissipation) the result 
is completely fixed by Galilean invariance and thermodynamic
relations. The constitutive equations are
\bea 
\label{g_sfl}
 {\bf g} &=& \rho_n{\bf v}_n + \rho_s {\bf v}_s\, , \\
 \Pi_{ij} &=& P\delta_{ij} + \rho_nv_{n,i}v_{n,j} 
    + \rho_s v_{s,i}v_{s,j} \, , \\
\label{j_sfl}
 {\bf j}^{\; \epsilon} &=&  \rho sT {\bf v}_n 
    + \left(\mu+\frac{1}{2}{\bf v}_s^2\right)
       \left(\rho_n{\bf v}_n+\rho_s {\bf v}_s\right)
    + \rho_n{\bf v}_n  {\bf v}_n\cdot \left({\bf v}_n-{\bf v}_s\right)
    \, , 
\eea
where $\rho_n$ and $\rho_s$ are the normal and superfluid
density of the system. The total density $\rho=\rho_n+\rho_s$
is the sum of the normal and superfluid contributions. The
ratio $\rho_s/\rho$ is a function of the temperature, the chemical 
potential, and the relative velocity $|{\bf v}_n-{\bf v}_s|$. This 
function, like the equation of state $P(\mu,T,|{\bf v}_n-{\bf v}_s|)$,
depends on microscopic details.  The conservation laws are given by 
equ.~(\ref{e_cons}-\ref{g_cons}). These equations have to be 
supplemented by an equation of motion for the superfluid velocity. 
Landau showed that Euler's equation for the superfluid velocity 
is given by \cite{Landau:fl}
\be 
\label{vs_euler}
\frac{\partial {\bf v}_s}{\partial t} 
 + ({\bf v_s}\cdot {\bm \nabla}) {\bf v}_s = -{\bm \nabla} \mu \,. 
\ee
Because ${\bf v}_s$ is irrotational 
we can rewrite the convective derivative on the LHS of equ.~(\ref{vs_euler}) 
as a total derivative, ${\bf v}_s\cdot {\bm \nabla} {\bf v}_s =
\frac{1}{2} {\bm \nabla} ({\bf v}_s^{\,2})$. 

 As in the case of a normal fluid we may consider dissipative
corrections to the constitutive equations. The form of these
terms is constrained by rotational and Galilean invariance, 
and by the second law of thermodynamics. The viscous corrections
to the energy momentum tensor are  
\bea
\label{del_pij_sfl}
\delta\Pi_{ij} &=& -\eta\left(\nabla_iv_{n,j}+\nabla_jv_{n,i}-
 \frac{2}{3}\delta_{ij}{\bm \nabla}\cdot {\bf v}_{n}\right) \\
 & & \mbox{}
 - \delta_{ij}\Big(
     \zeta_1 {\bm \nabla}\cdot 
           \left(\rho_s\left({\bf v}_{s}-{\bf v}_{n}\right)\right)
   + \zeta_2 \left({\bm \nabla}\cdot {\bf v}_{n}\right)
  \Big)\, .  \nonumber 
\eea
We observe that viscous shear stresses only arise from the 
normal component of the flow. In addition to the normal bulk
viscosity term proportional to $\zeta_2$ there is a second
contribution that involves the relative motion of the normal 
and superfluid components. Two additional bulk viscosities appear
in the dissipative correction to the RHS of equ.~(\ref{vs_euler}).
We replace ${\bm \nabla}\mu$ by ${\bm \nabla}(\mu+H)$ with
\be
\label{del_H_sfl}
 H = -\zeta_3 {\bm \nabla}\cdot
     \left(\rho_s\left({\bf v}_s-{\bf v}_n\right)\right)
   - \zeta_4 {\bm \nabla}\cdot {\bf v}_n
  \, .
\ee
Onsager's symmetry principle requires that $\zeta_4=\zeta_1$.
The dissipative correction to the energy current is $\delta 
j^{\epsilon}_i=v_{n,j}\delta\Pi_{ij}+\rho_s(v_{s,i}-v_{n,i})H
+Q_i$ where $Q_i=-\kappa\nabla_i T$ as in the case of a normal
fluid. 

 Superfluid hydrodynamics contains two velocity fields, the 
normal flow velocity ${\bf v}_n$ and the superfluid (irrotational)
flow velocity ${\bf v}_s$. This extra degree of freedom leads to 
an additional sound mode called second sound. The velocity of 
second sound depends strongly on temperature and vanishes at
the critical temperature where $\rho_s/\rho\to 0$. If thermal 
expansion can be neglected second sound is an oscillatory motion 
of the superfluid against the normal fluid which does not lead 
to any mass transport and can be viewed as a pure entropy wave.

\subsubsection{Relativistic fluids}
\label{sec_rel_fluid}

 In a relativistic fluid the equations of energy and momentum 
conservation can be written as a single equation 
\be
\label{rel_hydro}
 \partial_\mu T^{\mu\nu}= 0 \, ,
\ee
where $T^{\mu\nu}$ is the energy momentum tensor. In ideal fluid
dynamics the form of $T_{\mu\nu}$ is completely fixed by 
Lorentz invariance,
\be 
\label{T_ideal}
 T^{\mu\nu} = (\epsilon+P)u^\mu u^\nu + P \eta^{\mu\nu}\, ,
\ee
where $u^\mu$ is the fluid velocity ($u^2=-1$) and $\eta^{\mu\nu}=
{\rm diag}(-1,1,1,1)$ is the metric tensor. In a relativistic 
theory there need not be a conserved particle number. If a 
conserved particle number, for example baryon number, exists 
then there is a second hydrodynamic equation that express 
particle number conservation
\be 
\label{n_cons}
 \partial_\mu ( n u^\mu) = 0\, , 
\ee
where $n$ is the particle density. As in the non-relativistic
case the hydrodynamic equations have to be supplemented by an
equation of state $P=P(\epsilon)$ or $P=P(\epsilon,n)$. The 
four equations given in equ.~(\ref{rel_hydro}) can be split
into two sets using the longitudinal and transverse projectors 
\be 
\Delta^{||}_{\mu\nu}=-u_\mu u_\nu, \hspace{1cm}
\Delta_{\mu\nu} = \eta_{\mu\nu}+u_\mu u_\nu\, . 
\ee
With the help of the thermodynamic relations $d\epsilon=Tds$ 
and $\epsilon+P=sT$ the longitudinal equation is equivalent
to entropy conservation
\be 
\label{rel_s_cont}
 \partial_\mu (su^\mu)=0\, ,
\ee
and the transverse equation is the relativistic Euler equation
\be 
\label{rel_euler}
 Du_\mu = -\frac{1}{\epsilon+P}\nabla_\mu^\perp P\, , 
\ee
where $D=u\cdot \partial$ and $\nabla_\mu^\perp=\Delta_{\mu\nu}
\partial^\nu$. Comparison with equ.~(\ref{euler}) shows that the 
inertia of a relativistic fluid is governed by $\epsilon+P$.

The form of the dissipative terms depends on the precise definition
of the fluid velocity. A useful choice is to define $u^\mu$ by the 
requirement that in the local rest frame $T^{00}=\epsilon$ and 
$T^{0i}=0$. This definition is called the Landau frame \cite{Landau:fl}.
In this frame the dissipative correction to the energy momentum tensor 
in the rest frame has the same form as in the non-relativistic case, 
see equ.~(\ref{del_pij_nrfl}). 
We write the stress tensor as $T^{\mu\nu}=T^{\mu\nu}_{0}+
\delta^{(1)} T^{\mu\nu}+\delta^{(2)} T^{\mu\nu}+\ldots$, where 
$T^{\mu\nu}_{0}$ is the stress tensor of the ideal fluid given 
in equ.~(\ref{T_ideal}), $\delta^{(1)} T^{\mu\nu}$ is the 
first order viscous correction, etc. A covariant expression for
$\delta^{(1)}T^{\mu\nu}$ is
\be 
\delta^{(1)} T^{\mu\nu}= -\eta \sigma^{\mu\nu}
 - \zeta \Delta^{\mu\nu} \partial \cdot u \, 
\ee
where we have defined 
\be 
\label{sig_vis}
\sigma^{\mu\nu} = 
  \Delta^{\mu\alpha}\Delta^{\nu\beta} 
  \left(\partial_\alpha u_\beta+\partial_\beta u_\alpha 
  - \frac{2}{3}\eta_{\alpha\beta}\partial\cdot u \right) \, . 
\ee
The dissipative correction to the conserved particle current is
$j_\mu=nu_\mu+\delta j_\mu$ with
\be 
\label{del_j_mu}
\delta^{(1)} j_\mu= -\kappa \left(\frac{nT}{\epsilon+P}\right)^2
 \Delta_{\mu}^\perp \left(\frac{\mu}{T}\right)\, , 
\ee
where $\kappa$ is the thermal conductivity and $\mu$ is the 
chemical potential associated with the conserved density $n$. 
Alternatively, one can define the velocity via the conserved 
particle current (Eckart frame). In that case there is no dissipative 
contribution to $j_\mu$ and the thermal conductivity appears
in stress tensor. 

 The hydrodynamic equations determine the propagation of 
sound and diffusive modes. We consider the case without a 
conserved particle number. In this case all the modes can 
be found by considering correlation functions of the 
energy-momentum current $g^i=T^{0i}$. The longitudinal 
and transverse correlation functions are
\bea
\label{S_gg_L}
S_{gg}^L(\omega,{\bf k}) &=& 2sT \, 
  \frac{\Gamma_s\omega^2{\bf k}^2}
       {(\omega^2-c_s^2{\bf k}^2)^2 + (\Gamma_s\omega{\bf k}^2)^2} \, , \\
\label{S_gg_T}
S_{gg}^T(\omega,{\bf k}) &=& 
    \frac{2\eta{\bf k}^2}
         {\omega^2+(\frac{\eta}{sT}{\bf k}^2)^2}\, .
\eea 
As in the non-relativistic fluid we find a pair of sound waves, 
and a pair of diffusive shear modes. The sound attenuation length 
is given by
\be 
\label{Gam_s}
 \Gamma_s = \frac{\frac{4}{3}\eta+\zeta}{sT}\, , 
\ee
and the analog of the kinematic viscosity is the ratio $\eta/(sT)$. 

 An new issue that arises in viscous relativistic hydrodynamics
is the apparent lack of causality of the equations of motion. 
The problem can be seen by inspecting the linearized equation
for the diffusive shear mode. The equation is first order in time,
but second order in spatial gradients. As a result discontinuities
in the initial conditions can propagate with infinite speed. This 
is not really a problem of the hydrodynamic description -- the 
relevant modes are outside the domain of validity of hydrodynamics -- 
but the acausal modes cause difficulties in numerical implementations. 
To overcome these difficulties one can include second order gradient 
corrections in the stress tensor. The resulting theory is
called second order viscous hydrodynamics. One can shows that for
physically reasonable ranges of the second order coefficients the
theory is causal \cite{Romatschke:2009im}.
In general there are large number 
of second order terms. A complete classification of the second 
order terms in a relativistic conformal fluid was recently given 
in \cite{Baier:2007ix}. Conformal symmetry implies that $\zeta=0$ 
and $\delta^{(1)}T_{\mu\nu}=-\eta \sigma_{\mu\nu}$. The second order 
correction is 
\bea
\label{del_Pi_2}
 \delta^{(2)}T^{\mu\nu} &=&
 \eta\tau_{II} \left[ ^\langle D\sigma^{\mu\nu\rangle}
 +\frac{1}{3}\,\sigma^{\mu\nu} (\partial\cdot u) \right] \\
& & \hspace{0.2cm}\mbox{}
 +\lambda_1\sigma^{\langle\mu}_{\;\;\;\lambda} \sigma^{\nu\rangle\lambda}
 +\lambda_2\sigma^{\langle\mu}_{\;\;\;\lambda} \Omega^{\nu\rangle\lambda}
 +\lambda_3\Omega^{\langle\mu}_{\;\;\;\lambda} \Omega^{\nu\rangle\lambda}\, ,
\nonumber 
\eea
where $\sigma^{\mu\nu}$ is the first order shear tensor defined above,  
\be 
\label{sym_tr}
A^{\langle\mu\nu\rangle}= \frac{1}{2}\Delta^{\mu\alpha}
\Delta^{\nu\beta} \left( A_{\alpha\beta}+A_{\beta\alpha} - 
 \frac{2}{3}\Delta^{\mu\nu}\Delta^{\alpha\beta}A_{\alpha\beta}
\right)
\ee
denotes the transverse traceless part of $A^{\alpha\beta}$ and
\be 
\label{vort}
\Omega^{\mu\nu} = \frac{1}{2}\Delta^{\mu\alpha}
\Delta^{\nu\beta} \left(\partial_\alpha u_\beta - \partial_\beta u_\alpha
 \right)
\ee
is the vorticity. Equ.~(\ref{del_Pi_2}) defines four new second order 
transport coefficients, $\tau_{II}$ and $\lambda_{1,2,3}$. These 
coefficient can be determined using kinetic theory \cite{York:2008rr} 
or the AdS/CFT correspondence \cite{Baier:2007ix,Bhattacharyya:2008jc}.

 Equation (\ref{del_Pi_2}) is a constitutive relation that determines
the stress tensor in terms of thermodynamic variables. Formally, we 
may replace time derivatives by spatial derivatives using the lower
order equations of motion. Another option, inspired by the approach
of Israel and Stewart \cite{Israel:1979wp}, is to promote $\pi^{\mu\nu}
=\delta T^{\mu\nu}$ to a hydrodynamic variable. The equation of motion 
for $\pi^{\mu\nu}$ is 
\bea
\label{del_Pi_3}
 \pi^{\mu\nu} &=&
 -\eta\sigma^{\mu\nu} 
 -\tau_{II} \left[ ^\langle D\pi^{\mu\nu\rangle}
 +\frac{4}{3}\,\pi^{\mu\nu} (\partial\cdot u) \right] \\
& & \hspace{0.2cm}\mbox{}
 +\frac{\lambda_1}{\eta^2}
        \pi^{\langle\mu}_{\;\;\;\lambda} \pi^{\nu\rangle\lambda}
 -\frac{\lambda_2}{\eta}
        \pi^{\langle\mu}_{\;\;\;\lambda} \Omega^{\nu\rangle\lambda}
 +\lambda_3
        \Omega^{\langle\mu}_{\;\;\;\lambda} \Omega^{\nu\rangle\lambda}\, ,
\nonumber 
\eea
This equation describes the relaxation of $\pi^{\mu\nu}$ to the 
Navier-Stokes form $-\eta\sigma^{\mu\nu}$.
There are also a number of more phenomenological approaches that include 
some subset of higher order terms, for example the already mentioned
Israel-Stewart formalism \cite{Israel:1979wp} or the equations of 
Lindblom and Geroch \cite{Geroch:1990bw}, see \cite{Romatschke:2009im} 
for a review. We note that whatever formalism is used, a necessary 
condition for the applicability of second order hydrodynamics is that 
higher order corrections are small, $\delta^{(2)}T^{\mu\nu}\ll 
\delta^{(1)} T^{\mu\nu} \ll T^{\mu\nu}$.

\vspace*{0.3cm}
\noindent{\it Remarks:}
The second order formalism was initially developed for 
non-relativistic fluids by Burnett \cite{Burnett:1935,Burnett:1936},
see \cite{Garcia:2008} for a review. Higher order hydrodynamic
equations can be derived from kinetic theory by computing 
moments of the Boltzmann equation. This procedure is known 
as Grad's moment method \cite{Grad:1949}. It is not easy to find 
systems in which the second order theory provides a quantitative 
improvement over the Navier Stokes equation. An example is the 
work of Uhlenbeck, Foch and Ford on sound propagation in gases
\cite{Foch:1967,Foch:1970}. Finally, we note that relativistic 
superfluid hydrodynamics was formulated by Carter, Khalatnikov 
and Lebedev \cite{Khalatnikov:1982,Carter:1992}, see 
\cite{Son:2000ht,Son:2002zn,Mannarelli:2008jq} for more recent 
studies that emphasizes the connection to effective field theory. 

\subsection{Diffusion}
\label{sec_diff}

 An important diagnostic of the properties of a fluid is the 
diffusion of a dilute gas of impurities suspended in the fluid.
We will see, in particular, that if the fluid is composed
of quasi-particles then the diffusion of impurities and the shear 
viscosity, which is related to momentum diffusion, are closely 
linked. The two transport coefficients have the same dependence 
on the coupling constant, and their temperature dependence is 
the same up to kinematic factors. In a perfect fluid, however, 
this link may be broken: The diffusion constant goes to zero 
while the shear viscosity remains finite.

 We will assume that the number of impurity particles is conserved.
The number density satisfies the continuity equation 
\be 
\frac{\partial n}{\partial t} + {\bm \nabla}\cdot{\bf \jmath} = 0\, .
\ee
If the number density varies smoothly then the current ${\bf \jmath}$
can be expressed in terms of the thermodynamic variables. At leading 
order in derivatives of the density we can write $\vec\jmath=
-D{\bm \nabla}n$, where $D$ is the diffusion constant. Inserting 
this expression into the continuity equation gives the diffusion 
equation
\be 
\label{diff_equ}
\frac{\partial n}{\partial t}= D{\bm \nabla}^2 n\, . 
\ee
A more microscopic view of diffusion is provided by studying
the Brownian motion of an individual suspended particle. The
motion is described by a stochastic (Langevin) equation  
\be
\label{Langevin} 
 \frac{d{\bf p}}{dt} = -\eta_D {\bf p} +{\bf \xi}(t), 
 \hspace{1cm}
 \langle \xi_i(t)\xi_j(t')\rangle = \kappa\delta_{ij}\delta(t-t').
\ee
Here, ${\bf p}$ is the momentum of the particle, $\eta_D$ is the 
drag coefficient, and ${\bf \xi}(t)$ is a stochastic force. The 
coefficient $\kappa$ is related to the mean square momentum change 
per unit time, $3\kappa=\langle (\Delta{\bf p})^2\rangle/(\Delta t)$. 
The Langevin equation can be integrated to determine the mean 
squared momentum. In the long time limit ($t\gg \eta_D^{-1}$) the 
particle thermalizes and we expect that $\langle {\bf p}^{\, 2}\rangle 
= 3mT$. This requirement leads to the Einstein relation
\be
\label{Einstein} 
 \eta_D=\frac{\kappa}{2mT} . 
\ee
The relation between $\eta_D$ and the diffusion constant can be
determined from the mean square displacement. At late times
$\langle [\Delta {\bf x}(t)]^2\rangle = 6D|t|$ and 
\be 
\label{Einstein:2}
 D= \frac{T}{m\eta_D}=\frac{2T^2}{\kappa}.
\ee
A special case is the diffusion of large spherical 
particles suspended in a simple fluid. In this case the drag 
coefficient can be computed using the Navier-Stokes equation 
and the drag is related to the shear viscosity of the fluid, 
$\eta_D= 6\pi \eta a/m$, where $a$ is the radius of the particles. 
This leads to a relation between the diffusion constant and the 
shear viscosity, $D=T/(6\pi\eta a)$. 

\subsection{Dynamic universality}
\label{sec_dyn}

 In the vicinity of a second order phase transition fluctuations
of the order parameter relax slowly. This implies that order 
parameter fluctuations have to be included in the hydrodynamic 
description. The resulting hydrodynamic models describe universal 
features of transport phenomena near a continuous phase transition  
\cite{Hohenberg:1977ym}. Dynamic universality classes, like the 
well-known static ones, depend on the symmetries of the order 
parameter and the number of dimensions. Universal aspects of 
transport also depend on the nature of the order parameter, 
whether it is conserved or not, and on the presence of couplings 
(non-vanishing Poisson brackets) between the order parameter and 
the conserved fields. In this section we will briefly review the 
hydrodynamic description of a simple fluid near the liquid-gas 
endpoint. This theory is known as model H in the classification 
of Hohenberg and Halperin \cite{Hohenberg:1977ym}. We will see that 
critical fluctuations lead to a divergent shear and bulk viscosity 
at the liquid-gas endpoint. The hydrodynamic description of the 
superfluid-normal transition in liquid helium and dilute atomic 
gases is called model F. This model describes the divergence of
the heat conductivity at the superfluid transition.

 Near the critical point sound modes ($\omega\sim k$) are higher 
in energy than diffusive modes ($\omega\sim k^2$), and the 
longitudinal components of the momentum density ${\bf g}$ can be 
neglected. A minimal model that describes the coupling of the order 
parameter $\phi$ to the transverse momentum density ${\bf g}_T$ is 
\cite{Hohenberg:1977ym}
\bea 
\label{mod_H_1}
\frac{\partial\phi}{\partial t} &=&
 \lambda_0 {\bm \nabla}^2 \frac{\delta{\cal F}_T}{\delta\phi}
  - g_0 {\bm \nabla}\phi\cdot \frac{\delta{\cal F}_T}{\delta{\bf g}}
 +\zeta_\phi \,  , \\ 
\label{mod_H_2}
 \frac{\partial g_i}{\partial t} &=& P_{ij}^T
  \left[ \eta_0{\bm \nabla}^2 \frac{\delta{\cal F}_T}{\delta g_j}
        + g_0 (\nabla_j\phi)\frac{\delta{\cal F}_T}{\delta\phi}
        + \zeta_{g_j} \right]\, . 
\eea
where $P^T_{ij}=(\delta_{ij}+\nabla_i\nabla_j/{\bm \nabla}^2 )$ is a
transverse projector, $\zeta_\phi$ and $\zeta_{g_j}$ are random
forces, and the free energy ${\cal F}_T={\cal F}-{\cal F}_h$ is 
given by 
\bea
 {\cal F} &=& \int d^dx\, \left[ 
   \frac{1}{2}({\bm \nabla}\phi)^2 + \frac{r_0}{2} \phi^2 
 + u_0\phi^4 + \frac{1}{2} {\bf g}^2 \right] \, ,  \\
 {\cal F}_h &=& \int d^dx\, \left[
 h\phi+ {\bf A}\cdot {\bf g} \right] \, , 
\eea
where $h$ and ${\bf A}$ are external fields. The coefficients
$\lambda_0$ and $\eta_0$ are the bare values of the thermal 
conductivity and shear viscosity. Fluctuations cause the 
physical value of the zero frequency transport coefficients
to diverge near the critical point. In order to study the 
critical behavior of the bulk viscosity the longitudinal component
of ${\bf g}$ has to be included \cite{Onuki:1997}.

 In a normal fluid the only conserved charges are the particle density, 
the energy density, and the momentum density. The order parameter is a 
suitable linear combination of the energy density and the particle 
density. In QCD the hydrodynamic variables include the chiral condensate, 
and the conserved energy density, baryon density, and isospin
density. 
The QCD phase diagram is expected to have two critical 
points, one that corresponds to the endpoint of the nuclear 
liquid-gas phase transition, and another one that is related 
to the endpoint of the first order chiral phase transition
\cite{Stephanov:2004wx}. QCD hydrodynamics in the vicinity of the 
chiral critical point was analyzed by Son and Stephanov 
\cite{Son:2004iv} who argue that the chiral endpoint, like 
the nuclear liquid-gas endpoint, is correctly described by 
model H.
The values of the critical exponents can be determined using the 
epsilon expansion. The shear and bulk viscosity diverge with the 
correlation length $\xi$ as \cite{Hohenberg:1977ym,Onuki:1997}
\be 
\label{eta_zeta_uni}
 \eta \sim \xi^{x_\eta}\hspace{0.25cm}(x_\eta\simeq 0.06),  
  \hspace{1cm}
 \zeta \sim \xi^{x_\zeta}\hspace{0.25cm}(x_\eta\simeq 2.8).
\ee
The critical endpoint is in the same static universality class
as the Ising model and $\xi\sim t^{-0.63}$, where $t=(T-T_c)/T$.
We note that the divergence in the bulk viscosity is much stronger 
than the divergence in the shear viscosity. 
These results demonstrate that, while there is empirical 
evidence for the suggestion that the viscosity minimum is located at 
endpoint of the liquid-gas phase transition (see Table \ref{tab_eta}), 
this idea cannot be rigorously correct. Indeed, both $\eta/s$ and 
$\zeta/s$ diverge near the critical endpoint.

\subsection{Kubo relations and spectral functions}
\label{sec_kubo}

 Hydrodynamics is an effective description of the low energy, 
long wavelength response of a fluid. The transport coefficients 
appear as unknown constants in the hydrodynamic equations. These
constants can be extracted from experiment, or computed from a
more microscopic theory. The relationship between transport 
coefficients and correlation functions in a microscopic quantum
field theory is provided by Kubo relations. We have seen that 
hydrodynamics fixes the low energy and low momentum behavior
of the correlation functions of conserved charges, see 
equ.~(\ref{S_vv}-\ref{S_pp}). In the field theory these correlation 
functions can be computed using linear response theory. The
response is governed by the retarded correlation function. In the 
case of shear viscosity the relevant correlation function is 
\be
\label{G_ret}
G^{xy,xy}_R(\omega,{\bf k}) = -i\int dt\int d^3x\, 
  e^{i(\omega t-{\bf k}\cdot{\bf x})} \Theta(t)
  \langle \left[ T^{xy}({\bf x},t),T^{xy}(0,0)\right]\rangle\, , 
\ee
where $T^{\mu\nu}$ is the energy momentum tensor. The spectral 
function is defined by $\rho(\omega,{\bf k})=-2\,{\rm Im}\,G_R
(\omega,{\bf k})$. 
The imaginary part of the retarded correlator is a measure 
of dissipation, while the correlation function $S(\omega,{\bf k})$ 
(see Sect.~\ref{sec_nr_fluid}) is related to fluctuations. The relation
between these two functions is called the fluctuation-dissipation 
theorem \cite{Forster}. In the low frequency limit
$\rho(\omega,{\bf k})=(\omega/T)S(\omega,{\bf k})$.
Matching the correlation function from linear response theory to the 
hydrodynamic correlator gives the Kubo relation
\be 
\label{eta_kubo}
\eta = \lim_{\omega\to 0} \lim_{k\to 0} 
   \frac{\rho^{xy,xy}(\omega,{\bf k})}{2\omega}\, . 
\ee
The formula for the bulk viscosity involves the trace of the 
energy momentum tensor
\be 
\label{zeta_kubo}
\zeta = \frac{1}{9}\lim_{\omega\to 0} \lim_{k\to 0} 
   \frac{\rho^{ii,jj}(\omega,{\bf k})}{2\omega}\, ,
\ee
and analogous results can be derived for the thermal conductivity 
and diffusion constants. 

 The spectral function contains information about the physical
excitations that carry the response. We will discuss this issue
in more detail when we compare the strong coupling (AdS/CFT) and
weak coupling spectral functions in Sec.~\ref{sec_ads}. Dispersion 
relations connect the spectral function to correlation functions 
with different analyticity properties. The Matsubara (imaginary 
energy) correlation function is 
\be 
\label{G_E_w}
G_E(i\omega_n)= \int \frac{d\omega}{2\pi} \frac{\rho(\omega)}
 {\omega-i\omega_n}\, , 
\ee
where $\omega_n=2\pi nT$ is the Matsubara frequency. The imaginary
time correlation function is given by 
\be 
\label{G_E_tau}
G_E(\tau)= \int \frac{d\omega}{2\pi} K(\omega,\tau) \rho(\omega) \, , 
\ee
where the kernel $K(\omega,\tau)$ is defined by
\be 
\label{Ker}
K(\omega,\tau) = \frac{\cosh[\omega(\tau-1/(2T))]}{\sinh[\omega/(2T)]}
 =  \left[1+n_B(\omega)\right] e^{-\omega\tau}
      + n_B(\omega)e^{\omega\tau}\, ,
\ee
and $n_B(\omega)$ is the Bose distribution function. Equation
(\ref{G_E_tau}) is the basis of attempts to compute transport 
coefficients using imaginary time quantum Monte Carlo data
\cite{Karsch:1986cq,Meyer:2007ic,Meyer:2007dy,Sakai:2007cm}.
The idea is to compute $G_E(\tau)$ numerically, invert the 
integral transform in equ.~(\ref{G_E_tau}) to obtain $\rho(\omega)$,
and then extract transport coefficients from $\rho'(0)$. The difficulty 
is that $G_E(\tau)$ is typically only computed on a small number 
of points, and that the imaginary time correlator is not very sensitive 
to the slope of the spectral function at low energy. Many recent 
calculations make use of the maximum entropy method to obtain 
numerically stable spectral functions and reliable error estimates 
\cite{Aarts:2007wj,Aarts:2007va}. It was also observed that one can 
minimize the contribution from continuum states to the imaginary 
time Green function by studying the correlators of conserved charges, 
energy and momentum density, at non-zero spatial momentum 
\cite{Aarts:2006wt,Meyer:2008gt}. In more physical
terms this means that one is extracting the viscosity from the 
sound pole rather than the shear pole. In Table \ref{tab_eta_latt} 
we summarize some recent lattice QCD results on the shear and bulk 
viscosity in the high temperature phase of pure gauge QCD. We observe 
that the shear viscosity to entropy density ratio is close to the 
conjectured bound $1/(4\pi)$. The bulk viscosity is large in the 
vicinity of the phase transition but decreases quickly and becomes 
extremely small at $T=1.64T_c$. 

\begin{table}[t]
\begin{center}\begin{tabular}{|c|lll|}\hline
$T$       &  1.02 $T_c$   & 1.24 $T_c$ & 1.65 $T_c$ \\ \hline
$\eta/s$  &               &  0.102(56) & 0.134(33)  \\
$\zeta/s$ &  0.73(3)      &  0.065(17) & 0.008(7)     \\ \hline
\end{tabular}\end{center}
\caption{\label{tab_eta_latt}
Lattice QCD results for the ratio of shear and bulk viscosity 
to entropy density in a pure gluon plasma. The calculations
were performed for three different temperatures, given in units 
of the critical temperature $T_c$. Data from 
\cite{Meyer:2007ic,Meyer:2008gt}.}
\end{table}

\subsection{Kinetic theory: Shear viscosity}
\label{sec_kin}

 If the fluid can be described in terms of weakly interacting 
quasi-particles then the hydrodynamic variables can be written 
in terms of quasi-particle distribution functions $f_p({\bf x},t)$.
In the case of a non-relativistic fluid the energy current, momentum
current, and stress tensor are given by 
\bea
\label{j_eps_kin}
j^{\epsilon}_i({\bf x},t) &=& \int\frac{d^3p}{(2\pi)^3} 
   E_p v_{p,i} f_p({\bf x},t) \, ,\\
\label{g_i_kin}
g_i({\bf x},t) &=& \int\frac{d^3p}{(2\pi)^3} 
   m v_{p,i} f_p({\bf x},t) \, , \\
\label{Tij_kin}
\Pi_{ij}({\bf x},t) &=& \int\frac{d^3p}{(2\pi)^3} 
  m v_{p,i} v_{p,j} f_p({\bf x},t),
\eea
where $E_P$ is the quasi-particle energy, and $v_{p,i}=(\partial E_p)/
(\partial p_i)$ is the quasi-particle velocity. The equation of motion 
for $f_p({\bf x},t)$ is the Boltzmann equation 
\be
\label{Boltz_equ}
\frac{\partial f_p}{\partial t}+ {\bf v}\cdot
{\bm \nabla} f_p + {\bf F}\cdot {\bm \nabla}_{\! p}\, f_p= C[f_p]\, ,
\ee
where ${\bf F}$ is an external force and $C[f_p]$ is the 
collision term. In local thermal equilibrium the distribution 
function is determined by the local temperature, chemical potential,
and flow velocity. We have
\be 
 f^0_p({\bf x},t) = 
  \frac{1}{\exp((E_p-{\bf v}\cdot {\bf p}-\mu)/T)\mp 1}\, ,
\ee
where the $\mp$ sign corresponds to bosons/fermions. Transport 
coefficients characterize how the distribution function relaxes
to its equilibrium value if it is perturbed slightly away from 
it. We can write
\be
 f_p({\bf x},t) = f_p^0({\bf x},t) + \delta f_p({\bf x},t)
\ee
and linearize the Boltzmann equation in $\delta f_p$. In order 
to determine transport coefficients we also use a gradient
expansion of the local velocity, temperature and chemical 
potential and linearize the Boltzmann equation in the ``driving
terms'' $\nabla_iv_j$, $\nabla_iT$ and $\nabla_i\mu$. This 
procedure is known as the Chapman-Enskog method. In the next
section we will describe the method in the case of phonon
mediated transport in a superfluid, and then discuss some 
of the modifications that appear when studying high temperature 
Fermi and Bose gases as well as gauge theories. 

\begin{figure}[t]
\begin{center}
\raisebox{0.085\vsize}{a)}\;
\includegraphics[width=6cm]{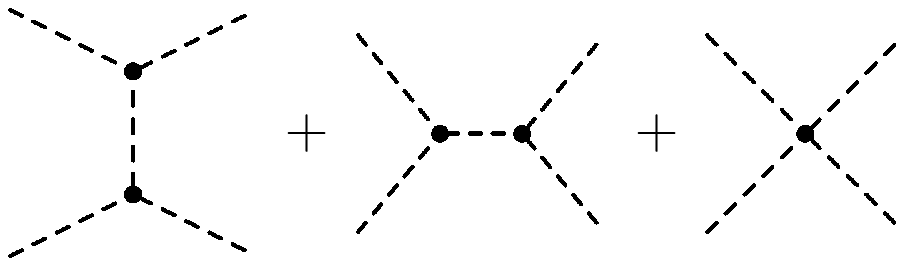}
\hspace*{1cm}
\raisebox{0.085\vsize}{b)}\;
\includegraphics[width=1.6cm]{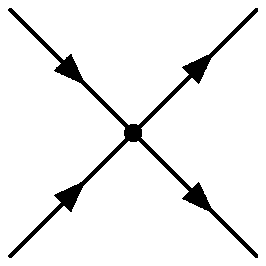}
\end{center}
\caption{\label{fig_kin_gb}
Leading order processes that contribute to the shear viscosity of 
the Fermi gas in the unitarity limit at low temperature (Fig.~a) 
and high temperature (Fig.~b). Dashed lines are phonon propagators 
and solid lines are fermion propagators. }
\end{figure}

\subsubsection{Phonons in dilute Fermi gases}
\label{sec_kin_ph}

 In the following we will concentrate on the shear viscosity of 
the low temperature, superfluid, phase of the dilute Fermi gas 
at unitarity. The calculation is similar to the computation of 
the shear viscosity of superfluid helium, but as explained in 
Sect.~\ref{sec_fl} the low energy effective theory of the dilute
Fermi gas is more tightly constrained. We discuss the shear viscosity 
of liquid helium, as well as the viscosity of the low temperature 
(chiral symmetry broken) phase of QCD in Sect.~\ref{sec_kin_gbs}. 
The stress tensor of a phonon gas is 
\be
 \Pi_{ij}({\bf x},t) = c_s^2\int\frac{d^3p}{(2\pi)^3} 
  \frac{p_i p_j}{E_p} f_p({\bf x},t)\, . 
\ee
In order to study the shear viscosity we write $\delta f_p=-\chi(p) 
f_p^{0}(1+f_p^{0})/T$ with
\be
\label{ansatz}
\chi(p) = g(p) \left(p_i p_j-\frac{1}{3}\delta_{ij}{\bf p}^2\right)
  \left(\nabla_i v_j+\nabla_j v_i
     -\frac{2}{3} \delta_{ij}{\bm \nabla}\cdot{\bf v}\right)\, . 
\ee
Inserting this ansatz into the energy momentum tensor gives
\be
\label{eta_ans}
\eta  = \frac{4c_s^2}{15 T } \int \frac{d^3 p}{(2\pi)^3}
         \frac{p^4}{2 E_p}f_p^{0}(1+f_p^{0}) g(p) \, . 
\ee
The function $g(p)$ is determined by the linearized Boltzmann 
equation. Linearizing the LHS of the Boltzmann equation
in derivatives of ${\bf v},\mu,T$ gives
\be
\label{LHSdf}
\frac{d f_p}{dt}  \simeq  
   c_s^2\frac{f_p^{0}(1+f_p^{0})}{2 E_p T} p_{ij} v_{ij}\, , 
\ee
where we have defined
\be 
p_{ij} =  p_i p_j -\frac{1}{3}\delta_{i j} {\bf p}^2, \hspace{0.5cm}
v_{ij} =  \nabla_i v_j+\nabla_j v_i
  -\frac{2}{3} \delta_{ij}{\bm \nabla}\cdot{\bf v}\, . 
\ee
The RHS of the Boltzmann equation contains the collision term $C[f_p]$.
In the present case the dominant contribution arises from binary 
$2\leftrightarrow 2$ collisions. The linearized collision term is 
\bea
\label{collisionA}
 C_{2\leftrightarrow2}[f_p] &\simeq& 
     \frac{1+f_p^{0}}{2E_p T}
    \int d\Gamma(k;k',p') (1+f_k^{0})f_{k'}^{0} f_{p'}^{0}
    |{\cal M}|^2\\
 & & \; \times
  \left[g(p) p_{ij} +g(k) k_{ij}-g(k') {k'}_{ij} -g(p'){p'}_{ij}
      \right]   v_{ij} \nonumber \\
 & \equiv & 
      \frac{f_p^0(1+f_p^{0})}{2E_p T} C_{ij}[g(p)] v_{ij}\nonumber, 
\eea
where ${\cal M}$ is the scattering matrix element,  
\be
d\Gamma(k; k',p') = \left( \prod_{q=k,k',p'}
 \frac{d^3q}{(2\pi)^3 2 E_q} \right)
 \times(2\pi)^4\delta^{(4)}(p+k-k'-p')
\ee
is the phase space, and we have defined the linearized collision 
operator $C_{ij}[g(p)]$. The linearized Boltzmann equation can 
now be written as
\be
\label{linearBoltzmann}
 C_{ij}[g(p)] = \frac{c_s^2}{T}\, p_{ij}. 
\ee
This result can be used to rewrite the relation for the viscosity 
in equ.~(\ref{eta_ans}) as 
\be
\label{eta_bol}
\eta= \frac{2}{5}\int\frac{d^3 p}{(2\pi)^3}
  \frac{f_p^0(1+f_p^{0})}{2E_p T}
       p_{ij} g(p) C_{ij}[g(p)]\, . 
\ee
The two relations equ.~(\ref{eta_ans}) and (\ref{eta_bol}) can be 
used to derive a variational estimate of the shear viscosity. We can
view equ.~(\ref{eta_ans}) as an inner product with measure 
$f^0(1+f^0)/(2E_p)$ and write 
\be
\label{eta_me}
 \eta = \frac{2}{5}\;\langle X|g\rangle \, , 
\ee
where $X=(c_s^2/T) p_{ij}$ and $g=g(p)p_{ij}$. The linearized 
Boltzmann equation is $C|g\rangle =|X\rangle$ and equ.~(\ref{eta_bol}) 
can be written as $\eta=\frac{2}{5}\,\langle g|C|g\rangle$. The linearized 
collision operator $C$ is a hermitean, negative semi-definite operator. 
The zero eigenvalues of $C$ correspond to the conservation laws for 
energy, momentum, and particle number. Consider a variational ansatz 
$|g_{var}\rangle$ for the exact solution $|g\rangle$ of the linearized 
Boltzmann equation. The triangle equality implies 
\be 
\label{eta_tri}
\langle g_{var}|C|g_{var}\rangle 
\langle g |C | g\rangle \geq
  \langle g_{var} | C | g\rangle^2 =
  \langle g_{var} | X \rangle^2 \, . 
\ee
Using $\eta=\frac{2}{5}\,\langle g|C|g\rangle$ we get  
\be 
\label{eta_var}
 \eta \geq \frac{2}{5}\;\frac{\langle g_{var}|X\rangle^2} 
                {\langle g_{var}|C|g_{var}\rangle^2} \, . 
\ee
This result is, of course, not a lower bound on the exact value 
of $\eta$, but it is a bound within the approximation that is used to 
compute the collision term. A popular choice for $g_{var}$ is the 
driving term $X$. This ansatz provides a good estimate in the case 
of non-relativistic particles interacting via short range interactions, 
as well as for gauge boson exchanges in QCD, but not in the case of 
phonon scattering\footnote{A detailed discussion of upper
and lower bounds on transport coefficients can be found in 
\cite{Jensen:1969,Smith:1989}. We also refer the reader to comparisons 
of the variational results with exact solutions of the Boltzmann equation
\cite{Brooker:1968}.}. 

  A systematic method for improving the variational estimate is based 
on orthogonal polynomials. We can construct a complete set of polynomials 
that are orthogonal with respect to the inner product defined in
equ.~(\ref{eta_me}). In non-relativistic physics these polynomials 
are known as Sonine polynomials \cite{Landau:kin} and suitable 
generalizations can been constructed for Bose and Fermi gases
\cite{Rupak:2007vp}. We now fix an integer $N$ and expand the solution 
of the linearized Boltzmann equation in the first $N$ polynomials. 
At finite $N$ solving the Boltzmann equation reduces to the problem
of inverting an $N\times N$ matrix. The solution provides a variational 
estimate for $\eta$ which becomes exact as $N\to\infty$. Convergence is 
usually quite fast.

  To complete the calculation of the shear viscosity we need to compute 
the scattering amplitude ${\cal M}$. The collision operator at leading 
order in $T/\mu$ is determined by the scattering amplitude at leading 
order in $q/\mu$, where $q=p,p',k,k'$. The amplitude is given by the 
diagrams in Fig.~\ref{fig_kin_gb}a with vertices and propagators 
determined by the effective lagrangian given in equ.~(\ref{ph_vert}). 
The expression for ${\cal M}$ is not very instructive and can be found 
in \cite{Rupak:2007vp}. The best estimate for $\eta$ is obtained by 
using $g(p)\sim p^{-1}$. We find
\be
\label{eta_ph}
\eta  = 9.3\times 10^{-6}\frac{\xi^5}{c_s^3}
  \frac{T_F^8}{T^5}\, , 
\ee
where $\xi\simeq 0.4$ is the universal parameter we introduced in 
equ.~(\ref{xi}). In the low temperature limit the entropy density
is dominated by the phonon contribution
\be
\label{s_ph}
s=\frac{2\pi^2}{45}\frac{T^3}{c_s^3}\, . 
\ee 
The ratio $\eta/s$ drops sharply with temperature. Extrapolating 
to $T=T_c\simeq 0.15T_F$ gives $\eta/s\sim 0.8$, with very large 
uncertainties. 

\subsubsection{Phonons and rotons in liquid helium, pions in QCD}
\label{sec_kin_gbs}

 The calculations of shear viscosity of liquid $^4$He below the
$\lambda$ point is similar to the computation of $\eta$ in the 
superfluid Fermi gas. The main difference is that close to $T_c$ 
it is important to include the roton contribution. Rotons form a 
dilute gas, and unlike phonons, 
their cross section is approximately constant. As a 
consequence the roton viscosity is independent of the roton 
density, see the discussion below equ.~(\ref{eta_mfp}). The 
typical roton momentum is determined by the roton minimum of 
the dispersion relation and depends only weakly on temperature. 
This implies that the roton viscosity is almost temperature 
independent. The value of the roton viscosity depend on the 
poorly known roton-roton interaction. A fit to experimental 
data for the shear viscosity below the lambda point gives 
$\eta_r\simeq 1.2\cdot 10^{-5}$ poise. The leading correction 
to the roton term comes from phonon-roton scattering. Landau 
and Khalatnikov find \cite{Khalatnikov:1965}
\be 
\label{eta_rot_ph}
\eta = \eta_r +\frac{A}{T^{1/2}} 
  \exp\left(\frac{\Delta}{T}\right)
  \frac{10+8\bar\Theta/\Theta_{ph}}{1+8\bar\Theta/\Theta_{ph}}\, , 
\ee
where $\Delta$ is the roton energy defined in equ.~(\ref{roton}), 
$A$ is a constant, and $\Theta/\Theta_{ph}$ is the ratio of the 
roton-roton and roton-phonon relaxation rates. This ratio scales
as $T^{4.5}\exp(\Delta/T)$. For $T<0.9$ K we can use $\bar\Theta
\gg\Theta_{ph}$ and the temperature dependence of the phonon-roton
term is governed by the $T^{-0.5}\exp(\Delta/T)$ term. For 
$T<0.7$ K phonon-phonon scattering dominates and the viscosity 
scales as $T^{-5}$, as in the previous section. At even smaller 
temperature, $T< 0.5$ K, phonon splitting, 
also known as Beliaev damping,
becomes important and the temperature dependence changes to
$\eta\sim T^{-1}$ \cite{Maris:1973}. The roton contribution to 
the entropy density is  
\be 
\label{s_r}
s_r = \frac{2(m^*)^{1/2}p_0^2\Delta}
           {(2\pi)^{3/2}T^{1/2}}
   \left( 1+\frac{3T}{2\Delta}\right)
   \exp\left(-\frac{\Delta}{T}\right)\, ,
 \ee
where $m^*$ and $p_0$ are given in equ.~(\ref{roton}). The 
phonon contribution is given by equ.~(\ref{s_ph}) with 
$c_s=238$ m/sec. If we push equ.~(\ref{eta_rot_ph}) and
(\ref{s_r}) to the limit of their applicability, $T\sim 2$ K, 
we find $\eta/s\sim 2$. 

 The computation of the shear viscosity in low temperature 
QCD also proceeds along similar lines. The analog of the 
phonon in QCD is the pion, and pion interactions are governed by the 
effective lagrangian given in equ.~(\ref{l_chpt_comp}). The pion 
is not massless, $m_\pi=139$ MeV. At very low temperature, $T\ll 
m_\pi$, the pion scattering amplitude is approximately constant 
and the viscosity is only weakly temperature dependent. At higher 
temperature we can set $m_\pi\simeq 0$ and the scattering amplitude 
is energy dependent. The main difference as compared to phonon 
scattering is that the four-pion interaction is of the form 
$(\phi\partial\phi)^2$ instead of $(\partial\phi)^4$, and that 
there is no three-pion interaction. As a consequence the $\pi\pi$ 
scattering matrix element scales as the second power of the 
external momenta. The pion entropy is given by equ.~(\ref{s_ph}) 
with $c_s=c/\sqrt{3}$ and an isospin degeneracy factor 3. An 
approximate calculation of the ratio $\eta/s$ gives 
\cite{Prakash:1993bt,Csernai:2006zz}
\be 
\label{eta_pi}
\frac{\eta}{s} = \frac{15}{16\pi}\frac{f_\pi^4}{T^4}
\ee
Variational solutions of the Boltzmann equation reported in 
\cite{Chen:2006iga} give $\eta/s$ ratios that are about five 
times larger. The first study of the shear viscosity of a pion
gas can be found in \cite{Gavin:1985ph}. More detailed investigations
of the viscosity of hadronic mixtures were published in 
\cite{Prakash:1993bt,Itakura:2008}.

\subsubsection{Non-relativistic atoms: Dilute Fermi gases and $^4$He}
\label{sec_kin_ferm}
 
 The shear viscosity of the dilute Fermi gas at high temperature 
is determined by binary scattering between the atoms. The 
$s$-wave scattering matrix is
\be 
{\cal M} =  \frac{4\pi}{m} \frac{1}{1/a+iq}\, ,
\ee
where $q$ is the relative momentum. In the unitarity limit $a\to\infty$ 
the scattering amplitude diverges as $1/q$ in the limit of small momenta. 
For $T\gg T_F$ the infrared divergence is effectively cut off by the 
thermal momentum $(mT)^{1/2}$. The viscosity in the high temperature 
limit is \cite{Massignan:2004,Bruun:2005}
\be
\label{eta_aa}
\eta = \frac{15}{32\sqrt{\pi}}(m T)^{3/2}.
\ee 
This result is based on the variational function $g(p)\sim 1$. 
Corrections due to more complicated distribution functions are 
small, $\Delta\eta/\eta<2\%$ \cite{Bruun:2006}. The scaling of
$\eta$ with temperature can be understood as a combination of 
the $T^{1/2}$ scaling of a dilute hard sphere gas (see 
Sect.~\ref{sec_intro}) with an extra factor $(mT)$ from the 
$1/q^2$ behavior of $|{\cal M}|^2$. The high temperature 
limit of the entropy density is that of a classical gas
\be
\label{s_boltz}
s=\frac{2\sqrt{2}}{3\pi^2} (m T_F)^{3/2}
  \left[\log\left(\frac{3\sqrt{\pi}}{4} \frac{T^{3/2}}{T_F^{3/2}}
    \right)  +\frac{5}{2}\right] .
\ee
Combining equ.~(\ref{eta_aa}) and (\ref{s_boltz}) gives $\eta/s\sim
x^{3/2}/\log(x)$ with $x=T/T_F$. The classical expression for the
entropy becomes unphysical (negative) for $T\simeq T_c$. Extrapolating
to $T\simeq 2T_c$ gives $\eta/s\simeq 0.5$. 

 The shear viscosity of helium is governed by scattering in the 
potential given in equ.~(\ref{V_vdw}). In the high temperature limit
the dominant contribution does not come from the Van der Waals tail, 
but from the repulsive short range contribution. For a potential 
of the form $V\sim r^{-\nu}$ the viscosity scales as $T^s$ with 
$s=\frac{1}{2}+\frac{2}{\nu-1}$ \cite{Chapman:1970}. In the case 
of a Lennard-Jones (6-12) potential this implies $\eta\sim T^{0.68}$. 
A somewhat better fit to the data is provided by 
\be
 \eta = \eta_0 \left(\frac{T}{T_0}\right)^{0.647}
\ee
with $\eta_0=1.88\cdot 10^{-5}$ Pa$\cdot$s and $T_0=273.15$ K. The 
entropy density is given by the classical result, equ.~(\ref{s_boltz}). 
For $T=10$ K we get $\eta/s\simeq 4$, and extrapolating to $T=4$ K
gives $\eta/s\simeq 1.5$.
Very accurate calculations that are based on realistic potentials
and include higher order terms in the density can be found in 
\cite{Aziz:1995}. These calculations are reliable down to about 
10 K.

\begin{figure}[t]
\begin{center}
\includegraphics[width=12cm]{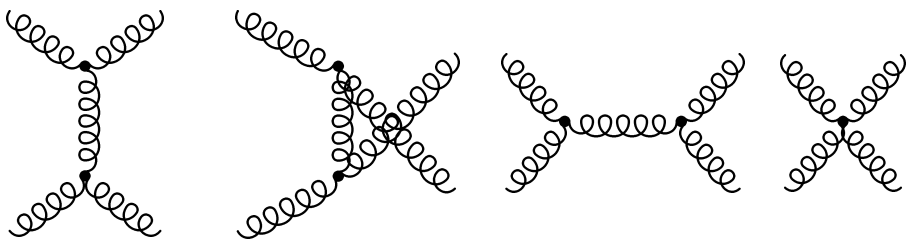}
\end{center}
\caption{\label{fig_kin_gg}
Leading order processes that contribute to the shear viscosity of a 
pure gluon plasma. The coefficient $k$ defined in equ.~(\ref{eta_qcd})
is determined by the $t$-channel diagram. The full leading order result, 
including the coefficient $\mu^*$, requires the remaining diagrams, as 
well as gluon bremsstrahlung from the external legs (not shown).}
\end{figure}

\subsubsection{Gauge fields in QCD}
\label{sec_kin_ym}

\begin{figure}[t]
\begin{center}
\includegraphics[width=7.00cm]{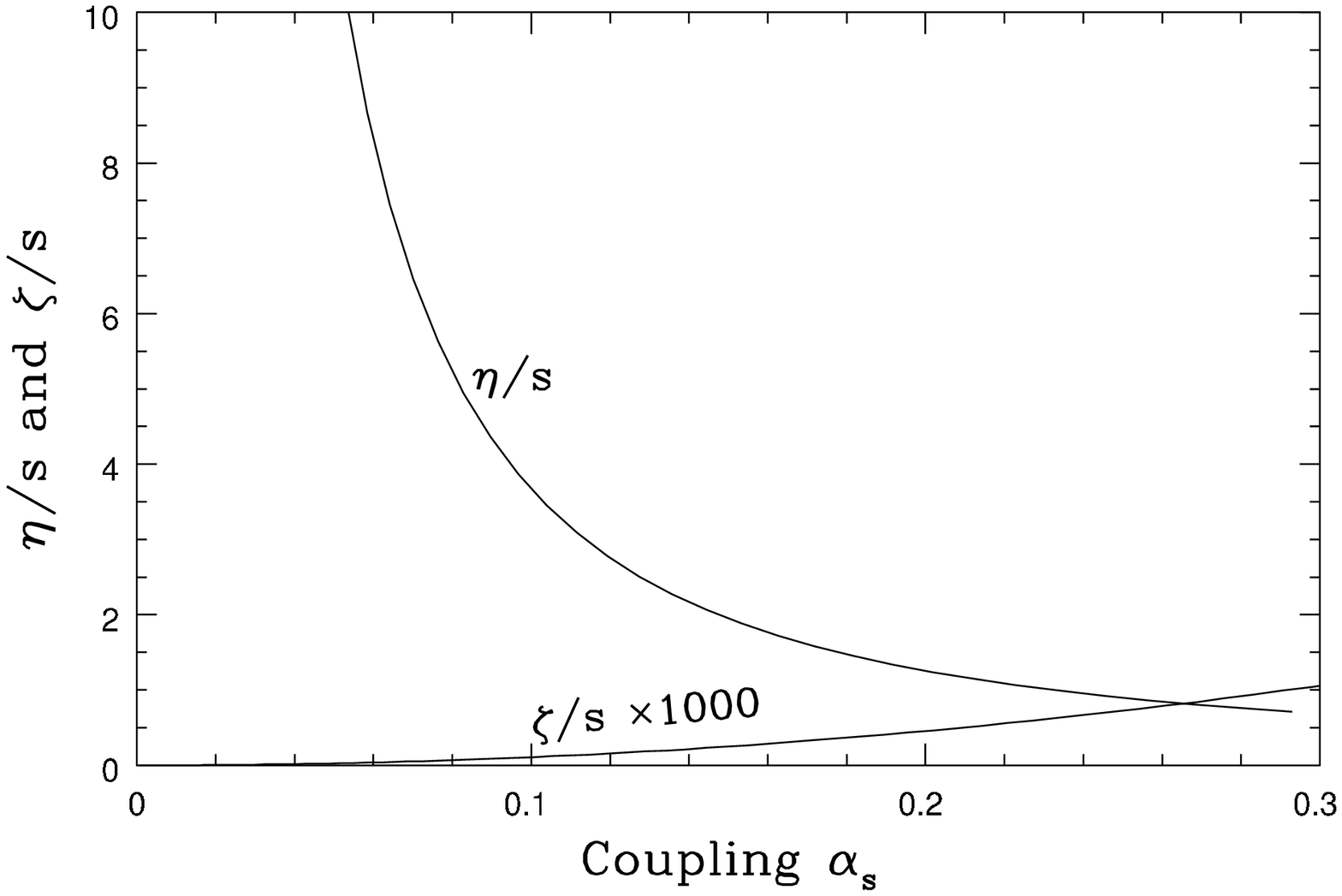}
\hspace*{0.3cm}
\includegraphics[width=6.00cm]{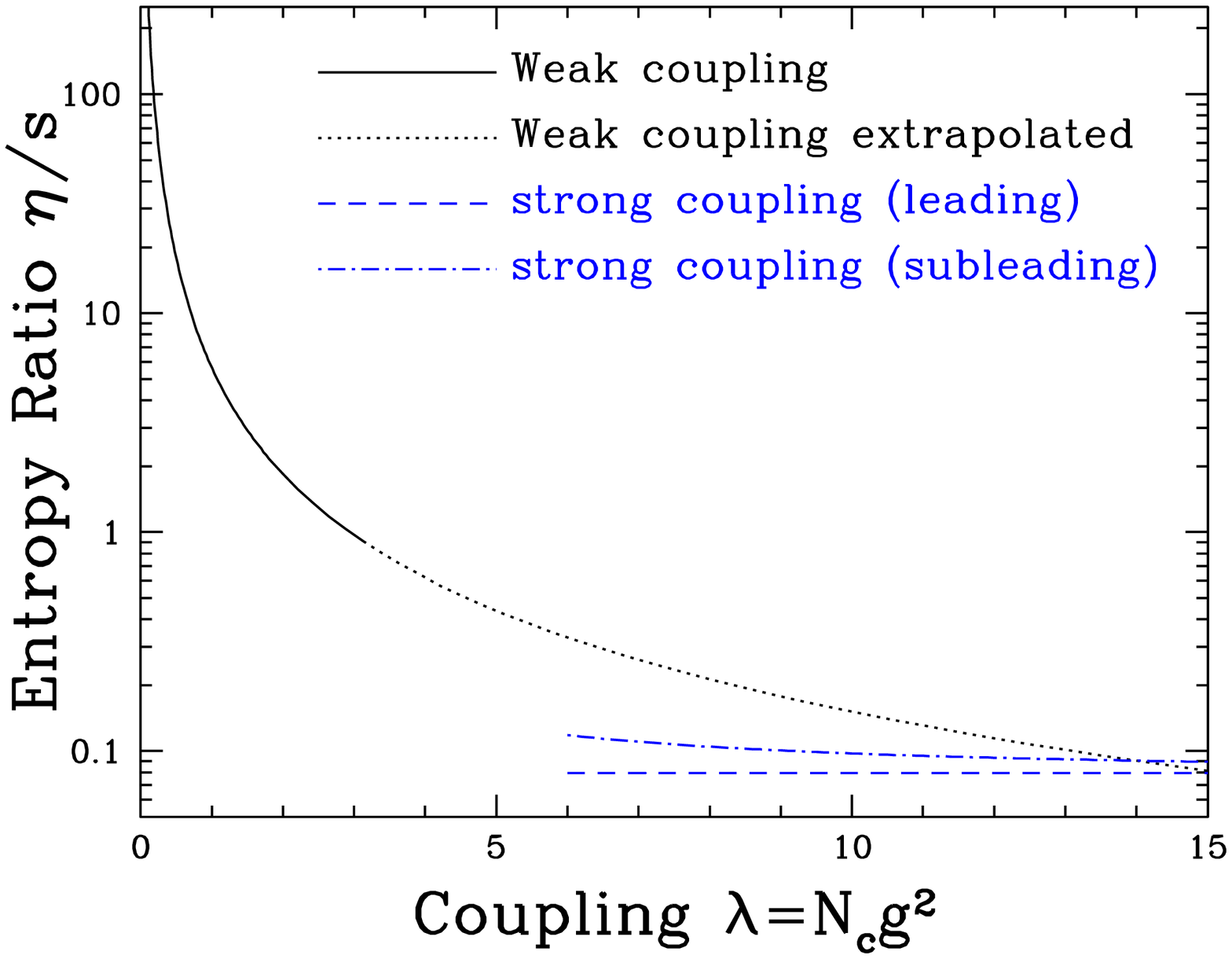}
\end{center}
\caption{\label{fig_qcd_eta}
Shear and bulk viscosity to entropy density ratio in QCD (left 
panel) and ${\cal N}=4$ supersymmetric Yang-Mills theory (right
panel). The left panel shows the shear and bulk viscosity to 
entropy density ratio in QCD with $N_f=3$ flavors as a function 
of the strong coupling constant $\alpha_s$, from \cite{Arnold:2006fz}.
The right panel shows the ratio $\eta/s$ in ${\cal N}=4$ SUSY Yang 
Mills theory as a function of the 't Hooft coupling $\lambda=
g^2N_c$. The solid line shows the weak coupling result, the dotted 
line is an extrapolation of the weak coupling result to the strong 
coupling regime, the dashed lined is $\lambda\to\infty$ result from 
the AdS/CFT correspondence, and dash-dotted line is the leading 
correction to the strong coupling result, from \cite{Huot:2006ys}.}
\end{figure}

 The shear viscosity of a quark gluon plasma is determined by binary
quark and gluon scattering. We first consider a pure gluon plasma. 
The leading order gluon-gluon scattering diagrams are shown in 
Fig.~\ref{fig_qcd_eta}. The squared tree level amplitude is 
\be 
|{\cal M}|^2 = \frac{9g^4}{2} \left(
  3 -\frac{ut}{s^2} - \frac{us}{t^2} - \frac{ts}{u^2} \right)\, , 
\ee
where $g$ is the gauge coupling and $s,t,u$ are the Mandelstam
variables. The differential cross section diverges for small momentum 
transfer $q$ as $1/q^4$. This is the standard Rutherford behavior, 
which arises from $t$-channel gluon exchange. In the calculation 
of the shear viscosity the differential cross section is weighted 
by an extra factor  
of $(1-\cos\theta)$, where $\theta$ is the scattering angle.
The quantity $\sigma_T=\int d\cos\theta\, (d\sigma)/(d\cos\theta)
(1-\cos\theta)$ is known as the transport cross section. The 
transport cross section diverges logarithmically at small $\theta$. 
This divergence is regulated by medium corrections to the gluon 
propagator, see equ.~(\ref{Pi_L},\ref{Pi_T}). Electric gluon
exchanges are screened at a distance $r_D\sim m_D^{-1}$, and 
the electric contribution to $\sigma_T$ is proportional to 
$g^4\log(m_D)$. There is no static magnetic screening, but 
gluons with energy $\omega$ are dynamically screened at a distance 
$r\sim (\omega m_D^2)^{-1/3}$. After integrating over energy the 
magnetic contribution also scales as $g^4\log(m_D)$. Combining
electric and magnetic $t$-channel exchanges gives
\cite{Baym:1990uj,Arnold:2000dr}
\be
\label{eta_qcd}
 \eta = k \frac{T^3}{g^4\log(\mu^*/m_D)} \, , 
\ee
where $k=27.13$. We will specify the coefficient $\mu^*$ below. This result 
corresponds to an optimized trial function $\chi(p)=A(p)p_{ij}v_{ij}$, 
but the simple approximation $A(p)\sim {\it const}$ agrees with the 
exact result to better than $1\%$. In order to compute the shear 
viscosity of a quark gluon plasma we have to include $t$-channel 
diagrams for quark-quark and quark-gluon scattering. The result 
is of the same form as equ.~(\ref{eta_qcd}) with \cite{Arnold:2000dr}
\be 
k(N_f) = (27.13,60.81,86.47,106.67),\hspace{0.5cm}
 (N_f=0,1,2,3)\, . 
\ee
Note that $k$ increases with $N_f$ faster than the total
number of degrees of freedom. This is related to the fact that 
quarks have smaller color charges than gluons, which implies
that quark-gluon scattering amplitudes are suppressed relative
to gluon-gluon amplitudes.  

 In order to make an absolute prediction for the shear viscosity 
we need to determine the constant $\mu^*$ in equ.~(\ref{eta_qcd}). 
This coefficient receives contributions from $s$ and $u$-channel 
gluon exchanges. These contributions are straightforward to include. 
A more difficult problem arises from the fact that $\mu^*$ is sensitive 
to soft ($q\sim m_D$) binary $2\to 2$ scattering followed by 
collinear $1\to 2$ splitting. The inverse mean free time for this 
process is given by $\tau^{-1} \sim g^4T^3/m_D^2\times g^2 \sim 
g^4 T$, comparable to the transport mean free time $\tau_{\it tr}^{-1}
\sim T^4/\eta\sim g^4 T$. Since the scattering angle is zero
collinear splitting does not directly contribute to shear viscosity, 
but it degrades the momentum and assists in randomizing the momentum 
distribution in subsequent binary collisions. 

 The difficulty with collinear splitting is that the formation 
time of the emitted gluon is of order $1/(g^2T)$.
This is the same order of magnitude as the quasi-particle life
time given in equ.~(\ref{gam_plas}), which implies that kinetic
theory is breaking down. Arnold, Moore, and Yaffe showed that 
if interference between subsequent gluon emission processes,
the Landau-Pomeranchuk effect, is taken into account an effective
Boltzmann equation with $2\to2$ and $1\to 2$ collision terms
can be derived \cite{Arnold:2002zm}. Arnold at al.~find 
\cite{Arnold:2003zc}
\be 
 \mu^*(N_f\!=\!0) =  2.765\, T \, . 
\ee 
They also show that $\mu^*$ is only weakly dependent on the number 
of flavors, $\mu^*(N_f\!=\!3)=2.957\, T$, and compute additional terms
in an expansion in inverse logarithms of $\mu^*/m_D$. 

 The entropy density of the quark gluon plasma is given by 
\be 
 s = \frac{2\pi^2}{45}\, 
  \left( 2(N_c^2-1)+\frac{7}{8} 4N_f \right)\, T^3 \, . 
\ee
Higher order corrections to the entropy density are large, but the 
situation in the regime $T\geq 2T_c$ can be improved using resummation 
schemes, see Fig.~\ref{fig_eos_qcd}. The resummed entropy differs 
from the free gas result by no more than 15\% for $T>2T_c$. The 
magnitude of higher order corrections to the viscosity is not known, 
but next-to-leading order results for the heavy quark diffusion 
constant suggest that higher corrections to transport coefficients are 
large \cite{CaronHuot:2008uh}. 

 The leading order QCD result is shown in Fig.~\ref{fig_qcd_eta}.
Clearly, $\eta/s$ is strongly dependent on the coupling, and 
without performing higher order calculations it is not clear 
what value of $\alpha_s$ one should use at a given temperature.
An interesting perspective is provided by exact results for 
$\eta/s$ in the strong coupling limit of ${\cal N}=4$ SUSY 
Yang-Mills theory, see Sect.~\ref{sec_ads}. These results can 
be compared to weak coupling calculations based on kinetic theory 
\cite{Huot:2006ys}. The weak coupling result for $\eta/s$ in the 
${\cal N}=4$ theory is smaller than the corresponding ratio in
QCD by a factor $\sim 1/7$. This is related to the fact that in 
the ${\cal N}=4$ theory all states are in the adjoint representation, 
and that the theory contains extra scalars. Both of these differences
lead to larger cross sections. 

 Weak and strong-coupling results for $\eta/s$ as a function
of the 't Hooft coupling $\lambda=g^2N_c$ in SUSY Yang-Mills theory 
are shown in the right panel of Fig.~\ref{fig_qcd_eta}. We observe 
that $\eta/s$ in the ${\cal N}=4$ theory reaches the strong coupling 
limit when extrapolated to a 't Hooft coupling $\lambda=g^2N_c\simeq 
12$. As discussed in Sec.~\ref{sec_susy} this is a large $N_c$ result.  
Naively extrapolating to $N_c=3$ the value $\lambda\simeq 12$ 
corresponds to $\alpha_s=g^2/(4\pi)\simeq 0.3$. We also note that
the value of 't Hooft coupling at which the weak coupling result
for $\eta/s$ reaches the strong coupling limit is larger than the 
coupling $\lambda\sim 5$ at which the corresponding expression for 
the entropy reaches the strong coupling limit $s/s_0=0.75$, see 
Fig.~\ref{fig_eos_qcd}. If we consider $s/s_0=0.8$ to be the 
``QCD-like'' point, then we should restrict ourselves to $\lambda<5$. 
In this case $\eta/s$ does not drop below $0.5$.

\subsection{Kinetic theory: Other transport properties}
\label{sec_kin_other}
\subsubsection{Bulk viscosity}
\label{sec_kin_zeta}

 Bulk viscosity measures the amount of energy dissipated
as a fluid is slowly expanded or compressed. In a conformally 
invariant system changing all the momenta and positions by a 
constant scale factor connects equilibrium states and the bulk 
viscosity must vanish. In kinetic theory bulk viscosity is 
typically sensitive to processes that change the particle
number or the composition of the system. The kinetic theory
prediction for bulk viscosity is proportional to the corresponding
relaxation time, and to deviations from conformality in the 
equation of state. Depending on the interplay between these two 
effects, the temperature dependence of the bulk viscosity can 
differ dramatically between different fluids, and between shear and
bulk viscosity. There are many fluids for which bulk viscosity 
is not an important source of dissipation, either because they
are approximately incompressible, like water, or because the 
fluid is compressible but approximately scale invariant, like
the QGP plasma. On the other, we have seen that bulk viscosity 
is the dominant source of dissipation near a second order phase
transition, see equ.~(\ref{eta_zeta_uni}).

 The Fermi gas at unitarity is exactly conformal and the bulk 
viscosity in the normal phase vanishes. In the low temperature 
phase conformal invariance requires $\zeta_1=\zeta_2=0$, but 
$\zeta_3$ can be non-zero \cite{Son:2005tj}. This coefficient 
was recently computed in \cite{Escobedo:2009}. The result is
sensitive to non-linearities in the phonon dispersion relation. 
If $1\to 2$ phonon splitting is kinematically allowed then $\zeta_3
\sim T^3$, where the constant of proportionality depends on the 
curvature of the dispersion relation. The bulk viscosity of 
liquid helium was calculated by Khalatnikov \cite{Khalatnikov:1965}. 
As in the case of shear viscosity the main contribution comes from 
phonons and rotons. Khalatnikov finds that $\zeta_2$, 
the bulk viscosity of the normal component, 
is about an order of magnitude bigger than $\eta$. The other two 
bulk viscosities, $\zeta_1$ and $\zeta_3$, 
involve motion of the normal fluid relative to the 
superfluid. They have different physical units, and cannot be 
directly compared to $\zeta_2$.
The linear combination that appears in the damping 
of second sound is $\alpha_\zeta=\zeta_2+\rho^2\zeta_3-2\rho\zeta_1$. 
At normal density there are significant cancellations between 
these terms and $\zeta_2\sim (\rho^2\zeta_3-2\rho\zeta_1)$. 
The bulk viscosity of helium vapor is small. Note that the bulk 
viscosity of most gases is dominated by internal excitations, 
like rotational and vibrational modes. 

 The bulk viscosity of a pion gas at low temperature was 
computed by Chen and Wang \cite{Chen:2007kx}. They find that the
bulk viscosity scales as $\zeta\sim T^7/f_\pi^4$ (up to logarithms). 
The bulk viscosity of the high temperature QGP phase was calculated 
by Arnold, Dogan, and Moore \cite{Arnold:2006fz}. The result is  
\be 
\label{zeta_qcd}
\zeta = \frac{A\alpha_s^2T^3}{\log(\mu^*/m_D)}\, , 
\ee
where $A=0.443$ and $\mu^*=7.14\, T$ in pure gauge QCD. In full QCD
with $N_f=3$ quark flavors $A=0.657$ and $\mu^*=7.77\, T$. We observe
that $\zeta$ scales as $\alpha_s^4\times\eta$. The trace anomaly 
$\epsilon-3P$ is proportional to $\alpha_s^2$, so bulk viscosity 
scales like the shear viscosity times the second power of the 
deviation from conformality. This is in agreement with an a
simple formula proposed by Weinberg \cite{Weinberg}, $\zeta
\sim (c_s^2-\frac{1}{3})^2\eta$. However, Weinberg's relation 
is known to be violated in some theories, see \cite{Jeon:1994if} 
for an example.

\subsubsection{Diffusion}
\label{sec_kin_diff}

 The diffusion of of impurities in liquid helium has been studied
in some detail. Of particular interest is the behavior of dilute
solutions of $^3$He in $^4$He. At low temperature the diffusion 
constant is determined by scattering off phonons and $D\sim 1/T^7$
\cite{Khalatnikov:1965,Bowley:2002}. At high temperature diffusion
is governed by scattering between atoms and $D\sim T^{1+s}$ with 
$s=\frac{1}{2}+\frac{2}{\nu-1}$ for a $1/r^{\nu}$ potential 
\cite{Chapman:1970}. We conclude that the temperature dependence 
of the diffusion constant is identical to that of the shear viscosity.
In case of the unitary Fermi gas one can make use of the 
fact that the number of spin up and spin down fermions is 
separately conserved, and study the diffusion of minority 
spin down particles in a background of majority spin up
fermions \cite{Bruun:2008}. This process contains important 
information about the interaction between the different spin
states, but it is not directly related to the viscosity of the spin 
balanced gas.

 The diffusion constant for heavy quarks in a quark gluon plasma
can be determined by computing the mean square momentum transfer 
per unit time, see Sec.~\ref{sec_diff}. For approximately thermal
heavy quarks the diffusion constant is dominated by heavy quark
scattering on light quarks and gluons, $qQ\to qQ$ and $gQ\to gQ$. 
As in the case of shear viscosity the most important Feynman 
diagrams involve $t$-channel gluon exchanges. Since the heavy quark
is slow the dominant interaction is electric gluon exchange and
the cross section is regularized by Debye screening. The leading
order result is \cite{Svetitsky:1987gq,Moore:2004tg}
\be 
\label{D_QCD}
 D=\frac{36\pi}{C_Fg^4T}\left[
 N_c\left(\log\left(\frac{2T}{m_D}\right)+c\right)
 +\frac{N_f}{2}\left(\log\left(\frac{4T}{m_D}\right)+c\right)
 \right]^{-1},
\ee
where $C_F=(N_c^2-1)/(2N_c)$ and $c=0.5-\gamma_E+\zeta'(2)/\zeta(2)$. 
We note that the diffusion constant has the same 
parametric dependence on the coupling as the shear viscosity.
The relaxation time $\eta_D^{-1}$ scales as $M_Q/(T^2g^4\log(g)$, 
which is larger by a factor $M_Q/T$ compared to the hydrodynamic 
relaxation time $\eta/(sT)\sim 1/(Tg^4\log(g))$. This is confirmed 
by numerical estimates, which give $\eta_D^{-1}\simeq 6.7/T\simeq 7$ 
fm for charm quarks at $T=200$ MeV \cite{Moore:2004tg}. For comparison, 
the hydrodynamic relaxation time is $\eta/(sT)\simeq 1$ fm (for 
$\eta/s\simeq 1$).

\subsubsection{Thermal conductivity}
\label{sec_kin_th}

  Thermal transport in superfluid helium is a complicated 
process. In a superfluid heat transport can take place by a process 
similar to internal convection where the superfluid moves relative 
to the normal fluid. Only the normal fluid carries entropy and 
as a result heat is carried along with the normal component. The
convective contribution to heat flow is controlled by the shear 
viscosity of the normal fluid. Within the normal fluid heat is 
carried by phonons and rotons. Khalatnikov showed that there is 
no heat transport in a gas of phonons with exactly linear 
dispersion relation \cite{Khalatnikov:1965}. The thermal conductivity
of the normal fluid is dominated by rotons and phonon-roton scattering.
This situation is somewhat similar to heat transport in a solid. 
At very low temperature, heat transport is ballistic and the 
entropy is carried by a net flow of phonons along the temperature 
gradient. At higher temperature non-linearities in the phonon
dispersion relation and the ``umklapp'' process play a role.

  The situation at high temperature is much simpler. Heat flow 
is a diffusive process, and the thermal conductivity is determined 
by scattering between atoms. A simple mean free path estimate
analogous to equ.~(\ref{eta_mfp}) is 
\be 
\label{kappa_mfp}
\kappa= \frac{1}{3}nc_p p l_{\it mfp}\, , 
\ee
where $c_p$ is the specific heat at constant pressure. This 
estimate suggests that the ratio of the shear and thermal 
diffusion constants, the Prandtl number 
\be 
\label{Prandtl}
 {\it Pr} = \frac{\eta c_p}{\kappa} \, ,
\ee
is close to one. At large $T$ the thermal conductivity of helium
scales as $T^s$ with $s=\frac{1}{2}+\frac{2}{\nu-1}$, as in the 
case of shear viscosity. The Prandtl number is approximately 
constant, ${\it Pr}\simeq 2.5$.

 Most studies of the thermal conductivity of a quark gluon plasma 
have focused on the regime of very high baryon density. In the 
limit $\mu\gg T$, where $\mu$ is the quark chemical potential, 
the thermal conductivity scales as $\kappa\sim \mu^2/\alpha_s^2$ 
\cite{Heiselberg:1993cr}. In the opposite limit $T\gg\mu$ there 
is an old relaxation time estimate $\kappa\sim T^4/(\alpha_s^2\mu^2)$ 
\cite{Danielewicz:1984ww}. Note that while $\kappa$ diverges as 
$\mu\to 0$, the dissipative contribution to the baryon current, 
equ.~(\ref{del_j_mu}), is finite. 

\section{Holography}
\label{sec_ads}

 In kinetic theory conserved charges are carried by well defined
quasi-particles. The time between collisions is long compared to 
the quantum mechanical scale, $\hbar/T$, and quantum mechanical
interference between scattering events is not important. In the
strong coupling limit quantum mechanical effects are large and 
quasi-particles lose their identity. A powerful new tool to study 
transport phenomena in this regime is the AdS/CFT correspondence
\cite{Maldacena:1997re,Witten:1998qj,Gubser:1998bc}.  

 The AdS/CFT correspondence is referred to as a holographic
duality -- it relates string theory on a certain higher dimensional 
manifold to four dimensional gauge theory on the boundary of this
space.
The correspondence is simplest if the field theory is strongly 
coupled. In this limit the string theory reduces to a classical 
gravitational theory. The holographic correspondence 
then implies that a four dimensional field theory is capable of
encoding gravity in five dimensions. 
The idea of a correspondence between field theories and higher dimensional
gravity originated from developments within string theory, but there
are precursors to the correspondence that come from the physics of 
black holes. It has been known for some time that black holes carry
entropy, and that the entropy is proportional to the area, and not 
the volume of the black hole. It was also known that the evolution
of black holes respects the second law of thermodynamics, 
and that it can be described by treating the event horizon as a
physical membrane with well defined transport properties like electric
conductivity and shear viscosity \cite{Thorne:1986}.

The best studied example of the AdS/CFT correspondence is the
equivalence between $\N=4$ Super Yang Mills theory (see Sect.~\ref{sec_susy})
and string theory on ${\rm AdS}_5 \times {\rm S}_5$. For our purposes 
the dynamics only involves ${\rm AdS}_5$. This is a 5-dimensional space, 
which in AdS/CFT terminology is called the {\it bulk}. The dual field 
theory exists on the {\it boundary} of this space, which is 3+1 
dimensional Minkowski space. The gauge gravity duality works as 
follows: Classical gravity equations of motion are solved in the 
$4+1$ dimensional curved geometry of ${\rm AdS}_5$. Fluctuations 
of gravitational fields in the bulk induces charges on the $3+1$ 
dimensional boundary. The dynamics of  $3+1$ dimensional boundary 
theory {\it is} the strongly coupled conformal field theory which 
we wish to study. Transport properties of the boundary theory can 
be determined by perturbing the boundary charges with an external 
field which then propagates into the bulk. The response of the induced
charges to the applied field determines the transport coefficients.
For each conserved charge of the field theory there is a corresponding
field in the gravitational theory. The field corresponding
to the stress tensor $T^{\mu\nu}$ is the graviton $h^{\mu\nu}$,
and  the field corresponding to the conserved $R$ charge current
$J^{\mu}_R$ is the five dimensional Maxwell field $A^{\mu}$.

 The AdS/CFT setup is analogous to a parallel plate capacitor.
Electromagnetic fields in the bulk, the space between the plates,
induce surface charges on the boundary. Fluctuations of the
bulk field create fluctuations of the surface charges, and
correlation functions of the surface charges can be related
to normal modes of the bulk field. What is remarkable about the
AdS/CFT correspondence is that the gravitational theory in the
bulk defines a local field theory on the boundary, and that
there are classical gravitational field configurations that
correspond to field theories at finite temperature. These
configurations can  be used to study dissipative phenomena 
in the boundary field theory.

 The gravitational field configuration relevant to field theories
at finite temperature is an ${\rm AdS}_5$ black hole. In the black
hole geometry the gravitational field is non zero as we approach the
boundary of the $4+1$ dimensional space. This gravitational field is
balanced by a non-zero stress tensor in the boundary field theory --
the gravitational setup corresponds to the dynamics of a field theory
with a non-zero density matrix.  
The event horizon of the back hole spans three spatial dimensions
in the bulk and radiates at the Hawking temperature $T_H$. The black
hole fills ${\rm AdS}_5$  with a bath of gravitational radiation, and 
the temperature of the heat bath is identified with the temperature
of the boundary field theory. The dynamics of graviton propagation
in the black hole background determines stress tensor correlators
at finite temperature in the boundary field theory. These correlators
determine the shear viscosity according to Kubo formulas.

 There is a vast amount of literature on the AdS/CFT correspondence.
A detailed review with extensive references is \cite{Aharony:1999ti}, 
and more pedagogical reviews can be found in 
\cite{Petersen:1999zh,D'Hoker:2002aw,Maldacena:2003nj}. Reviews with 
an emphasis on transport phenomena are \cite{Son:2007vk,Rangamani:2009xk}.
Here we will concentrate on a few selected issues that are relevant
to this review. First, we will explain the calculation of the shear
viscosity and the spectral weights of strongly coupled fluids. Then 
we will comment on the conjectured viscosity bound, and the calculation
of other transport properties. Finally, we will review the derivation 
of higher order fluid dynamics using holography, and summarize some
recent attempts to extend the correspondence to non-relativistic 
theories.

\subsection{The equation of state from holography}
\label{sec_ads_eos}

${\rm AdS}_5\times {\rm S}_5$ is the product of five dimensional 
Anti-deSitter Space (AdS) and a five-sphere. Anti-deSitter space 
is a simple solution of the source free Einstein equation with 
a negative cosmological constant. Note that, on large scales, 
our universe is a approximately a four dimensional deSitter 
space. The geometry of ${\rm AdS}_5\times {\rm S}_5$ is described
by the metric
\be
\label{vacuumads}
 ds^2 = \frac{r^2}{\Rads^2}\left( -dt^2 + d\x^2 \right) 
   + \frac{\Rads^2}{r^2}\, dr^2
   + L^2 d\Omega_5^2 \, . 
\ee
Here, $d\Omega_5^2$ is the metric of the five-sphere, and 
$(t,{\bf x},r)$ are the coordinates on ${\rm AdS}_5$. The coordinate 
$r$ is referred to as the ``radial'' ${\rm AdS}_5$ coordinate. The 
limiting value $r\to\infty$ is the ``boundary'' of ${\rm AdS}_5$. A 
fixed $r$ slice of ${\rm AdS}_5$ is a $3+1$ dimensional flat Minkowski 
space, but the five dimensional space is curved, with a constant negative 
curvature. $\Rads$ is the corresponding curvature radius. We require that 
$\Rads$ is large compared to the string length $\ell_s$ which guarantees 
the validity of the classical approximation. In the AdS/CFT correspondence 
$\Rads$ is related to the coupling constant of the $\N=4$ gauge theory, 
$\lambda \equiv g^2_{YM} N_c$, through the relation $(\Rads/\ell_s)^4
= \lambda$. The classical approximation to the gravitational theory
is reliable if the field theory is strongly coupled. The classical
fields can be expanded in $S_5$ spherical harmonics. At strong coupling
higher harmonics are separated by a large gap, and we will ignore
the $S_5$ from now on.

 The metric of an ${\rm AdS}_5$ black hole is
\be
 ds^2 = \frac{r^2}{\Rads^2} \left( -f(r) dt^2 + d\x^2  \right) 
+ \frac{\Rads^2}{f(r) r^2 }\, dr^2\, , 
\ee
where $f(r) = 1 - (r_0/r)^4$. The black hole horizon is a $3+1$ 
dimensional surface at $r=r_0$. The horizon radius is related 
to the Hawking temperature of the black hole by $r_{0} \, \hbar/\Rads^2 
= \pi T_H$. This formula is an example of a general radius-energy 
relation in the AdS/CFT correspondence. A modification of the AdS 
geometry at radius $r$ corresponds to a modification of the field
theory at an energy scale $r\hbar/L^2$. It is convenient to perform
a change variables $u\equiv (r_0/r)^2$ and write the metric as
\be
\label{bh_son}
 ds^2 = \frac{(\pi T \Rads)^2}{u}  \left(-f(u) dt^2 + d\x^2 \right) + 
\frac{\Rads^2}{4 u^2 f(u)} du^2\, ,
\ee
where $f(u)=1-u^2$. Now the horizon 
is at $u=1$. The boundary limit is found by evaluating all quantities 
at $u=\epsilon$ and then taking the boundary limit $\epsilon \rightarrow 0$.

 As discussed in the introduction to this section, the modified 
metric implies that there is an induced stress tensor at the boundary, 
$u=\epsilon$. This is an important point, and we will compute the 
induced stress tensor in two different ways. First, we will determine 
it by varying the action with respect to the boundary metric. This 
is the standard method by means of which one can determine the source 
of a given gravitational field. The only unusual ingredient is the 
fact that induced stress tensor is located on the boundary. We will 
provide an alternative derivation based on the analogy with the induced 
surface charge in electrodynamics below.

 The boundary metric $g_{\mu\nu}$  is related to the metric of the 
five dimensional theory $G_{\mu\nu}$ by the AdS scale factor
\be
   g_{\mu\nu} \equiv \frac{u}{(\pi T \Rads)^2 } \, G_{\mu\nu} \, .
\ee
Here and below Greek letters denote four dimensional indices $(x^\mu) 
= (t,x,y,z)$ while Roman letters denote five dimensional indices $(x^M) 
= (x^\mu,u)$.  Near the boundary the metric can be written 
\be
  g_{\mu\nu} = g_{\mu\nu}^{o}  +  u^2  {\mathcal B}_{\mu\nu}  + O(u^4)\, ,
\ee
where $g_{\mu\nu}^o$ is interpreted as the metric of $\N=4$ gauge theory. 
Usually $g_{\mu\nu}^o$  is simply $\eta_{\mu\nu}$. We will see that 
the coefficient of $u^2$ determines the induced stress tensor on the 
boundary. 

The induced stress tensor is 
\be
\label{variation}
  \llangle T_{\mu \nu}\rrangle = 
  \lim_{\epsilon \rightarrow 0} \left. 
  \frac{-2}{\sqrt{-g}} \frac{\delta S}{\delta {g}^{\mu\nu} } 
       \right|_{u=\epsilon}  \, ,
\ee
where $\sqrt{-g} = (-{\rm det}\, g_{\mu\nu})^{1/2}$. The action is a sum 
of the Einstein-Hilbert action,  the Gibbons-Hawking-York boundary term,
and counter terms which are needed to render the action finite in 
the limit $u\rightarrow 0$,
\bea
  S &\equiv& S_{EH} + S_{GH} + S_{CT} \, .  
\eea
The Einstein-Hilbert action is 
\be
 S_{EH} =\frac{1}{2\kappa_5^2}\int_{\mathcal M} d^5x \, 
    \sqrt{-g} \left(\mathcal{R} + 2\Lambda\right) \, , 
\ee 
where ${\mathcal R}$ is the Ricci scalar and $\Lambda = 6/\Rads^2$ 
is the cosmological constant. The five dimensional Newton constant 
$1/\kappa_5^2$ is related to the number of colors in the field theory,
$1/\kappa_5^2 = N_c^2/(4\pi^2 \Rads^3)$. The Gibbons-Hawking-York 
\cite{Gibbons:1976ue,York:1972sj} boundary action is 
\be
 S_{GH}= \frac{1}{2\kappa_5^2} 
    \int_{\partial {\mathcal M}} d^4x \sqrt{-\gamma} \,2K \, ,
\ee
where we have defined the boundary metric
\be
\gamma_{\mu\nu} = \left. G_{\mu\nu} \right|_{u=\epsilon}  \, ,
\ee
and $K$ is the trace of the extrinsic curvature\footnote{More
explicitly, $K=G^{\mu\nu} \nabla_\mu n_\nu $ with $n^{M}$ an 
outward directed normal to the boundary of the AdS space, $n^{M} 
= - \sqrt{G^{55}} \delta^{5M}$. Note that $K_{\mu\nu} = \nabla_{\mu} 
n_{\nu} = -n_{u} \Gamma^{u}_{\mu\nu} =  n^u \partial_u G_{\mu\nu}$.}.  
The boundary term guarantees that the variation of the action with
respect to the 5-dimensional metric gives the Einstein equations 
in the bulk provided the variation vanishes on the boundary. Without
the boundary action one also has to require that derivatives of the 
variation vanish on the boundary, see \cite{Wald:1984}. Finally, 
the counter term 
\be
 S_{CT} = -\frac{6}{\Rads} 
     \int_{\partial {\mathcal M}} d^4x \sqrt{\gamma} \, , 
\ee
is needed to render the action finite in the limit $u\rightarrow 0$. 
Notice that the counter term is independent of temperature. With these 
definitions, the variation relates the stress tensor to the extrinsic 
curvature 
\be
\label{ex_curvature}
  \llangle T_{\mu\nu} \rrangle =-\frac{1}{\kappa_5^2} 
   \lim_{u\rightarrow 0} \frac{(\pi T L)^2}{u} 
    \left[ K_{\mu\nu}  - K \gamma_{\mu\nu}  
        +  \frac{3}{L} \gamma_{\mu\nu} \right] \, .
\ee
Substituting the black hole metric equ.~(\ref{bh_son}) and using the 
definition of the extrinsic curvature we have  
\be
\label{Tmunu}
   \llangle T_{\mu\nu} \rrangle = {\rm diag} (\epsilon,p,p,p)\, ,   
     \qquad \frac{\epsilon}{3} = p=\frac{N_c^2}{8\pi^2}  (\pi T)^4 \, .
\ee
We find that $\epsilon=3P$, as expected for a scale 
invariant theory. We can compare the coefficient of $T^4$
to its value in the non-interacting theory. ${\cal N}=4$ SUSY QCD 
has $8(N_c^2-1)\simeq 8N_c^2$ bosonic and fermionic degrees 
of freedom. The contribution of a massless fermion to the 
pressure is 7/8 of that of a massless boson, see equ.~(\ref{P_SB}).
Equation (\ref{Tmunu}) shows that the pressure in strongly 
coupled  ${\cal N}=4$ SUSY QCD is three quarters of the 
Stefan Boltzmann value.

 We can also obtain equ.~(\ref{ex_curvature}) in 
analogy with the induced surface charge on a capacitor plate. 
Consider a plate that spans the $x$-$y$ plane. The surface 
charge density is related to the jump of electric field  
across the  plate
\be
  \sigma = \Big[ E^z \Big] \, , 
\ee
where $\left[E^z\right] = E^z_{+} - E^z_{-}$  notates the jump.
The analogous formulas in the gravitational theory  are known 
as  junction conditions \cite{Misner:1974qy}. Integrating the 
Einstein equations across a Gaussian pill box relates the 
surface stress $\tau^{\mu}_{\phantom{\mu}\nu}$ to the jump in 
extrinsic curvature
\be
  \tau^{\mu}_{\phantom{\mu} \nu} = -\frac{1}{\kappa_5^2} 
     \Big[ K^{\mu}_{\phantom{\mu}\nu}  
         - K \delta^{\mu}_{\phantom{\mu}\nu} \Big] \, .
\ee
Thus the particular combination of extrinsic curvature plays an 
analogous role to the normal electric field, i.e. a combination of
$-K_{\mu\nu}= n_{u} \Gamma^{u}_{\mu\nu} $ is the analog of 
${\bf n}\cdot {\bf E}$. If we have a semi-infinite metal block 
with surface charge density $\sigma$, then the outgoing electric 
field  is related to the surface charge $E^{z} =\sigma$. By analogy, 
we associate the outgoing flux of extrinsic curvature at $u=\epsilon$
with the stress tensor in the gauge theory
\be
   \sqrt{-g} T^{\mu}_{\phantom{\mu}\nu} 
        = -\frac{1}{\kappa_5^2} \sqrt{\gamma} 
         \left( K^{\mu}_{\phantom{\mu} \nu} 
              - K\delta^{\mu}_{\phantom{\mu}\nu} \right) \, .
\ee
Then taking the boundary limit $u\rightarrow 0$, we tentatively 
define the stress
\be
  T^{\mu}_{\phantom{\mu}\nu}  = 
     -\frac{1}{\kappa_5^2} \lim_{u \rightarrow 0} 
      \frac{(\pi T L)^4}{ u^2} 
       \left( K^{\mu}_{\phantom{\mu}\nu} 
            - K \delta^{\mu}_{\phantom{\mu}\nu} \right) \, .
\ee
Substituting the black hole AdS metric into this expression gives a
divergent result. Nevertheless, the difference between this stress and
the stress determined with the vacuum AdS metric equ.~(\ref{vacuumads}) 
is finite
\be
  \llangle T^{\mu}_{\phantom{\mu}\nu} 
      \rrangle - \llangle T^{\mu}_{\phantom{\mu}\nu} \rrangle_{\rm vacuum}   
   = -\frac{1}{\kappa_5^2} \lim_{u \rightarrow 0} 
     \frac{(\pi T L)^4}{u^2} 
       \left( K^{\mu}_{\phantom{\mu}\nu} 
            - K \delta^{\mu}_{\phantom{\mu}\nu}   
      + \frac{3}{L} \delta^{\mu}_{\phantom{\mu}\nu} \right) \, .
\ee
After lowering the indices of $K^{\mu}_{\phantom{\mu}\nu}$ with 
$\gamma_{\mu\nu} = \left[ (\pi T L)^2/u \right] \, \eta_{\mu\nu}$, this 
equation is the same as derived previously in equ.~(\ref{ex_curvature}).

\subsection{Shear viscosity from holography}
\label{sec_ads_cor}

 In the previous section we computed the average stress tensor on 
the boundary, $\llangle T^{\mu\nu}(\x,t) \rrangle$, which is a 
one point function of the conformal field theory. By Kubo's formula, 
equ.~(\ref{eta_kubo}), the shear viscosity can be related to a retarded 
two-point function. We will determine this function using linear response 
theory. Momentarily ignore the fifth dimension and consider turning on 
a time varying gravitational field $h_{xy}^{o}(\omega)$ in the usual 
four dimensional field theory. This time varying gravitational field 
induces a deviation  from the equilibrium stress tensor  in the same 
way that a time varying electric field induces a net current. According 
to linear response theory, the expectation value of the stress energy 
tensor is
\be
   \llangle T_{xy}(\omega) \rrangle_{h_{xy}^o }  = 
    T^{\rm eq}_{xy}(\omega)  + G_{R}(\omega)\, h_{xy}^o(\omega) \, ,
\ee 
where $T^{\rm eq}_{xy}= (\epsilon + p) u_{x}u_{y} + p g_{xy} = 
p h_{xy}^o(\omega)$ is the equilibrium stress tensor, and $G_{R}(\omega)$ 
is the equilibrium retarded correlator defined in equ.~(\ref{G_ret}). 
Kubo's formula dictates the functional form of this correlator in the 
small frequency limit, $G_R(\omega) = -i\omega \eta$. Thus the average 
stress tensor in the presence a time varying gravitational field is 
\be
\label{txyfromhxy}
   \llangle T_{xy}(\omega) \rrangle_{h_{xy}^o}  
    = p h_{xy}^{o} - i\omega \eta \, h_{xy}^o(\omega)  \, .
\ee
Now consider the small fluctuations of the metric field $H_{xy}(\omega, u)$ 
around the black hole metric equ.~(\ref{bh_son}) of the five dimensional 
theory. The equation of motion  for the gravitational fluctuation
is found by linearizing the Einstein equations  
\be
 {\mathcal R}_{MN}  -\frac{1}{2} G_{MN} 
    \left( {\mathcal R} + 2\Lambda\right) = 0 \, .
\ee
After a modest amount of algebra, the ${\mathcal R}_{xy}$ equation 
becomes an equation for $h_{xy} \equiv u H_{xy}(\omega,u)/(\pi T L)^2$
\be
\label{lingrav}
h_{xy}''(\omega,u) - {1+u^2\over uf} h_{xy}'(\omega,u) 
      + {\omega^2 \over  (2\pi T)^2 u f^2}
  h_{xy}(\omega,u) = 0\, ,
\ee
where the primes denote derivatives with respect to $u$. This is a 
second order linear differential equation with regular singular 
points in the physical domain at the horizon $u=1$ and 
the boundary $u=0$. Solving equ.~(\ref{lingrav}) near the black 
hole horizon $u=1$, we determine that the fluctuation
of the metric is a linear 
combination  of two solutions, $h_{xy}(\omega, u) \sim (1-u)^{\mp 
i\omega/4\pi T}$. These solutions describe the  gravitational 
wave propagating into $(-)$ and out of $(+)$ the black hole, respectively.   
The infalling solution is the physically relevant retarded solution. 
Near the boundary $u\rightarrow 0$ (or $r\rightarrow \infty$) the 
gravitational field is also a linear combination of two solutions
\be
\label{bndry_lin}
  h_{xy}(\omega,u) = h_{xy}^o(\omega) \left(1 + \ldots\right)    
  +   {\mathcal B}(\omega) \, u^2\left(1 + \ldots\right) \, ,
\ee
where $\ldots$ denotes terms that vanish as $u \to 0$. The two 
modes are called the non-normalizable mode and the normalizable 
mode. The non-normalizable mode is constant as $r\rightarrow\infty$  
while the normalizable mode falls as $1/r^4$. Inserting the metric 
perturbation equ.~(\ref{bndry_lin}) into equ.~(\ref{ex_curvature}) 
the average stress tensor is 
\be
\label{brelation}
  \llangle T_{xy}(\omega) \rrangle  = p \, 
    h_{xy}^o(\omega) +   (\epsilon + p)\, {\mathcal B}(\omega)\, ,    
\ee
with the previously defined energy density and pressure, equ.~(\ref{Tmunu}). 
We observe that the coefficient of the non-normalizable mode, $h_{xy}^o$,
can be interpreted as the external gravitational field applied to the
gauge theory,  while the coefficient of the normalizable mode,
${\mathcal B}(\omega)$ is proportional to the induced stress 
tensor in the boundary theory.

\begin{figure}
\begin{center}
\includegraphics[height=2.5in]{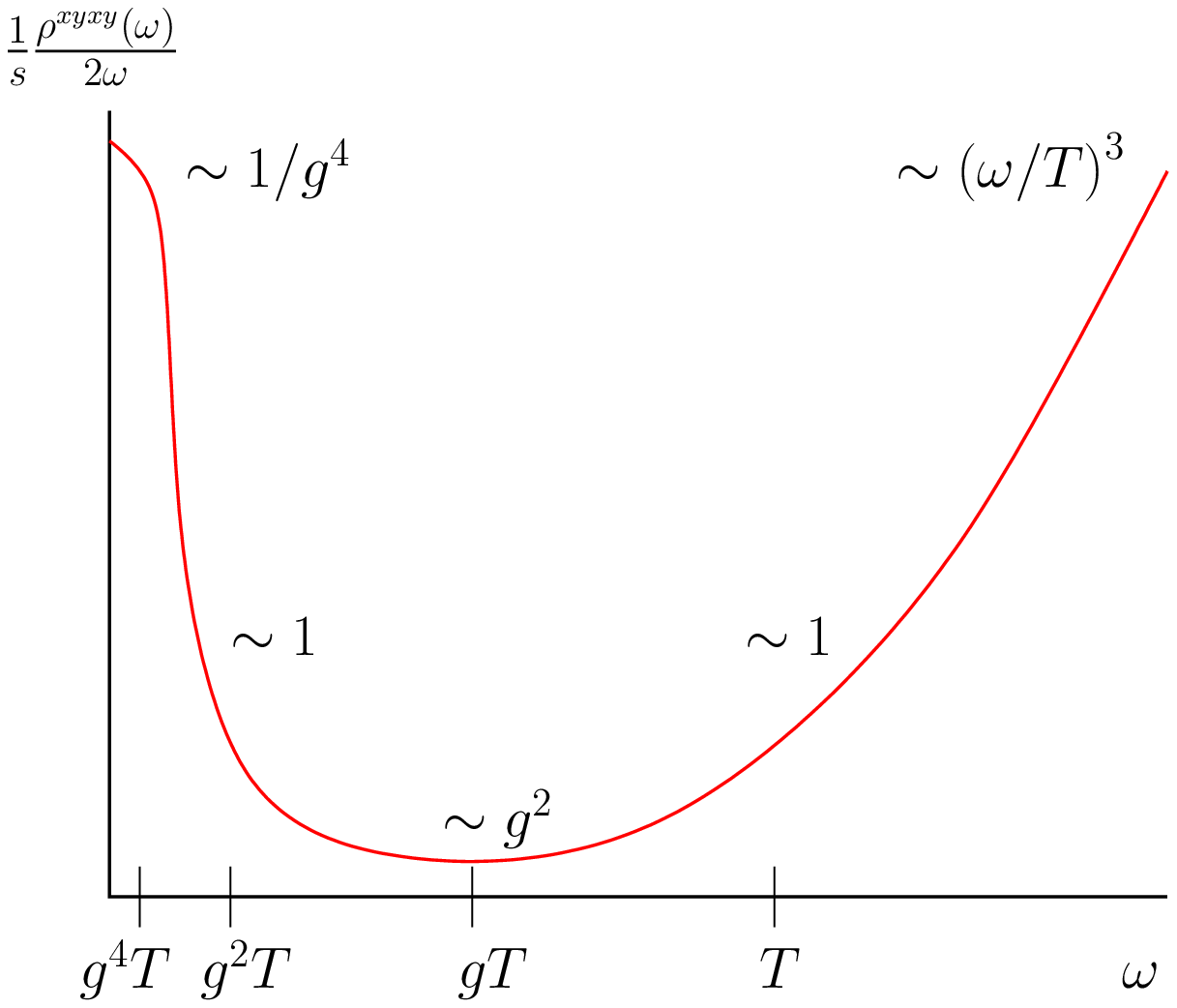}
\hspace{0.2in}
\includegraphics[height=2.5in]{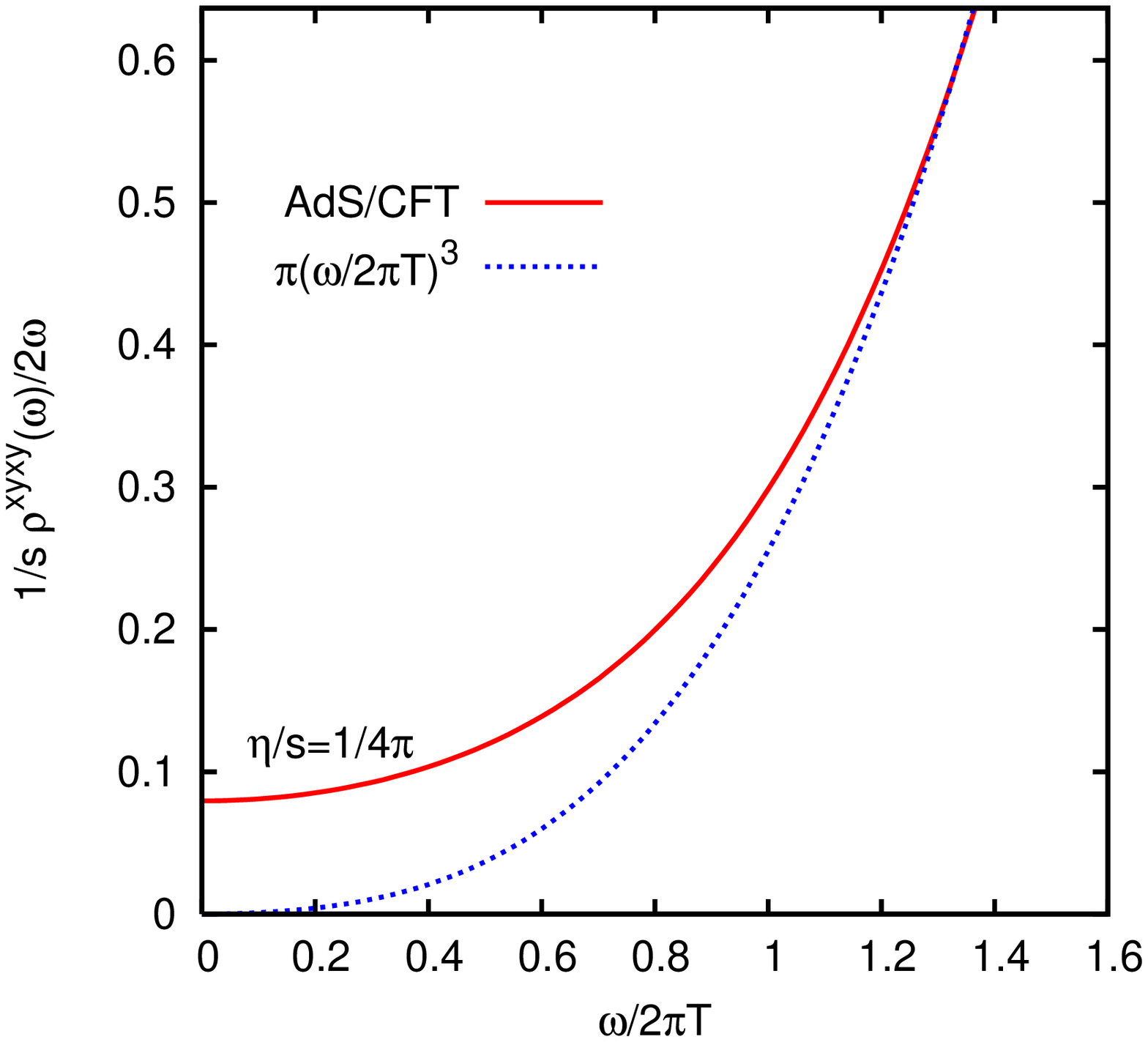}
\end{center}
\caption{Spectral function $\rho^{xyxy}(\omega,{\bf k}\! =\! 0)$ 
associated with the correlation function of the $xy$ component of 
the energy momentum tensor. The spectral function is normalized
to entropy density $s$. Left panel (Fig.~(a)): Schematic picture 
of the spectral density in weak coupling QCD or SUSY Yang Mills theory 
\cite{Aarts:2002cc,Moore:2008ws}. Right panel (Fig.~(b)): Spectral 
density in strong coupling SUSY Yang-Mills theory calculated using 
the AdS/CFT correspondence, from \cite{Teaney:2006nc}.  
\label{fig_rho_Txy}}
\end{figure}

  For an arbitrary value of ${\mathcal B}(\omega)$ the general linear 
combination of solutions near the boundary would approach a linear 
combination of the infalling and outgoing solutions near the horizon. 
Thus the coefficient ${\mathcal B}(\omega)$ should be adjusted so that 
only the infalling solution $(1-u^2)^{-i\omega/4\pi T}$ is present near 
$u=1$. In general the required ${\mathcal B}(\omega)$ has to be determined 
numerically. For small $\omega$ however, a straightforward calculation 
shows that to linear order in $\omega$ the solution which is infalling 
at the horizon is
\bea
h_{xy} &=& h_{xy}^{o}(\omega)\,  
    (1-u)^{-i\omega/4\pi T} 
    \left[ 1- \frac{i\omega }{4\pi T} \log(1 +u) + O(\omega^2) \right]  \, . 
\eea
Expanding this functional form near the boundary we find ${\mathcal B}
(\omega) = -i\omega/(4\pi T)$. Then using $\epsilon+p=s T$ and comparing 
the functional forms in equ.~(\ref{txyfromhxy}) and equ.~(\ref{brelation}) 
we conclude that $\llangle T_{xy}(\omega)\rrangle  = p h_{xy}^o -i\omega 
\eta h_{xy}^o$ with
\be
  \frac{\eta}{s}=  \frac{1}{4\pi}   \, .
\ee
Remarkably, the strong coupling limit of the shear viscosity is 
small and independent of the coupling. The difference as compared 
to the weak coupling result becomes even clearer if one considers 
the spectral function. As described in Sect.~\ref{sec_kubo} the 
Kubo formula relates the shear viscosity to the zero energy limit 
of the stress-energy spectral function. In weak coupling QCD 
the spectral function has a narrow peak near zero energy
which reflects the fact that momentum transport is due 
to quasi-particles that are almost on-shell. The height of 
the transport peak is governed by the kinetic theory result
for the shear viscosity. Kubo's formula implies that 
$\rho(\omega)/\omega\sim T^3/g^4$ as $\omega\to 0$. The width 
can be reconstructed from the $f$-sum rule
\be
\label{f_sr}
T \int_0^\Lambda \frac{d\omega}{\omega} \rho^{xyxy}(\omega)  =
 \frac{T(\epsilon+P)}{5}  \, , 
\ee
where $g^4T \ll \Lambda \ll g^2T$. Since the height of the 
transport peak is $T^3/g^4$, the width must be $g^4T$. The 
high energy part of the spectral density can be computed from 
the one-loop correlation function. The result is $\rho(\omega)
\sim \omega^4$. A schematic picture of the spectral function 
is shown in Fig.~\ref{fig_rho_Txy}(a). 

 In the strong coupling limit the width of the transport peak 
becomes bigger, and the height becomes smaller. In $\N=4$ SUSY 
Yang Mills the infinite coupling limit can be determined 
as outlined above \cite{Teaney:2006nc,Kovtun:2006pf}. Specifically, 
the spectral function may be found by determining $G_{R}(\omega)$ 
from the numerical coefficient ${\mathcal B}(\omega)$. The result 
is shown in Fig.~\ref{fig_rho_Txy}(b). Clearly, the transport peak 
has completely disappeared, and there is no possibility of a 
quasi-particle interpretation of momentum transport. Whether the 
spectral function of the quark gluon plasma near $T_c$ looks more 
like Fig.~\ref{fig_rho_Txy}(a) or (b) will have to be settled by 
numerical calculations on the lattice, see Sect.~\ref{sec_kubo}. 
It is interesting to note that numerical calculations of the shear 
viscosity, which require the determination of the zero 
energy limit of $\rho(\omega)/\omega$, are easier in the case of 
strong coupling than they are for weak coupling. 

\subsection{The KSS bound}
\label{sec_kss}

 The calculation of the shear viscosity has been extended to other 
strongly coupled field theories with gravitational duals. It was
discovered that within a large class of theories the strong coupling 
limit of $\eta$ depends on the theory, but the ratio $\eta/s$ does not. 
This observation can be understood using Kubo's formula and the 
optical theorem. The optical theorem implies that the imaginary 
part of a correlation function can be related to the total cross
section. As a consequence, the shear viscosity can be expressed
in terms of the total graviton absorption cross section 
\cite{Policastro:2001yc},
\be 
\eta = \frac{\sigma_{\rm abs}(0)}{2\kappa_5^2}\, . 
\ee
The low energy limit of $\sigma_{\rm abs}$ is equal to the area $A$ 
of the event horizon, and the entropy density is given by the 
Hawking-Bekenstein formula, $s=A/(4G)$ where $G=\kappa_5^2/(8\pi)$.
The ratio $\eta/s$ is independent of $A$ and $\kappa_5$. More formal 
arguments for the universality of $\eta/s$ in the strong coupling 
limit of field theories with holographic duals were given in 
\cite{Buchel:2003tz,Son:2007vk,Iqbal:2008by}. Corrections to the 
infinite coupling limit of ${\cal N}=4$ SUSY Yang Mills theory were 
studied in \cite{Buchel:2004di,Buchel:2008ac,Buchel:2008sh,Myers:2008yi}. 
The result is
\be 
\label{eta_s_l32}
\frac{\eta}{s} = \frac{1}{4\pi} \left\{ 
 1 +  \frac{15\zeta(3)}{\lambda^{3/2}} + \ldots \right\} \, . 
\ee
The first correction is positive, as one would expect from the 
fact that $\eta/s\to \infty$ as $\lambda\to 0$. 
Based on these observations, Kovtun, Son, and Starinets (KSS) 
conjectured that 
\be
\label{KSS}
\frac{\eta}{s}\geq \frac{1}{4\pi}
\ee 
is a universal bound that applies to all fluids \cite{Kovtun:2004de}. 
There is no proof of this conjecture, and a number of authors have 
attempted to construct counter examples. One possibility is a weakly 
interacting non-relativistic fluid with an exponentially large 
number of species or internal degrees of freedom, and therefore 
a very large entropy \cite{Kovtun:2004de,Cohen:2007qr,Dobado:2007tm}.
These systems are unusual because the time scale for thermal 
equilibration vastly exceeds the time scale for momentum 
equilibration, and because the fluid is not stable on very 
long time scales  \cite{Son:2007xw}. More recently, it was 
realized that theories with holographic duals described by
higher derivative gravity may violate the KSS bound
\cite{Brigante:2007nu,Brigante:2008gz,Kats:2007mq,Buchel:2008vz}.
An explicit example was constructed by Kats and Petrov 
\cite{Kats:2007mq}. They showed that in ${\cal N}=2$ SUSY
$Sp(N_c)$ gauge theory with a certain combination of matter fields
\be 
\frac{\eta}{s} = \frac{1}{4\pi}\left( 1-\frac{1}{2N_c} \right)\, ,  
\ee
up to corrections of $O(\lambda^{-3/2})$. For $\lambda^{3/2}\gg
N_c\gg\lambda\gg 1$ we find a violation of the KSS bound in a
controlled calculation. However, there are bounds on the 
coefficients of higher derivative terms, and a modified bound
on $\eta/s$ may yet exist \cite{Buchel:2008vz}.

\subsection{Other transport properties}
\label{sec_holog_other}

 There has been a large amount of work on applications of the 
AdS/CFT correspondence to transport properties other than the
shear viscosity. Here we briefly summarize some results relevant 
to this review. ${\cal N}=4$ SUSY Yang Mills theory has a conserved 
R-charge (see Sect.~\ref{sec_susy}), and we can study transport in 
the presence of a finite R-charge density. Son and Starinets find
that the shear viscosity and entropy density depend on the density, 
but the ratio $\eta/s$ does not \cite{Son:2006em}. They also 
determine the thermal conductivity 
\be 
\kappa = \frac{8\pi^2T}{\mu^2}\, \eta \, ,
\ee
as well as the R-charge diffusion constant. The heavy quark
diffusion constant was calculated in 
\cite{Herzog:2006gh,CasalderreySolana:2006rq,Gubser:2006bz}.
The result is 
\be 
D=\frac{2}{\pi T} \frac{1}{\sqrt{\lambda}},
\ee
which depends on the value of the coupling $\lambda$, and goes to 
zero in the strong coupling limit. The functional dependence on 
$\lambda$ is unusual from the point of view of perturbation theory, 
but typical of other AdS/CFT results. We also note that in the
strong coupling limit the ratio of the heavy quark diffusion coefficient
to the kinematic viscosity $\eta/(sT)$ goes to zero, whereas this 
ratio is independent of the coupling in the perturbative limit

 The bulk viscosity of ${\cal N}=4$ SUSY Yang Mills theory vanishes,
 but non-conformal deformations of the original AdS/CFT correspondence 
have been studied. Buchel proposed that in holographic models there 
is a lower bound on the bulk viscosity, $\zeta\geq 2(\frac{1}{3}-c_s^2)
\eta$, where $c_s$ is the speed of sound \cite{Buchel:2007mf}. Note that 
the weak coupling formula involves the square of $(\frac{1}{3}-c_s^2)$.
Gubser et al.~considered a number of model geometries tuned to reproduce 
the QCD equation of state, and find that $\zeta/s$ has a maximum near 
the critical temperature where $\zeta/s\simeq 0.05$ \cite{Gubser:2008sz}.
Larger values of $\zeta/s$ near $T_c$ have been suggested based
on lattice data for the QCD trace anomaly \cite{Karsch:2007jc}.

\subsection{Hydrodynamics and holography}
\label{sec_holog_hydro}

 Up to this point we have used the AdS/CFT correspondence to 
calculate the transport coefficients that appear in first order 
hydrodynamics. However, AdS/CFT can be used to compute the full 
correlation function, and not just the hydrodynamic limit. An example 
is the spectral function shown in Fig.~\ref{fig_rho_Txy}, 
and similar calculations have been performed in other channels
as well. In this section we wish to discuss how the stress tensor
of the fluid relaxes to the Navier-Stokes form. This process can 
be described by the second order terms introduced in 
Sect.~\ref{sec_rel_fluid}. We will follow the method outlined 
in~\cite{Bhattacharyya:2008jc}.

 A static fluid at temperature $T$ corresponds to a black hole 
with a Hawking temperature $T_H=r_0/(\pi L^2)$. First we switch 
to Eddington-Finkelstein coordinates, defining 
a new time coordinate $v= t + \int^r dr\, 
L^2/fr^2$. Then the metric is regular at the event horizon of the 
black hole, 
\be
 ds^2 = 2 \, dv\,  dr  +  \frac{r^2}{\Rads^2}\left[- f(r)  dv^2 
    + r^2  d\x^2 \right] \, .
\ee
Introducing four dimensional coordinates $x^{\mu} = (x^0,x^1,x^2,x^3) 
= (v,\x)$, a vector $u^{\mu} = (1,{\bf 0})$ characterizing the local 
rest frame, and a scale parameter $b$ characterizing the temperature 
the metric becomes
\be
 ds^2 = - 2 u_{\mu} dx^{\mu} dr 
        +  \frac{r^2}{\Rads^2} \left[ - f(b r) u_{\mu} u_{\nu} 
               dx^{\mu} dx^{\nu} 
        + r^2 P_{\mu\nu} dx^{\mu} dx^{\nu} \right] \, , 
\ee
where $P_{\mu\nu}=u_\mu u_\nu +\eta_{\mu\nu}$. The basic idea is 
to promote the variables $u^{\mu}$ and $b$ to slowly varying 
functions of $x^{\mu}$. The metric is then 
\bea
 ds^2 &=& - 2 u_{\mu}(x) dx^{\mu} dr  \nonumber \\
      & &\quad  + \frac{r^2}{L^2}\left[ -  f(b(x) r) u_{\mu}(x) u_{\nu}(x) 
               dx^{\mu} dx^{\nu} 
        + r^2 P_{\mu\nu}(x) dx^{\mu} dx^{\nu}  \right] \nonumber  \\
      & & \qquad    + \mbox{ corrections due to gradients} \, .
\eea
Variations in $u_\mu$ and $b$ correspond to fluctuations in 
the local fluid velocity and temperature. Substituting this form 
into the Einstein equations, the corrections to the metric are 
determined order by order in the gradients of $u^{\mu}(x)$ and 
$b(x)$. These metric corrections lead to deviations of the boundary 
stress tensor from an ideal fluid  of precisely the form required 
by hydrodynamics. Up to second order we can write 
\be
  T^{\mu\nu} = T^{\mu\nu}_{0} + \delta^{(1)}T^{\mu\nu} 
    + \delta^{(2)}T^{\mu\nu}  + \ldots \, ,  
\ee
and each term has physical significance. At zeroth order
\be
T^{\mu\nu}_{0} = \frac{N_c^2}{8\pi^2} (\pi T)^4 
  \left(\eta^{\mu\nu} + 4 u^{\mu} u^{\nu} \right) \, , 
\ee
which shows that $\epsilon=3P$ and that the pressure is $3/4$ of 
the Stefan Boltzmann value. At first order 
\be
\delta^{(1)} T^{\mu\nu} = -\frac{N_c^2}{8\pi^2} 
          (\pi T)^3 \sigma^{\mu\nu} \, ,
\ee
where $\sigma^{\mu\nu}$ is defined as in equ.~(\ref{sig_vis}). This
results shows that $\eta = N_c^2 \pi T^3/8$. Combined with the 
zeroth order stress tensor we find $\eta/s=1/4\pi$, in agreement 
with previous results. Finally, at second order
\bea
\delta^{(2)}T^{\mu\nu}
  &=& \eta\tau_{II} \left[ ^\langle D\sigma^{\mu\nu\rangle}
 +\frac{1}{3}\sigma^{\mu\nu} (\partial\cdot u) \right] \\
& & \hspace{0.2cm}\mbox{}
 +\lambda_1\sigma^{\langle\mu}_{\;\;\;\lambda} \sigma^{\nu\rangle\lambda}
 +\lambda_2\sigma^{\langle\mu}_{\;\;\;\lambda} \Omega^{\nu\rangle\lambda}
 +\lambda_3\Omega^{\langle\mu}_{\;\;\;\lambda} \Omega^{\nu\rangle\lambda}\, ,
\nonumber
\eea
where $D=u\cdot\partial$, and the vorticity $\Omega_{\mu\nu}$ as well 
as the transverse traceless tensor $A^{\langle \mu\nu\rangle}$ are 
defined in Sect.~\ref{sec_rel_fluid}. The form of $T^{\mu\nu}_{(2)}$ 
agrees with the general second order result for a conformal relativistic 
fluid derived in \cite{Baier:2007ix}. The second order coefficients are
\be
\tau_{\Pi} = \frac{2 - \ln 2}{\pi T}\, , 
       \qquad   \lambda_1= \frac{2\eta}{\pi T}\, ,   
       \qquad   \lambda_2 = \frac{2\eta\ln 2}{\pi T}\, , 
       \qquad   \lambda_3 = 0\, .
\ee
We observe that the relaxation times are of order $(\pi T)^{-1}$,
the shortest time scale characterizing the plasma.

\subsection{Non-relativistic AdS/CFT correspondence}
\label{sec_nrcft} 

 Given the role that the AdS/CFT correspondence has played
in improving our understanding of conformal relativistic fluids
it is natural to ask whether the correspondence can be extended
to non-relativistic scale invariant fluids like the dilute Fermi 
gas at unitarity. There has recently been significant progress in 
constructing holographic duals for non-relativistic field theories
\cite{Son:2008ye,Balasubramanian:2008dm,Herzog:2008wg,Maldacena:2008wh}.

 The basic idea proposed in \cite{Son:2008ye,Balasubramanian:2008dm}
can be explained by looking at the metric of $d+2$ dimensional 
flat space
\be 
 ds^2 = \eta_{\mu\nu} dx^\mu dx^\nu 
      = -2dx^+dx^- + dx^idx^i\, ,
\ee 
where we have introduced light cone coordinates $(x^+,x^-,x^i)$
with $X^\pm=(x^0\pm x^{d+1})/\sqrt{2}$ and $i=1,\ldots,d$. 
Consider the massless Klein-Gordon equation in this space.
In light cone coordinates
\be 
\label{KG_lc}
 \left( -2 \frac{\partial}{\partial x^-}
           \frac{\partial}{\partial x^+} 
        + \sum_{i=1}^{d} {\partial^2}{\partial x_i^2}
  \right) \phi(x) = 0\, . 
\ee
If the $x^-$-direction is compactified, then the corresponding 
momenta become discrete. We may write the lowest mode as 
$\phi(x)\sim e^{-imx^-}\psi(x^+,x_i)$ and the equation for $\psi$ 
becomes the non-relativistic Schr\"odinger equation
\be 
  \left( 2im \frac{\partial}{\partial x^+} 
        + {\bf \nabla}^2 \right) \psi(x^+,x_i) = 0\, , 
\ee
where $x^+$ plays the role of time. The symmetry group of this
equation is known as the Schr\"odinger group $Sch(d)$. The 
generators of the Schr\"odinger algebra include temporal and 
spatial translations, rotations, Galilean boosts, non-relativistic
dilatations (which scale space and time by different factors, 
${\bf x}\to s{\bf x}$ and $t\to s^2t$), a special conformal 
transformation (which scales $t\to t/(1+\lambda t)$ and ${\bf x}\to 
{\bf x}/(1+\lambda t)$), and the mass operator \cite{Hagen:1972pd}.

 The goal is to extend this construction to spaces that 
are asymptotically Anti-deSitter. The specific proposal in  
\cite{Son:2008ye,Balasubramanian:2008dm} is that the $Schr(d)$ 
symmetry of a non-relativistic $d+1$ dimensional conformal field 
theory can be mapped onto the isometries of the $d+3$ dimensional 
metric 
\be 
ds^2 = r^2\left( -2dx^+dx^- - \beta^2 r^2(dx^+)^2 +(dx^i)^2\right)
   + \frac{dr^2}{r^2},
\ee
which reduces to the metric of $AdS_{d+3}$ for $\beta\to 0$. 
This metric can be realized in string theory by starting from 
$AdS_5\times {\cal X}_5$, where ${\cal X}_5$ is a generalized
sphere called an Einstein-Sasaki manifold, and by applying a 
certain series of transformations that preserve solutions of
the Einstein equations \cite{Herzog:2008wg,Maldacena:2008wh,Adams:2008wt}. 
The resulting field theory is a $2+1$ dimensional field theory with 
infinitely many bosonic and fermionic fields, and an unusual 
equation of state $P\sim T^4/\mu^2$ \cite{Adams:2008wt}. 
This is still quite far from the $3+1$ unitary Fermi gas, but 
the theory provides an explicit realization of a non-relativistic
fluid which satisfies $\eta/s=1/(4\pi)$. 

 The hydrodynamics of a holographic fluid with Schr\"odinger 
symmetry was studied in more detail in \cite{Rangamani:2008gi}.
An interesting observation that was made in this paper is that 
the light cone reduction of a viscous relativistic stress tensor 
automatically leads to a $\vec\nabla T$ term in the non-relativistic 
energy current. The thermal conductivity is completely fixed 
by the shear viscosity and the equation of state, 
\be 
\kappa = 2\eta\frac{\epsilon+P}{\rho T}\,  . 
\ee
This result can be expressed in terms of the Prandtl number
${\it Pr}=c_p\eta/\kappa$, see equ.~(\ref{Prandtl}). Using the 
equation of state of a non-relativistic conformal fluid we find 
${\it Pr}=1$. The Prandtl number of many gases is indeed close 
to one, see Sect.~\ref{sec_kin_th}, but at strong coupling there 
is no obvious reason for the relation ${\it Pr}=1$ to hold.   

\section{Experimental determination of transport properties}
\label{sec_exp}

 In this section we will review experimental determinations
of transport properties of liquid helium, cold atomic gases, 
and the quark gluon plasma. We will focus on shear viscosity, 
since it is the main focus of this review, and since it is 
the only transport property for which good data is available 
for all three systems. 

 Liquid helium can be produced in bulk, and transport properties
can be measured using methods that were developed for classical
fluids. Cold atomic gases are produced in optical or magneto-optical
traps. These traps typically contain $10^5-10^6$ atoms. Hydrodynamic
behavior is observed when the trapping potential is modified, or 
if the local density or energy density is modified using laser
beams. The quark gluon plasma can only be created for brief periods 
in collisions of ultra-relativistic heavy ions. The system typically 
contains on the order of $10^3-10^4$ quarks and gluons, and lasts 
for about 10 fm/c ($3\cdot 10^{-23}$ sec). Hydrodynamic behavior
may take place during the expansion of the system and is reflected 
in the momentum spectra of particles in the final state.

\begin{figure}
\begin{center}
\includegraphics[width=8cm]{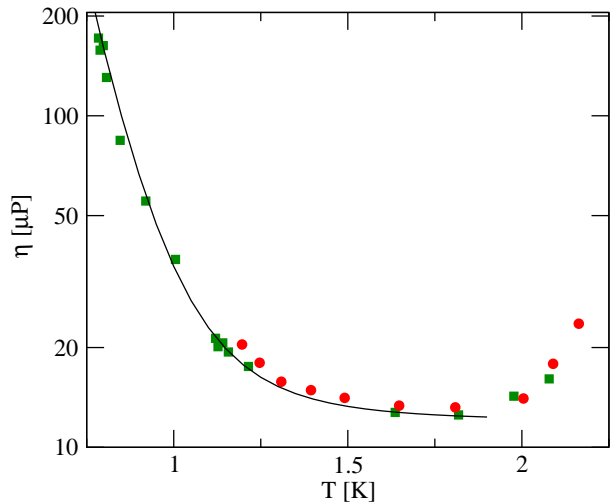}
\end{center}
\caption{\label{fig_eta_he}
Viscosity of $^4$He at atmospheric pressure as a function of 
temperature. Data taken from Woods and Hollis-Hallett (green squares)
\cite{Woods:1963} and Heikkila and Hollis-Hallett (red circles) 
\cite{Heikkila:1955}. This figure was adapted from \cite{Wilks:1966}. 
The solid line shows the theory of Landau and Khalatnikov. The 
viscosity minimum corresponds to $\eta/n\simeq 0.5$ and $\eta/s
\simeq 1.9$.}
\end{figure}

\subsection{Liquid helium}
\label{sec_heII}

 There are a number of techniques for measuring the viscosity 
of fluids. Three popular instruments are: 

\begin{enumerate} 
\item Capillary viscometers are based on Poiseuille flow. 
Poiseuille's formula states that the flow through a pipe
is inversely proportional to the shear viscosity, and 
proportional to the pressure drop as well as the fourth power 
of the diameter. 

\item Rotation viscometers measure the torque on a rotating 
cylinder or disk. The torque per unit length exerted by a pair 
of coaxial infinitely long cylinders is proportional to the shear 
viscosity and the difference between the angular velocities, and 
proportional to the ratio $R_1R_2/(R_1^2-R_2^2)$, where 
$R_{1,2}$ are the two radii. 

\item Vibration viscometers determine the damping of an 
oscillating sphere or plate. These devices have many advantages
but the data are more difficult to interpret, because the damping 
depends not only on the viscosity, but also on the density of the 
fluid.

\end{enumerate} 

 Initial measurements of the viscosity of superfluid liquid helium 
lead to an apparent contradiction between the results obtained 
using different methods. Capillary flow viscometers indicated 
vanishing viscosity  below $T_c$ \cite{Kapitza:1938}, oscillatory 
viscometers showed a drop of the shear viscosity \cite{Keesom:1938}, 
and experiments with rotation viscometers yielded a rise in viscosity 
below $T_c$ \cite{Woods:1963}. The contradictions can be resolved 
using superfluid hydrodynamics. The flow through a narrow capillary 
is entirely a superflow, and not sensitive to viscosity. Oscillation 
viscometers measure the product of viscosity and normal density, which 
drops with temperature. Modern measurements confirm the rise of 
viscosity below $T_c$ which is predicted by the phonon-roton theory,
see Fig.~\ref{fig_eta_he}.
The minimum viscosity of helium at normal pressure occurs just below
the $\lambda$ point where $\eta\simeq 1.2\cdot 10^{-5}$ Poise. The 
minimum of $\eta/s$ occurs at higher temperature, close to the 
liquid gas phase transition. Recent measurements confirm the (weak) 
divergence of the shear viscosity at the critical endpoint of the 
liquid gas phase transition predicted by dynamical universality 
\cite{Agosta:1987}. Experiments also find the expected (much stronger) 
divergence of the heat conductivity near the lambda point \cite{Lipa:1996}. 

 Once the shear viscosity and the heat conductivity are determined
sound attenuation experiments can be used to measure the bulk 
viscosity ($\zeta_2$ in the superfluid phase) \cite{Putterman:1974}. 
Below the $\lambda$ point $\zeta_2\simeq 10^{-4}$ Poise. Damping of 
second sound determines a linear combination of $\zeta_1$ and 
$\zeta_3$ in the superfluid phase \cite{Hanson:1954}, but the 
remaining linear combination is poorly constrained. 

\subsection{Cold atomic gases}
\label{sec_trap}

 Dilute Bose or Fermi gases are studied using optical traps that 
provide an approximately harmonic confinement potential 
\be 
\label{V_trap}
V(x) = \frac{m}{2}\sum_i \omega_i^2 x_i^2 .
\ee
The equilibrium density $n_0$ can be determined from the equation
of hydrostatic equilibrium, ${\bm \nabla}P_0=-n_0{\bm \nabla}V$. 
Using the Gibbs-Duhem relation $dP=nd\mu+sdT$ we can see
that this equation is solved by $n_0(x)=n(\mu(x))$, where $n(\mu)$
is the equilibrium density as a function of the chemical potential, 
and $\mu(x)=\mu-V(x)$. This result is known as the local density
(or Thomas-Fermi) approximation, introduced by Thomas and Fermi
in connection with the structure of heavy atoms. 
For dilute fermions at unitarity the equation of state at zero 
temperature is given by equ.~(\ref{l_sfl_pert}) and
\be 
\label{ThomasFermi}
n_0({\bf r}\,) = n_0(0) \left( 1-\sum_i \frac{x_i^2}{R_i^2}
 \right)^{1/\gamma}, \hspace{0.5cm}
 R_i^2 = \frac{2\mu}{m\omega_i^2} ,
\ee
where $\mu$ is the chemical potential and $\gamma=2/3$. The chemical 
potential is related to the Fermi energy by the universal parameter 
$\xi$ introduced in equ.~(\ref{xi}). Transport properties of strongly
interacting dilute Fermi gases can be extracted from a variety of 
experiments, free expansion from a deformed trap (elliptic flow) 
\cite{oHara:2002}, 
damping of collective oscillations 
\cite{Kinast:2004,Kinast:2004b,Bartenstein:2004,Kinast:2005,Altmeyer:2006}, 
sound propagation \cite{Joseph:2006}, and
expansion out of rotating traps \cite{Clancy:2007}.
In the following we shall concentrate on damping of collective
oscillations, as these experiments have been most carefully 
analyzed \cite{Kavoulakis:1998,Gelman:2004,Schafer:2007pr,Turlapov:2007}. 

We consider small oscillations around the equilibrium density,
$n=n_0+\delta n$. Since the damping is small, the motion is approximately
described by ideal hydrodynamics. The compressibility at constant 
entropy is  
\be
\left(\frac{\partial P}{\partial n}\right)_{S} = 
 (\gamma+1)\frac{P}{n} \ .
\ee
From the linearized continuity and Euler equation we get 
\cite{Heiselberg:2004}
\be 
\label{lin_eu}
m\frac{\partial^2{\bf v}}{\partial t^2} = 
 -\gamma\left({\bm \nabla}\cdot{\bf v}\right)
        \left({\bm \nabla} V\right)
 -{\bm \nabla}\left({\bf v}\cdot{\bm \nabla} V\right),
\ee
where we have dropped terms of the form $\nabla_i\nabla_j{\bf v}$
that involve higher derivatives of the velocity. This equation has
simple scaling solutions of the form $v_i=a_ix_i \exp(i\omega t)$ 
(no sum over $i$). Inserting this ansatz into equ.~(\ref{lin_eu})
we get an equation that determines the eigenfrequencies $\omega$. The
experiments are performed using a trapping potential with axial 
symmetry, $\omega_1=\omega_2=\omega_0$, $\omega_3=\lambda\omega_0$.
In this case we find one solution with $\omega^2 = 2\omega_0^2$ 
and two solutions with \cite{Heiselberg:2004,Stringari:2004,Bulgac:2004}
\bea
\label{w_rad}
 \omega^2 &=& \omega_0^2\Bigg\{ \gamma+1+\frac{\gamma+2}{2}\lambda^2 
   \\
 & & \hspace{1.5cm}\mbox{} 
  \pm \sqrt{ \frac{(\gamma+2)^2}{4}\lambda^4
           + (\gamma^2-3\gamma-2)\lambda^2 
           + (\gamma+1)^2 }\Bigg\}. \nonumber 
\eea
In the limit of a very asymmetric trap ($\lambda\to 0$) the 
eigenfrequencies are $\omega^2=2\omega_0^2$ and $\omega^2=(10/3)
\omega_0^2$. The mode $\omega^2=(10/3)\omega_0^2$ is a radial breathing 
mode with ${\bf a} = (a,a,0)$ and the mode $\omega^2=2\omega_0^2$ 
corresponds to a radial quadrupole ${\bf a} = (a,-a,0)$.

 The prediction of ideal hydrodynamics for the frequency of the radial 
breathing mode agrees very well with experimental results \cite{Kinast:2004}. 
Damping of collective modes is due to viscous effects. The dissipated 
energy is given by
\bea
\dot{E} &=& - \int d^3x\, \Bigg\{\frac{\eta(x)}{2}\, 
  \left(\nabla_iv_j+\nabla_jv_i-\frac{2}{3}\delta_{ij}
      {\bm \nabla}\cdot {\bf v} \right)^2  \\
 & & \hspace{2.5cm}\mbox{}
   + \zeta(x)\, \big( {\bm \nabla}\cdot{\bf v}\big)^2 
   + \frac{\kappa(x)}{T} ({\bm \nabla} T)^2\Bigg\}
  \nonumber , 
\eea
where $\eta(x),\zeta(x)$ and $\kappa(x)$ are the local shear viscosity, 
bulk viscosity, and thermal conductivity. In the unitarity limit the 
system is scale invariant and the bulk viscosity in the normal phase 
vanishes. In the superfluid phase 
there are three bulk viscosities, $\zeta_1,\zeta_2,\zeta_3$, 
see equ.~(\ref{del_pij_sfl}) and (\ref{del_H_sfl}). Scale invariance
implies $\zeta_1=\zeta_2=0$, see Sec.~\ref{sec_kin_zeta}, and the 
contribution of $\zeta_3$ vanishes if ${\bf v}_s={\bf v}_n$.
For isentropic oscillations 
$\delta T \sim (\delta n/n)T$. The solutions of equ.~(\ref{lin_eu})
satisfy $\delta n(x)\sim n_0(x)$. This implies that there are no 
temperature gradients, and that thermal conductivity does not 
contribute to dissipation. 

 We conclude that damping is dominated by shear viscosity. The
energy dissipated by the radial scaling solutions is 
\be 
\overline{\dot{E}} = -\frac{2}{3}
  \left( a_x^2+a_y^2-a_xa_y \right) \int d^3x\, \eta(x),
\ee
where $\overline{E}$ is a time average. The damping rate is given by 
the ratio of the energy dissipated to the total energy of the collective 
mode. The kinetic energy is 
\be 
 E_{kin} = \frac{m}{2}\, \int d^3x\, n(x) {\bf v}^{\,2} 
 = \frac{mN}{2} \left( a_x^2+a_y^2 \right) 
  \langle x^2 \rangle. 
\ee
In the case of a harmonic trapping potential the average 
$\langle x^2\rangle$ can be extracted using a virial theorem, 
$E=2N\langle V\rangle$ \cite{Thomas:2005}. The damping rate is 
\be 
\label{edot_e}
-\frac{1}{2} \frac{\overline{\dot{E}}}{E}   = 
 \frac{2}{3}\, \frac{a_x^2+a_y^2-a_xa_y}{a_x^2+a_y^2}\,
 \frac{\int d^3x\, \eta(x)}{mN \langle x^2\rangle} \, , 
\ee
where the factor 1/2 takes into account that the experiments
measure an amplitude, not energy, damping rate. We note that 
the second factor on the RHS is $1/2$ for the radial breathing 
mode and $3/2$ for the radial quadrupole mode. This dependence provides
an important check for the assumption that damping is dominated by 
shear viscosity. Also note that if the shear viscosity is proportional
to the density or the entropy density then $\dot{E}/E$ scales as 
$N^{-1/3}$. Near the surface the density is small and $\eta(x)$
will approach the Boltzmann limit, which is independent of density,
see equ.~(\ref{eta_aa}). This is a problem, because the total 
volume of the system is infinite (at non-zero temperature the 
density has an infinite range tail). This difficulty is related
to the breakdown of hydrodynamics near the surface of the cloud. 
An elegant solution to the problem is to include a finite relaxation 
time $\tau_\eta(r)=\eta/(n(r)k_BT)$ which diverges in the low density 
limit \cite{Bruun:2007}. 

\begin{figure}
\begin{center}
\includegraphics[width=8cm]{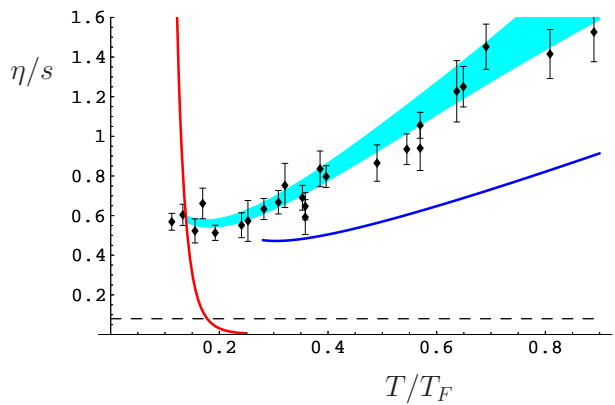}
\end{center}
\caption{\label{fig_eta_s}
Viscosity to entropy density ratio of a cold atomic gas in the unitarity 
limit, from \cite{Schafer:2007pr}. This data points are based on the 
damping data published in \cite{Kinast:2005} and the thermodynamic data 
in \cite{Kinast:2005b,Luo:2006}. The light blue band is an estimate 
of the systematic uncertainty due to the breakdown of hydrodynamics
near the surface of the cloud. The solid red and blue line show the
low and high temperature limits of $\eta/s$, see equ.~(\ref{eta_ph}) 
and (\ref{eta_aa}). The dashed line shows the conjectured viscosity 
bound $\eta/s=1/(4\pi)$. }
\end{figure}

 In order to compare with the proposed viscosity bound we will assume 
that the shear viscosity is proportional to the entropy density, $\eta(x)
=\alpha s(x)$. Note that in general $\alpha$ is a function of $T/T_F$
and varies across the trap ($T_F$ depends on density). This means that 
we will extract an average value of $\alpha=\eta/s$. We can write 
\be 
\label{eta_s}
\frac{\eta}{s} 
 =\frac{3}{4} \xi^{1/2} (3N)^{1/3} 
 \left(\frac{\bar\omega\Gamma}{\omega_\perp^2}\right)
 \left(\frac{E}{E_{T=0}}\right)
 \left(\frac{N}{S}\right), 
\ee
where $\Gamma/\omega_\perp$ is the dimensionless damping rate,
$\bar\omega=\omega_\perp^{2/3}\omega_z^{1/3}$ is mean trap
frequency, $(S/N)$ is the entropy per particle, and $E/E_{T=0}$ 
is the equilibrium energy of the cloud in units of the zero temperature 
value. Fig.~\ref{fig_eta_s} shows $\eta/s$ extracted from the experimental 
results of the Duke group \cite{Kinast:2005}. The entropy per particle 
was also taken from experiment \cite{Luo:2006}. Similar results are 
obtained if the entropy is extracted from quantum Monte Carlo data. 
We observe that $\eta/s$ in the vicinity of the transition temperature 
is about 1/2. We also note that the extracted shear viscosity roughly 
agrees with the high temperature, Fermion quasi-particle, kinetic 
theory result. The low temperature, phonon dominated, result is not 
seen in the data, presumably because the phonon mean free path is bigger 
than the system size. 

 There are many caveats that one should keep in mind regarding this 
analysis. First, we assume that shear viscosity is the only source 
of dissipation. There is some evidence for this assumption from 
comparisons of the damping rate of different collective modes 
\cite{Riedl:2008}. On the other hand, the dependence of the damping 
rate on particle number predicted by viscous hydrodynamics has never 
been demonstrated. Second, hydrodynamics can only be applied in a 
relatively narrow temperature regime $T<(2-3)T_c$. For higher 
temperatures the observed frequencies cross over from hydrodynamic 
behavior to a weakly collisional Boltzmann gas. This means that the 
kinetic theory prediction for the shear viscosity, equ.~(\ref{eta_aa}), 
is reliable but the frequency of the collective mode is too large
for hydrodynamics to be applicable. Finally, there is an issue 
that is specific to the scaling flows ($v_i\sim a_{ij} x_j$) 
considered here. Since the velocity field is linear in the 
coordinates, the second derivative of the velocity vanishes. This 
means that the viscous term in the Navier Stokes equation, $\rho 
\dot v_i \sim \nabla_j [\eta (\nabla_i v_j +\ldots)]$, is only 
sensitive to the density dependent part of the viscosity. But for 
a dilute gas the viscosity is expected to be density independent,
see equ.~(\ref{eta_aa}), so the dilute limit can not be verified 
using experiments that involve scaling flows. 

 There is clearly a need for additional experimental constraints. 
The first indication of almost ideal hydrodynamic behavior was
the observation of elliptic flow by O'Hara et al.~\cite{oHara:2002}.
The experiment showed that if the trapping potential is removed
the gas expands rapidly in the transverse direction while remaining 
nearly stationary in the axial. This is a consequence of the much
larger pressure gradient in the short direction. The ideal hydrodynamics 
of this experiment was worked out in \cite{Menotti:2002} but the 
effects of viscosity have not been carefully studied, in part because 
the data were taken at a single temperature. More recently Clancy et 
al.~studied the expansion of a gas cloud with an initial velocity field 
corresponding to a scissors mode \cite{Clancy:2007}. This is an interesting 
system, because the initial velocity field is irrotational ($\nabla\times 
{\bf v}=0$) but carries angular momentum. If the trapping potential 
is removed then the transverse size will grow initially, but if the 
gas remains irrotational then angular momentum conservation will
force the transverse expansion to slow down (and the rotation to speed up) 
before the transverse and axial radii become equal \cite{Edwards:2002}.  
This phenomenon was observed in the experiment, and an initial analysis 
leads to values of $\eta/s$ close to $1/(4\pi)$ \cite{Clancy:2008,Thomas:2009}.
This result is very important, but some of the caveats mentioned
above still apply. 

\subsection{The quark gluon plasma at RHIC}
\label{sec_qgp}


 Cold quantum fluids can be studied in conditions that are very 
close to equilibrium. The quark gluon plasma, on the other hand, 
can only be created in relativistic heavy ion collisions. In 
these collisions the initial state is very far from equilibrium,
and the system size is limited by the size of the heaviest stable 
nuclei. The applicability of hydrodynamics is not clear a priori. 
In this section we will summarize some of the evidence that has
been obtained from experiments at the Relativistic Heavy Ion 
Collider (RHIC). These experiments indicate that local equilibration
takes place, that nearly ideal fluid dynamics is applicable, and that 
the shear viscosity to entropy density ratio near $T_c$ is within a 
factor of a few of the KSS bound. 

 The collision energy in Au+Au collisions at RHIC is $100$ GeV per 
nucleon, and the nuclei are Lorentz contracted by a factor of $\gamma 
\simeq 100$. The transverse radius of a Au nucleus is approximately 
6 fm/c and on the order of 7000 particles are produced overall. The 
motion of the particles is relativistic, and the duration of a
heavy ion event is $\tau \sim 6$ fm. In order for hydrodynamics
to be applicable this time has to be large compared to the
equilibration time.

 The main observables are the spectra $dN/d^3p$ of produced 
particles. For momenta less than 2 GeV the spectra roughly follow 
Boltzmann distributions with a characteristic temperature close
to the QCD critical temperature. The first hint that the system 
is behaving collectively is the existence of radial flow. Heavy 
particle spectra have apparent temperatures that are larger than 
the temperatures extracted from light particles. This can be 
understood if there is a collective transverse expansion velocity 
$v_\perp$ which boosts the observed transverse momenta by an 
amount $p_\perp \sim m v_\perp$.  

\begin{figure}
\begin{center}
\includegraphics[width=6.5cm]{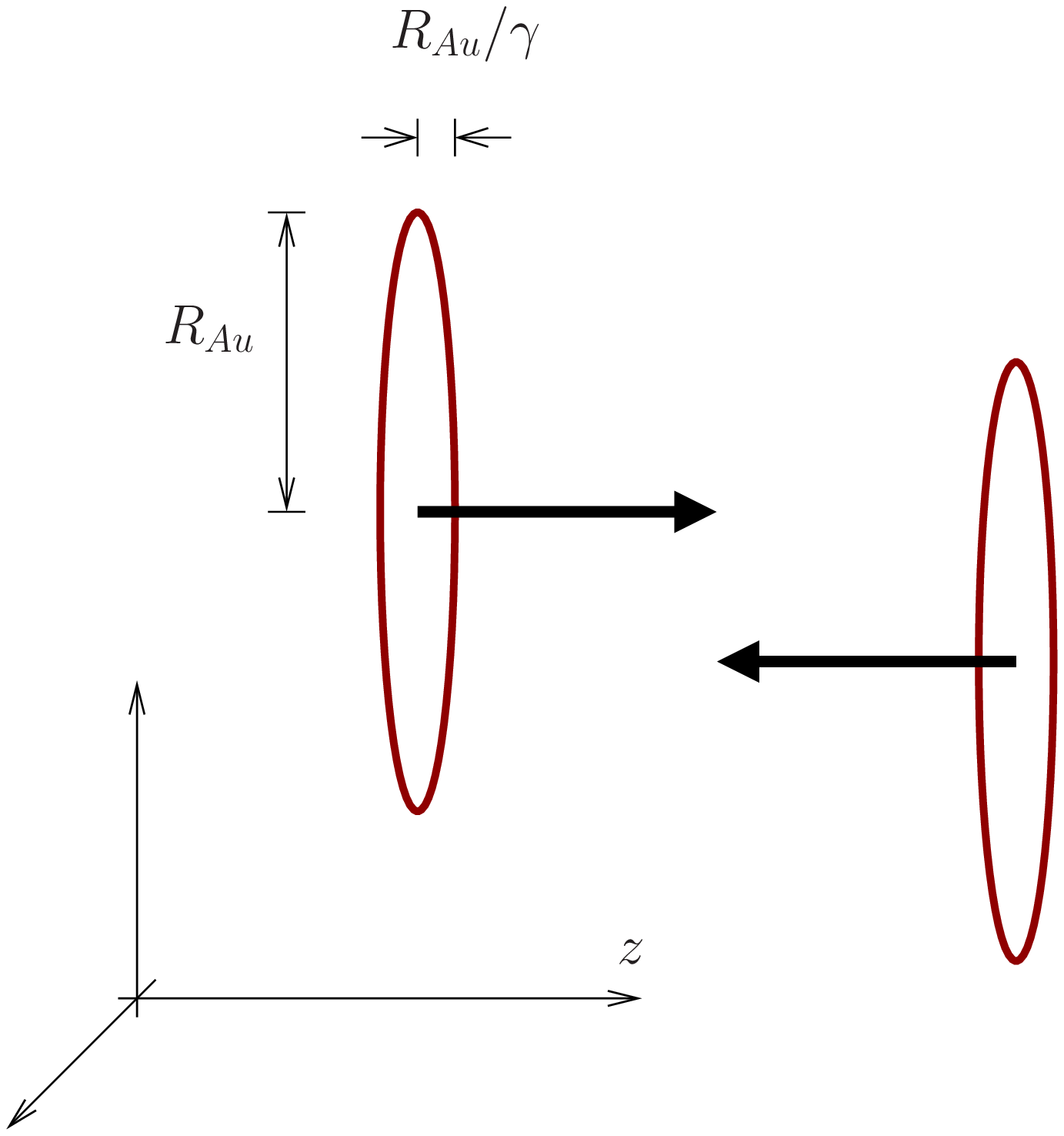}
\hspace{1cm}
\includegraphics[width=6.5cm]{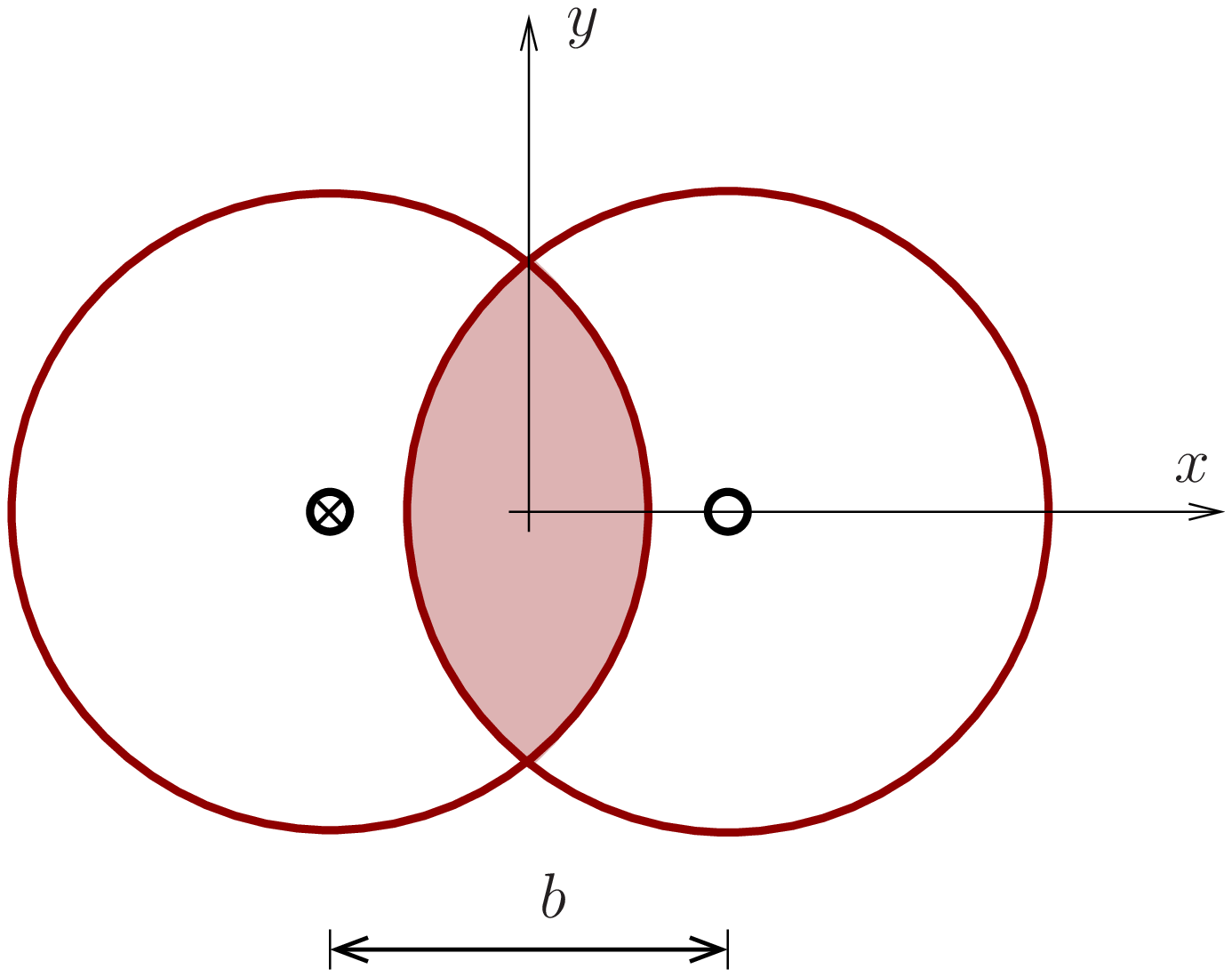}
\end{center}
\caption{Geometry of a high energy heavy ion collision. The left 
panel shows the collision of two Lorentz contracted gold nuclei. 
The beam direction is the $z$-axis. The right panel shows the 
same collision in the transverse plane. The impact parameter is 
along the $x$-axis, and the remaining transverse direction is 
the $y$-axis.}
\label{fig_coll_geom}
\end{figure}

More dramatic evidence for hydrodynamics is provided by the observation of 
elliptic flow in non-central heavy ion collisions. The centrality 
of the collision is characterized by the impact parameter $b$, the 
transverse separation of the two nuclei. The magnitude of $b$ can 
be determined experimentally by selecting events with a given 
multiplicity of produced particles. The uncertainty in the impact 
parameter determination is small except in very peripheral bins 
\cite{Miller:2007ri}. The direction of the impact parameter 
can be determined on an event by event basis using the azimuthal
dependence of the spectra. Imagine that the impact parameter 
direction is already known. This defines a coordinate system where 
$z$ is along the beam axis and $x$ is along the impact parameter 
direction, see Fig.~\ref{fig_coll_geom}.
We write $(p_x,p_y,p_z)=(p_\perp\cos(\phi),p_\perp
\sin(\phi),p_z)$, and the particle distribution can be expanded
in Fourier components of $\phi$, 
\be 
\label{v_2}
 \left. p_0\frac{dN}{d^3p}\right|_{p_z=0} = v_0(p_\perp)
 \Big( 1 + 2v_2(p_\perp)\cos(2\phi)  
         + 2v_4(p_\perp)\cos(4\phi) +\ldots \Big) . 
\ee
For a typical mid-central collision with $b\simeq6\,{\rm fm}$ the  
$v_2$ harmonic, called the elliptic flow coefficient, is approximately 
$6\%$. In an actual event the reaction plane can be determined (in 
principle) by plotting the distribution in $\phi$ relative to an 
arbitrarily chosen axis, and then requiring that the distribution has 
a maximum at $\phi=0$. This intuitive method to determine the reaction 
plane forms the basis of the event plane method. The result can be 
corrected for $v_2$ fluctuations and additional correlations 
among the produced particles. Current analyses are not based on the
event plane method but use two, four, and higher particle cumulants 
-- see \cite{Voloshin:2008dg} and references therein for a complete 
review. The scaling of these cumulants with multiplicity demonstrates 
that one can reliably extract collective flow down to small system sizes. 
These measurements provide a unique opportunity to study the approach 
to hydrodynamic behavior in a controlled fashion \cite{Gombeaud:2007ub}.

 Elliptic flow represents the collective response of the system 
to pressure gradients in the initial state. At finite impact parameter
the initial state has the shape of an ellipse, with the short axis
along the $x$ direction, and the long axis along the $y$ direction. 
This implies that pressure gradients along the $x$ axis are larger
than along the $y$ axis. Hydrodynamic evolution converts the 
initial pressure gradients to velocity gradients in the final state. 
Elliptic flow is a direct measure of collectivity. In particular, 
if the nucleus nucleus event were a simple superposition of proton 
proton collisions then the particle distribution would be azimuthally 
symmetric.

\subsubsection{The Bjorken model}
\label{sec_bj}

 The application of hydrodynamics to relativistic heavy ion 
collisions goes back to the work of Landau \cite{Landau:1953} 
and Bjorken \cite{Bjorken:1983}. Bjorken discovered a simple 
scaling solution that provides a natural starting point for 
more elaborate solutions in the ultra-relativistic domain. 
Consider two highly relativistic nuclei moving with equal but
opposite momenta in the $z$ direction. In the relativistic
regime the natural variable to describe the motion in the
$z$ direction is the rapidity
\be 
 y = \frac{1}{2}\log\left(\frac{E+p_z}{E-p_z}\right).
\ee
At RHIC the energy of the colliding nuclei is 100+100 GeV per 
nucleon, and the separation in rapidity is $\Delta y=10.6$. 
Bjorken suggested that the two highly Lorentz contracted nuclei 
pass through each other and create a longitudinally expanding 
fireball in which particles are produced. In the 
original model the number of produced 
particles is independent of rapidity, and the subsequent evolution 
is invariant under boosts along the $z$ axis. The evolution in proper 
time is the same for all comoving observers. The flow velocity is 
\be 
\label{u_bj}
 u_\mu = \gamma(1,0,0,v_z)= (t/\tau,0,0,z/\tau), 
\ee
where $\gamma$ is the boost factor and $\tau=\sqrt{t^2-z^2}$ 
is the proper time. The velocity field (\ref{u_bj}) solves the 
relativistic Euler equation (\ref{rel_euler}). In particular, 
there is no longitudinal acceleration. The remaining 
hydrodynamic variables are determined by entropy conservation.
Equ.~(\ref{rel_s_cont}) gives
\be 
\frac{d}{d\tau}\left[\tau s(\tau) \right] = 0 
\ee
and $s(\tau)=s_o\tau_o/\tau$. For an ideal relativistic gas 
$s\sim T^3$ and $T\sim 1/\tau^{1/3}$. Typical parameters
at RHIC are $\tau_o\simeq (0.6-1.6)$ fm and $T_o\simeq (300-425)$ MeV. 
The combination $\tau_oT_o^3$ is constrained by the final 
multiplicity, but individually $\tau_o$ and $T_o$ are not 
well constrained.  
 We note that the corresponding initial
temperature is significantly larger than the critical temperature
for the QCD phase transition. 

The temperature drops as a function of $\tau$ and eventually 
the system becomes to dilute for the hydrodynamic evolution
to make sense. At this point, the hydrodynamic description 
is matched to kinetic theory, 
\be 
\label{T_hyd_kin}
 T_{\mu\nu}^{\it hydro} \equiv T_{\mu\nu}^{\it kin} = 
  \int d\Gamma\, p_\mu p_\nu f({\bf x},{\bf p},t)\, .
\ee
For an ideal fluid the distribution function is parameterized
by the local temperature and flow velocity 
\be 
 f({\bf x},{\bf p},t) = \sum_i 
  \frac{d_i}{\exp(p\cdot u/T)\pm 1} \, ,
\ee
where $i$ labels different particle species, and $d_i$ are the 
corresponding degeneracies. Finally, the observed particle spectra 
are given by 
\be 
\label{CF}
\left( p_0\frac{dN}{d^3p}\right)_i = \frac{1}{(2\pi)^3}
 \int d\Sigma_\mu \, p^\mu f_i({\bf x},{\bf p},t)\, , 
\ee
where $\Sigma_\mu$ is the normal vector on the ``freezeout'' 
hypersurface, the surface on which the matching between the 
hydrodynamics and kinetic descriptions is performed.

 In order to quantitatively describe the observed particle 
distributions several improvements of the simple Bjorken 
model are necessary. First, one has to include the transverse
expansion of the system \cite{Baym:1984sr}. Transverse expansion 
becomes important at a proper time $\tau \sim R/c_s$, where $R$ 
is the (rms) size of the nucleus, and $c_s$ is the speed of sound. 
At very late times the expansion becomes three dimensional,
\be 
 s(\tau) \sim \frac{1}{\tau^3}\, , 
\ee
and $T\sim 1/\tau$. Transverse expansion is caused by
transverse pressure gradients. These gradients are sensitive
to the initial energy density of the system. One simple model
for the initial energy density (or entropy density) in the 
transverse plane is the Glauber model. In the  Glauber model  
the entropy  density is 
\bea
\label{Glauber}
 s({\bf x}_\perp,b) &\propto& 
   \,   T_A({\bf x}_\perp+{\bf b}/2) \Big[1 - \exp\left(-\sigma_{NN} 
   \,   T_A({\bf x}_\perp-{\bf b}/2) \right)\Big] \nonumber  \\
 & & +
   \,  T_A({\bf x}_\perp-{\bf b}/2) \Big[1  - \exp\left(-\sigma_{NN}
   \,  T_A({\bf x}_\perp+{\bf b}/2) \right) \Big] \, ,
\eea
where ${\bf b}$ is the impact parameter
\be
\label{T_A}
T_A({\bf x}_\perp) = \int dz \, \rho_A({\bf x}) 
\ee
is the thickness function, and $\sigma_{NN}(\sqrt{s})$ is the 
nucleon-nucleon cross section. Here, $\rho_A({\bf x})$ is the 
nuclear density. The idea behind the Glauber model is that the
initial entropy density is proportional to the number of nucleons 
per unit area which actually collide. Other variants exist. For 
instance, one can  distribute the energy density according to the 
number of binary nucleon-nucleon collisions -- see Ref.~\cite{Kolb:2001qz} 
for a comparison. A more sophisticated theory of the initial energy 
density is provided by the Color Glass Condensate (CGC) 
\cite{McLerran:1993ni,Hirano:2005xf}. This model leads to somewhat 
steeper initial transverse energy density distributions.

 Gradients in the transverse pressure lead to transverse 
acceleration and generate collective transverse flow. The
collective expansion leads to a blue-shift of the transverse 
momentum spectra of produced particles. For an azimuthally 
symmetric source with temperature $T$ and radial flow 
velocity $u^r(r)$ we get \cite{Schnedermann:1993ws}
\be 
\label{schnecki}
\left( p_0\frac{dN}{d^3p}\right)_i =
 \frac{2d_i}{(2\pi)^2} r m_\perp 
 \int rdr\, K_1\left( \frac{m_\perp u^\tau}{T} \right)
            I_1\left( \frac{p_\perp u^r}{T} \right)\, , 
\ee
where $m_\perp=\sqrt{p_\perp^2+m_i^2}$ is the transverse ``mass'',
and $K_1(z),I_1(z)$ are generalized Bessel functions. Using 
the asymptotic form of the Bessel functions one can show
that the spectrum has the form 
\be
\frac{dN}{d m_\perp^2} \sim 
   \exp\left(-\frac{m_\perp}{T_{\it eff}}\right), 
   \hspace{0.5cm}
   T_{eff} = T \sqrt{\frac{1+v_r}{1-v_r}}\, . 
\ee
This effect of transverse flow on the spectra is seen 
in the data. At RHIC, transverse velocities at freezeout reach 
$0.6\,c$. At finite impact parameter the initial energy density 
in the transverse plane is not azimuthally symmetric. The pressure 
gradient along the direction of the impact parameter is larger 
than the gradient in the orthogonal direction. The resulting 
anisotropy of the transverse flow can be characterized by the 
elliptic flow parameter $v_2$ defined in equ.~(\ref{v_2}).
The observed elliptic flow is remarkable because $v_2$ is a 
very direct measure of transverse pressure. The observed radial 
flow is proportional to the radial pressure and the expansion
time, but elliptic flow has to be generated early, when the 
system is still deformed. 

\subsubsection{Estimates of viscous corrections}
\label{sec_bj_vsic}

 We are now in a position to discuss the role of dissipative
effects. We begin with the effect of shear and bulk viscosity 
on the Bjorken solution. The scaling flow given in equ.~(\ref{u_bj})
is a solution of the relativistic Navier-Stokes equation. 
Viscosity only modifies the entropy equation. We get 
\be 
\label{bj_ns}
\frac{1}{s}\frac{ds}{d\tau} = -\frac{1}{\tau} 
 \left( 1 - \frac{\frac{4}{3}\eta+\zeta}{sT\tau}\right)\, .
\ee
We observe that dissipation is governed by the sound attenuation
length $\Gamma_s$, see equ.~(\ref{Gam_s}). The applicability 
of the Navier-Stokes equation requires that the viscous 
correction is small \cite{Danielewicz:1984ww}
\be 
\label{DG}
\frac{\eta}{s}+\frac{3}{4}\frac{\zeta}{s}
\ll \frac{3}{4}(T\tau)\, . 
\ee
For the Bjorken solution $T\tau\sim \tau^{2/3}$ grows with time, 
and this condition is most restrictive during the initial phase. 
Using $\tau_o=1$ fm and $T_o=300$ MeV gives $\eta/s<0.6$. For 
a three dimensional expansion $T\tau$ is independent of time.
At very late time the fluid is composed of hadrons, or pre-formed
hadronic resonances. In that case $\eta\sim T/\sigma$, where 
$\sigma$ is a hadronic cross section. Then, for a three dimensional
expansion, the viscous correction $\eta/(sT\tau)$ grows with proper 
time as $\tau^2$. This shows that the system has to freeze out 
at late time. 

It is instructive to study the viscous contribution to the stress 
tensor in more detail. At central rapidity we have (for $\zeta=0$)
\be 
T_{zz} = P -\frac{4}{3}\frac{\eta}{\tau}, 
\hspace{0.5cm}
T_{xx}=T_{yy} =  P + \frac{2}{3}\frac{\eta}{\tau} \, .
\ee
This means that shear viscosity decreases the longitudinal 
pressure and increases the transverse one. In the Bjorken
scenario there is no acceleration, but if pressure gradients
are taken into account shear viscosity will tend to increase
transverse flow. A similar effect will occur at finite impact 
parameter, 
see Fig.~\ref{fig_coll_geom}. Shear viscosity 
reduces the pressure along the $x$-direction, and increases the 
pressure in $y$-direction. As a consequence there is less
acceleration in the $x$-direction, and elliptic flow is suppressed.
This is the basic observation that motivates attempts to extract 
shear viscosity from the observed elliptic flow. 

 Viscosity modifies the stress tensor, and via the matching 
condition (\ref{T_hyd_kin}) this modification changes the distribution 
functions of produced particles. In Ref.~\cite{Teaney:2003kp} a simple 
quadratic ansatz for  the leading correction $\delta f$ to the 
distribution function was proposed
\be 
\label{del_f}
\delta f = \frac{3}{8} \frac{\Gamma_s}{T^2}f_0(1+f_0) 
   p_\alpha p_\beta  \nabla^{\langle \alpha} u^{\beta\rangle} \, ,
\ee
where $\nabla^{\langle \alpha} u^{\beta\rangle}$ is a symmetric
traceless tensor, see equ.~(\ref{sym_tr}). This form  summarizes 
the results of more involved kinetic calculations \cite{Arnold:2000dr}. 
The modified distribution 
function leads to a modification of the single particle spectrum. 
For a simple Bjorken expansion and at large $p_\perp$ we find
\be 
\label{del_N}
 \frac{\delta (dN)}{dN_0} = \frac{\Gamma_s}{4\tau_f} 
    \left( \frac{p_\perp}{T} \right)^2 \,, 
\ee
where $\tau_f$ is the freezeout time. We observe that the 
dissipative correction to the spectrum is controlled by the 
same parameter $\Gamma_s/\tau$ that appeared in equ.~(\ref{bj_ns}). 
We also note that the viscous term grows with $p_\perp$. At RHIC 
transverse momentum spectra are in agreement with hydrodynamic 
predictions out to transverse momenta several times larger than 
the temperature. In equ.~(\ref{del_N}) this is partially 
compensated by the fact that $\tau_f/\tau_o$ is a large number, 
but typically the requirement $\delta(dN)/(dN_0)<1$ provides a
more stringent bound on $\eta/s$ than equ.~(\ref{DG}). We 
can also include transverse expansion and study the leading 
dissipative correction to $v_2$ \cite{Teaney:2003kp}. The 
viscous correction tends to reduce $v_2$ and grows with $p_\perp$. 
At $p_\perp=1$ GeV an estimate similar to equ.~(\ref{del_N}) 
gives $(\delta v_2)/v_2 \sim 1$ for $\Gamma_s/\tau_f\sim 0.2$. Using 
$\tau_f\sim 5$ fm and $T_f=160$ MeV this translates into $\eta/s\leq 0.6$.

\subsubsection{Hydrodynamic Simulations}
\label{sec_thic_sim}

 There have been a number of recent numerical studies devoted to extracting 
the shear viscosity of the quark gluon plasma
\cite{Dusling:2007gi,Romatschke:2007mq,Song:2007ux,Luzum:2008cw,Huovinen:2008te}.
Here, we will follow the work of Dusling and Teaney \cite{Dusling:2007gi},
and refer the reader to the recent review by Heinz \cite{Heinz:2009xj}
for a more detailed comparison between different strategies for 
implementing relativistic viscous hydrodynamics for heavy ion collisions
at RHIC. 
In order to respect causality in viscous hydrodynamics we have 
to use a second order formalism. This means that the strains 
$\delta T^{\mu\nu}$ are promoted to dynamical fields which relax on 
a collisional time scale to the Navier-Stokes form rather than being 
specified instantaneously by the constitutive relations. Such relaxation 
processes are second order in the hydrodynamic expansion, see 
Sec.~\ref{sec_rel_fluid}. Dusling and Teaney considered a $2+1$ 
dimensional boost invariant flow and used a second order fluid model 
studied by \"Ottinger \cite{Ottinger:1998}. 
This model is formulated 
along the lines of equ.~(\ref{del_Pi_3}), and the dynamical strains 
are parametrized by an additional field $\mathring{c}^{\mu\nu}$,
\be
\label{generic-constitute}
 \delta T^{\mu\nu} = -P(\epsilon)  \alpha \,\mathring{c}^{\mu\nu}  \, .
\ee
The relaxation equation for $\mathring{c}^{\mu\nu}$ is written 
in terms of a tensor $c^{\mu\nu}$. This tensor satisfies the 
constraint
\be
\label{constraint}
c_{\mu\nu}u^\nu=u_\mu \, , 
\ee 
and is decomposed as 
\bea
c_{\mu\nu} &=& -u_\mu u_\nu + \mathring{c}_{\mu\nu}
                         + \overline{c}_{\mu\nu} \, , \\
\overline{c}_{\mu\nu} &=& 
  \frac{1}{3}\, \left(c^{\;\;\lambda}_\lambda-1\right)
                \left(\eta_{\mu\nu}+u_\mu u_\nu\right)\, .
\eea
The field $c_{\mu\nu}$ obeys the relaxation equation 
\be
\label{eq:cevol}
u^\lambda(\partial_\lambda c_{\mu\nu}
         -\partial_\mu c_{\lambda\nu}
         -\partial_\nu c_{\mu\lambda})
   =-\frac{1}{\tau_0}\,\overline{c}_{\mu\nu}
    -\frac{1}{\tau_2}\,\mathring{c}_{\mu\nu}\, .
\ee
The constraint equ.~(\ref{constraint}) is preserved under time evolution.
In the limit that the relaxation times $\tau_0$ and $\tau_2$ are small 
the evolution equation leads to the following solution for $c^{\mu\nu}$ 
in the local rest frame
\be
\label{cijequ}
c^{ij} = \tau_2 \left( \partial^i u^j + \partial^j u^i
                        -\frac{2}{3}\delta^{ij}\partial_k u^k\right)
   +\frac{2}{3}\, \tau_0\delta^{ij}\partial_k u^k   \, . 
\ee
Comparing with the Navier-Stokes equation we see that
\be
\eta=\tau_2 \, P \, \alpha  \, , \hspace{1cm}
\zeta=\frac{2}{3}\, \tau_0 \, P \, \alpha\, .
\ee
Dusling and Teaney used these relations (with $\alpha=0.7$) to 
fix $(\tau_0,\tau_2)$ in terms of $\eta$ and $\zeta$, and studied 
the sensitivity to the parameter $\alpha$. They considered a simple 
conformal equation of state $P=\epsilon/3$, and set $\zeta=0$.

\begin{figure}
\begin{center}
\includegraphics[width=7.5cm]{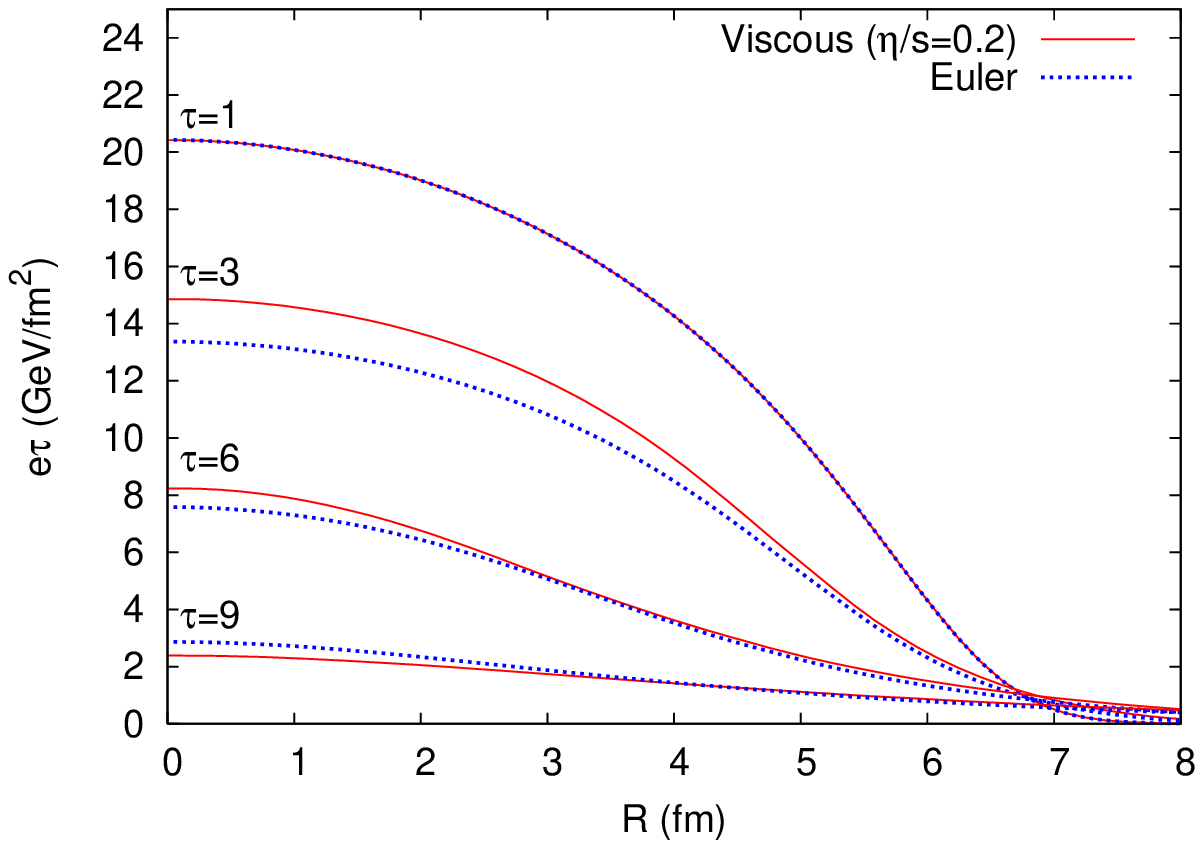}
\hspace{0.0in}
\includegraphics[width=7.5cm]{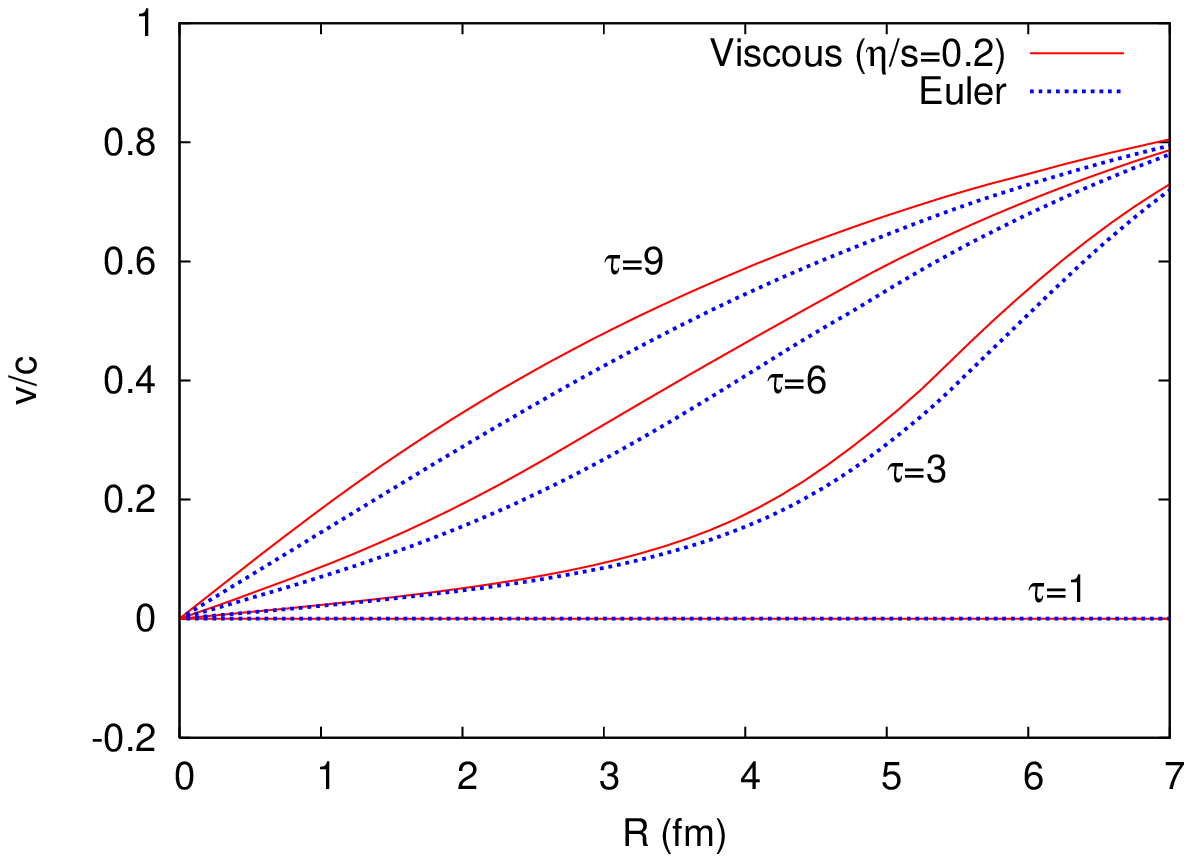}
\end{center}
\caption{Plot of the energy density per unit rapidity $e\tau$ (left) 
and of the transverse velocity (right) at times of $\tau=1,3,6,9$ fm/c, 
for $\eta/s=0.2$ (solid red line) and for ideal hydrodynamics (dotted 
blue line), from \cite{Dusling:2007gi}.}
\label{fig_v1d}
\end{figure}

An advantage of \"Ottinger's approach is that equ.~(\ref{eq:cevol}) is 
relatively simple to solve. One just evolves the spatial components of 
$c_{ij}$ and then uses the constraints (\ref{constraint}) to solve for 
the time components $c_{00}$ and $c_{0i}$. As hydrodynamics is universal, 
any fluid model can be recast in terms of the first and second order 
formalism described in Sec.~\ref{sec_rel_fluid}. For the \"Ottinger 
fluid model, expanding out the equations of motion to second order  
leads to the relation
\bea
\label{del_Pi_ott}
 \pi^{\mu\nu} &=&
 -\eta\sigma^{\mu\nu} 
 -\tau_2 \left[ ^\langle D\pi^{\mu\nu\rangle}
 +\frac{4}{3}\,\pi^{\mu\nu} (\partial\cdot u) \right] \\
& & \hspace{0.2cm}\mbox{}
 + \frac{\tau_2}{\eta}
        \pi^{\langle\mu}_{\;\;\;\lambda} \pi^{\nu\rangle\lambda}
 + \tau_2
        \pi^{\langle\mu}_{\;\;\;\lambda} \Omega^{\nu\rangle\lambda}
 -\frac{2}{3}\, \tau_2 \pi_{\mu\nu} (\partial \cdot u) \, .
\nonumber 
\eea
The last term in this expression differs from the general result 
in equ.~(\ref{del_Pi_ott}), which indicates that at second order 
in gradients this model contains terms that break conformal symmetry. 

The hydrodynamic equations were solved for several fixed values of the
shear viscosity to entropy density ratio $\eta/s$. At the initial 
time $\tau_o = 1$ fm the entropy per participant is adjusted and 
closely corresponds to the results of full hydrodynamic simulations
\cite{Teaney:2001av,Kolb:2000fh,Huovinen:2001cy}. The maximum initial
temperature is $T_o = 420$ MeV at an impact parameter $b=0$. The 
initial components of the stress tensor are set to the Navier Stokes 
values. 

\begin{figure}
\begin{center}
\includegraphics[width=8cm]{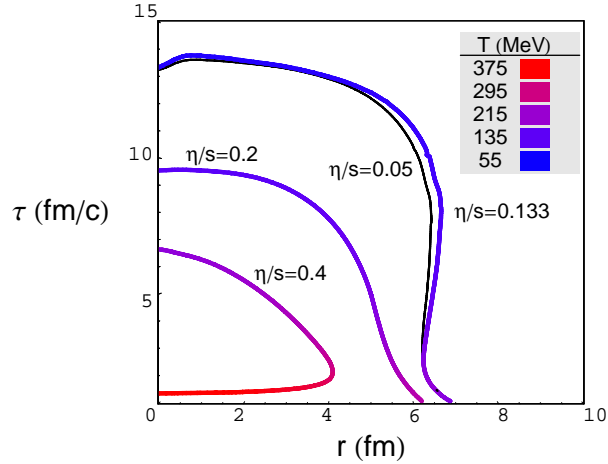}
\end{center}
\caption{Location of freezeout surfaces for for central Au-Au collisions.
The surfaces are determined by the condition $\frac{\eta}{p}\,(\partial\cdot 
u)=0.6$, slightly larger than the value 0.5 discussed in the text.  
Different surfaces correspond to different values of $\eta/s$. The 
shading corresponds to the freezeout temperature. The thin solid black
curve shows the contour set by $\frac{\eta}{p}\,(\partial\cdot u)=0.225$
for comparison.  }
\label{fig_fosurf}
\end{figure}
 
 Numerical results are shown in Fig.~\ref{fig_v1d}. The effect of 
viscosity is twofold. The longitudinal pressure is initially reduced 
and the viscous case does less longitudinal $PdV$ work as in the 
simple Bjorken expansion.  This means that at early times the energy 
per rapidity decreases more slowly in the viscous case. The reduction 
of longitudinal pressure is accompanied by a larger transverse pressure.
This causes the transverse velocity to grow more rapidly. The larger 
transverse velocity causes the energy density to deplete faster at 
late times in the viscous case. The net result is that a finite viscosity,
even as large as $\eta/s=0.2$, does not integrate to give major deviations 
from the ideal equations of motion.

Freezeout occurs when the viscous terms become large compared to the 
ideal terms.
Roughly, the system begins to break apart when  
\be
\label{fr_out}
\frac{\eta}{P} \, \partial\cdot u \sim 1.
\ee
This combination of parameters can be motivated from kinetic theory.  
The pressure is of order $P \sim \epsilon \llangle v_{\rm th}^2\rrangle$  
where $\llangle v_{\rm th}^2\rrangle $ is the typical quasi particle 
velocity and $\epsilon$ is the energy density. The viscosity is of 
order $\eta \sim \epsilon\llangle v_{\rm th}^2 \rrangle \tau_{R}$ 
where $\tau_R$ is the relaxation time. Thus the freezeout condition 
is $\frac{\eta}{P}\, (\partial\cdot u) \sim \tau_R (\partial\cdot u) 
\sim 1$.
Ideally there should be an overlap region where both 
viscous hydrodynamics and kinetic theory are valid. In this region 
hydrodynamics can be systematically coupled to a kinetic description
in order to correctly model the breakup. In simulations of heavy
ion collisions the size of the overlap region is small, and the 
breakdown of hydrodynamics is typically modeled via a freezeout 
surface. In practice the freezeout surface was chosen to satisfy
$(\eta/P) \, \partial\cdot u  = 0.5$, where the precise number 
on the right hand side is simply an educated guess based on 
examining the output of second order hydrodynamic simulations. 
Typical freezeout surfaces which satisfy this criterion are 
shown in Fig.~\ref{fig_fosurf}. 
We note that hydrodynamics breaks down both 
at late and also at early times. The latter is most clearly see from 
the $\eta/s=0.4$ curve. If $\eta/s$ is very small then freezeout
will occur at very late proper time. In the following, we shall
therefore use a simpler criterion $\chi\equiv \frac{1}{T}(\partial
\cdot u)={\rm const}$ which is independent of $\eta/s$. Taking 
$\chi=3$ roughly corresponds to the $\eta/s=0.2$ freezeout surface
in Fig.~\ref{fig_fosurf}.

 Finally, we can compute the spectra of produced particles. 
We follow the procedure outlined above, see equ.~(\ref{del_f}), 
and write the distribution function as $f=f_0 + \delta f$ with 
\be
\delta f = \frac{1}{2(\epsilon + P) T^2}\, f_0(1 + f_0) \, 
p^{\mu} p^{\nu} \delta T_{\mu\nu}  
\label{eq:df} \, .
\ee
The spectrum is determined by integrating the distribution function 
over the freezeout surface as in equ.~(\ref{CF}), and the elliptic flow 
parameter $v_2$ is computed from the definition in equ.~(\ref{v_2}). 
A comparison with the data obtained by the STAR collaboration is shown 
in Fig.~(\ref{fig_v2sum}). There are several curves here and we will 
go through them one by one:

\begin{figure}
\begin{center}
\includegraphics[width=7.5cm]{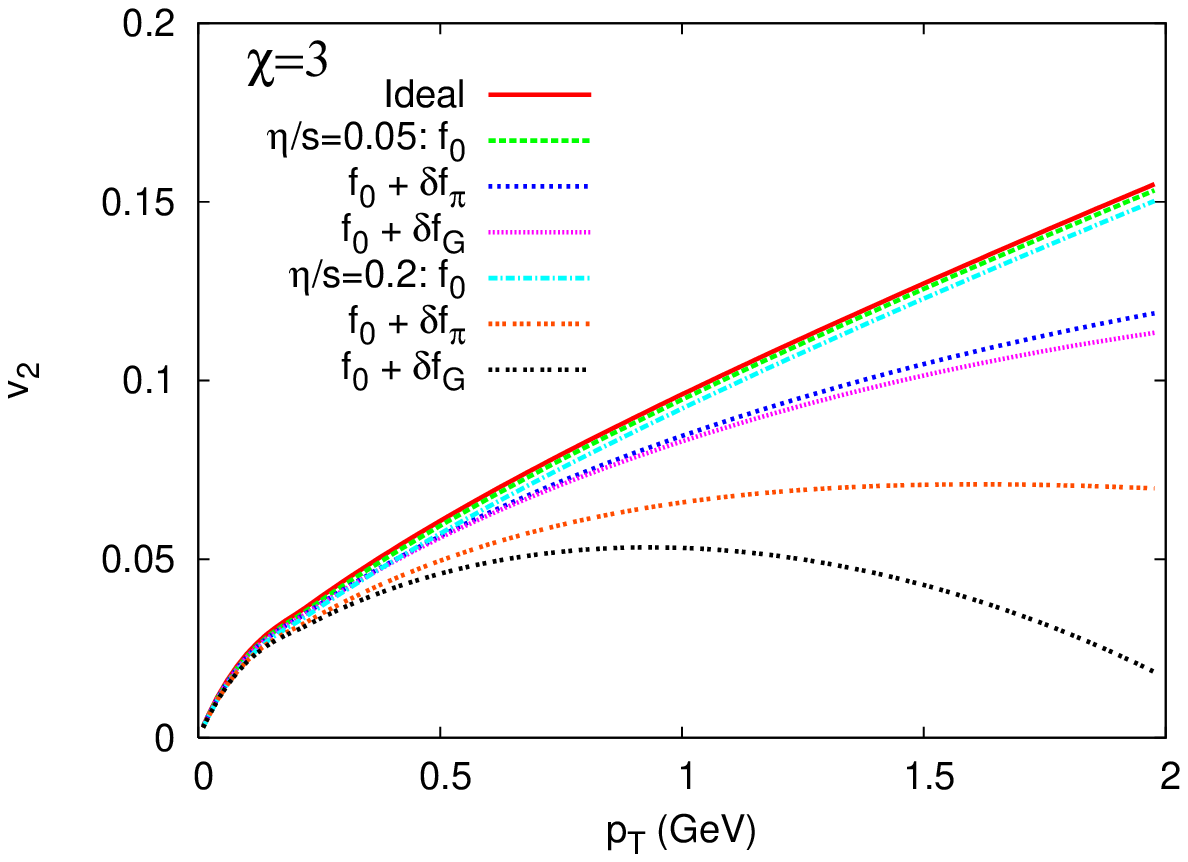}
    \hspace{0.0in}
\includegraphics[width=7.5cm]{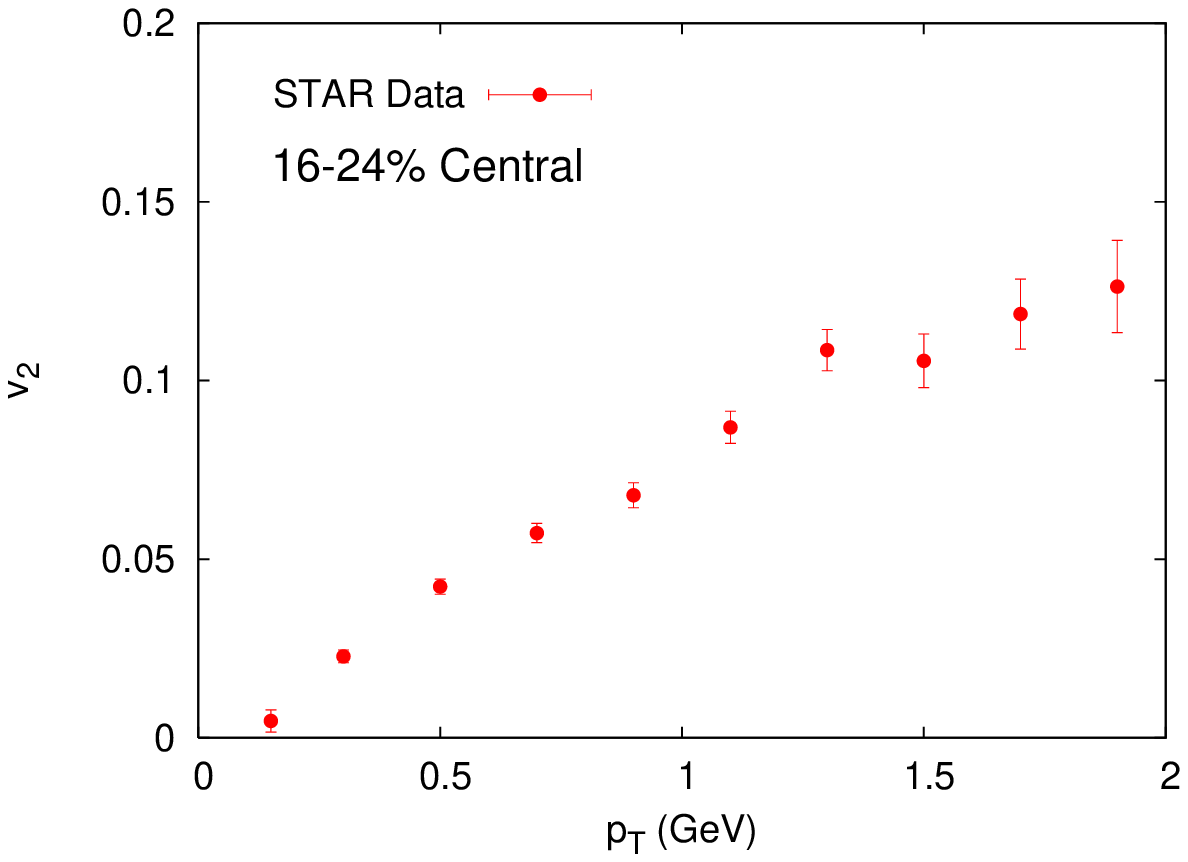}
\end{center}
\caption{Left: $v_2(p_T)$ for massless Bose particles for simulations using 
$\eta/s=0, 0.05, 0.2$ at an impact parameter of $b=6.5\,{\rm fm}$\,. Right: 
Four-particle cumulant data as measured in Au-Au collisions at $\sqrt{s}=200$ 
GeV for a centrality selection of 16\% to 24\% \cite{StarData}.}
\label{fig_v2sum}
\end{figure}

\begin{itemize} 
\item For the two different values of $\eta/s$, 0.05 and 0.2, there are 
three curves each. Our best estimates for the elliptic flow as a function 
of $p_T$ are labeled $f_0 + \delta f_\pi$ and are shown by the blue ($\eta/
s=0.05$) and orange ($\eta/s=0.2$) lines.  

\item To disentangle what part of the viscous modification is due to the
distribution function $\delta f$ and what part of it is due to changes in the
flow, we also compute $v_2(p_T)$ with $f_0$ only. We see that the effect on
$v_2$ from viscous modifications of the flow is relatively minor. This may 
be the largest obstacle to reliably extracting the shear viscosity from the 
heavy ion data. However, it is  important to realize that the modification 
to the distribution function reflects the viscous correction to the stress 
tensor itself. In hydrodynamic simulations the $p_T$ integrated $v_2$ tracks 
the asymmetry of the stress tensor \cite{Kolb:1999it}. In the context of
viscous hydrodynamics this is nicely illustrated in Fig.~8 of
Ref.~\cite{Luzum:2008cw}.  Nevertheless, in contrast to the atomic
physics experiments discussed in Sect.~\ref{sec_trap},  the observed viscous
corrections to the elliptic flow do not reflect a resummation of secular terms
in the gradient expansion. 

\item Finally instead of showing the spectrum computed with  $\delta 
T^{\mu\nu}$, we show $v_2(p_T)$ computed with velocity gradients directly.
For this purpose $\delta T^{\mu\nu}$ in equ.~(\ref{eq:df}) is replaced by 
$-\eta \sigma^{\mu\nu}$ where $\sigma^{\mu\nu}$ is computed from the flow 
velocities. $v_2(p_T)$ computed in this way is denoted by $f_0 + \delta 
f_G$ in the figure and corresponds to the magenta and black curves. 
To first order in gradients this is an identity and the difference 
between $f_\pi$ and $f_G$ is a measure of the magnitude of second order
terms.  For the smallest viscosity $\eta/s = 0.05$ the differences are 
quite small, but the effect is more noticeable for $\eta/s=0.2$.
\end{itemize} 

We conclude that for small values of $\eta/s\lsim 0.2$ the gradient 
expansion is working. There are, however, a number of issues that 
have to be considered in order to extract reliable values for $\eta/s$:

\begin{itemize}
\item The constraints on $\eta/s$ are sensitive to the initial 
values for the transverse energy density. In particular, using 
color glass initial conditions produces higher transverse pressure
gradients, and allows for values of $\eta/s$ about twice larger 
than Glauber model initial conditions \cite{Luzum:2008cw}.

\item  Near the edge of the nucleus the gradient expansion breaks down
completely. It is important to quantify the extent to which the effects 
of the edge propagate into the interior and invalidate the hydrodynamic 
description. One way to do this is by comparing the results of hydrodynamic
simulations to kinetic theory. For strongly coupled plasmas kinetic theory  
is not an appropriate description of the microscopic interactions. However,
hydrodynamics is independent of the microscopic details. Thus extrapolating
kinetic theory into the strongly coupled domain is good way to construct a
model which gracefully transitions  from a hydrodynamic description in the
interior to a kinetic description near the edge.  There are several important
developments in this direction \cite{Huovinen:2008te,Xu:2008av,Bouras:2009nn}.

\item Viscous effects in the hadronic phase are very important
\cite{Teaney:2001av,Hirano:2005xf}. These effects can be taken 
into account by coupling the hydrodynamic evolution to a 
hadronic cascade. 

\item Effects due to bulk viscosity may reduce both radial and 
elliptic flow \cite{Song:2008hj,Denicol:2009am}. Bulk viscosity 
is likely to be much smaller than shear viscosity in the plasma phase, 
but it is expected to grow near the phase transition.  The magnitude 
of this growth is not clear. 

\end{itemize}

 Some of these effects have yet to be carefully studied. However, 
even if one conservatively assumes that all uncertainties tend
to increase the bound on $\eta/s$, one still has to conclude that 
for a shear viscosity of $\eta/s > 0.4$ it will be impossible to 
reproduce the observed flow.  The question now is whether it will
be possible to describe the large set of available data on 
energy, impact parameter, rapidity, transverse momentum, and 
species dependence of flow using viscous hydro, and whether it 
is possible to extract a reliable value, with controlled error 
bars, for $\eta/s$ of the quark gluon plasma.  

\section{Summary and outlook}
\label{sec_sum}
\subsection{Summary}

In this review we summarized theoretical and experimental 
information on the behavior of nearly perfect fluids. We
characterized the ``perfectness'' in terms of the shear 
viscosity of the fluid. Shear viscosity is special because

\begin{itemize}

\item shear viscosity is the ``minimal'' transport property 
of a fluid. 
Thermal transport, diffusion, or conductivities 
require the presence of conserved charges, and bulk viscosity 
vanishes in the case of scale invariant fluids.
The ``perfectness'' of any fluid can be 
characterized by the dimensionless ratio $\eta/s$. If factors 
of $\hbar$ and $k_B$ are reinstated the dimensionless measure 
of fluidity is $[\eta/\hbar]/[s/k_B]$.

\item a small shear viscosity is uniquely associated with
strong interactions. Other transport coefficients may vanish
even if the interaction strength remains weak. For example, 
bulk viscosity vanishes if the theory is conformal, and the
diffusion constant goes to zero at the a localization phase
transition, but the shear viscosity of a weakly coupled fluid
is always large. 

\end{itemize}

Note that the reverse of the last statement is not true: The 
viscosity can be large, even if the interaction is strong. One
possible scenario is that the interaction is so strong that it leads 
to the breakdown of a continuous symmetry, and the emergence of a 
new set of weakly coupled quasi-particles. Note that the ${\cal N}=4$ 
super-conformal fluid is special in this regard: There is no symmetry 
breaking in the strong coupling limit. Another example is the 
liquid-gas phase transition. There are no weakly coupled quasi-particles, 
but the viscosity diverges because of critical fluctuations. 

 We summarized the main theoretical approaches to transport
coefficients: kinetic theory, holography, and non-perturbative 
approaches based on the Kubo formula.

\begin{itemize}
\item Kinetic theory applies whenever the fluid can be described
in terms of quasi-particles. For many fluids this is the case
in both the high and the low temperature limit. Typically, at 
high temperature the quasi-particles are the ``fundamental''
degrees of freedom (quarks, gluons, atoms) whereas the low
temperature degrees of freedom are composite (phonons, rotons, 
pions). In the regime in which kinetic theory applies the 
ratio $\eta/s$ is always parametrically large. This leads to
a characteristic ``concave'' temperature dependence of $\eta/s$.
Kinetic theory is useful in constraining the location of the 
viscosity minimum (usually, near the crossover between the 
high and low temperature regimes). Also, despite the weak
coupling restriction, kinetic theory is quantitatively 
accurate for many quantum fluids in the whole temperature 
range covered by experiment.

\item Holography is a new method for studying the transport
behavior of quantum fluids. It is most useful in the strong 
coupling limit of certain model field theories (like ${\cal N}=4$ 
super-conformal field theory), but the range of field theories 
that have known holographic duals has grown significantly over 
the years. More importantly, holographic dualities have led
to important new insights into the transport properties of 
strongly coupled fluids. This includes the proposed universal
bound on $\eta/s$, consistent higher order hydrodynamic
theories, computations of the spectral function associated
with the shear and other transport modes, etc. Holography 
has also been used to study specific solutions to the 
hydrodynamic equations, like the wake of a moving heavy quark, 
or the approach to equilibrium after the collision of two 
highly Lorentz contracted sources.

\item The Kubo formula connects non-perturbative calculations
of equilibrium correlation functions with non-equilibrium transport 
coefficients. Possible non-perturbative approaches include
Euclidean lattice, large-N, exact renormalization group, and
Dyson-Schwinger calculations. Significant progress has been 
made in computing the shear viscosity of the quark gluon plasma
on the lattice. The results are close to the proposed bound. 
Future calculations will answer the question whether transport
phenomena in the QCD plasma can be understood in terms of 
quasi-particles. 

\end{itemize}

 Finally, we summarized the experimental situation for 
the three most strongly coupled fluids that can be prepared
in the laboratory. 

\begin{itemize}
\item Liquid helium has been studied for many years and 
its shear viscosity is well determined. The minimum value
of $\eta/s$ is about 0.8 and is attained near the endpoint 
of the liquid gas phase transition. The ratio $\eta/n$ has 
a minimum closer to the lambda transition. Even though 
$\eta/s<1$ the transport properties of liquid helium can 
be quantitatively understood using kinetic theory. 

\item Strongly interacting cold atomic Fermi gases were first 
created in the laboratory in 1999. These systems are interesting 
because the interaction between the atoms can be controlled,
and a large set of hydrodynamic flows (collective oscillations,
elliptic flow, rotating systems) can be studied. Current
experiments involve $10^5-10^6$ atoms, and the range of 
temperatures and interaction strengths over which hydrodynamic
behavior can be observed is not large. There are also some
difficulties in extracting the viscosity that are related
to the nature of the flow profiles that have been studied. 
A conservative estimate is $\eta/s<0.5$. 

\item The quark gluon plasma has been studied in heavy ion
collisions at a number of facilities, AGS (Brookhaven), SPS
(CERN), RHIC (Brookhaven). Almost ideal hydrodynamic behavior
was observed for the first time in 200 GeV per nucleon (in the 
center of mass) Au on Au collisions at RHIC. These experiments
are difficult to analyze - the initial state is very far from
equilibrium and not completely understood, final state interactions 
are important, and the size and lifetime of the system are not
very large. Important progress has nevertheless been made in 
extracting constraints on the transport properties of the 
quark gluon plasma. A conservative bound is $\eta/s<0.4$, but 
the the value of $\eta/s$ that provides the best fit to the 
data is smaller, $\eta/s\sim 0.1$.

\end{itemize}

\subsection{Outlook}

 Much work remains to be done in order to advance theoretical
methods for predicting the transport properties of strongly 
coupled quantum fluids, to improve the determination of transport 
coefficients of these fluids, and to discover new nearly perfect fluids.

 However, in addition to that, we also want to understand
what nearly perfect are fluids are like, in particular 
whether they can be understood using quasi-particles and 
the tools of kinetic theory. There are several avenues for
addressing this question:

\begin{itemize}
\item Quantum Monte Carlo calculations can be used to determine
the spectral function of the energy momentum tensor. If energy 
and momentum are carried by quasi-particles then the spectral 
function has a peak at low energy. If, on the other hand, the fluid 
is AdS/CFT-like then there are no quasi-particles and no peak in 
the spectral functions. Current calculations in lattice QCD seem 
to prefer the AdS/CFT picture \cite{Meyer:2007ic,Meyer:2008gt}, but 
the issue is far from settled. 

\item Detailed simulations of kinetic equations can be used to
study the crossover from kinetic behavior (the Knudsen limit) to 
hydrodynamic behavior, see \cite{Huovinen:2008te,Xu:2008av,El:2008yy,Bouras:2009nn}
for work in the context of QCD and \cite{Massignan:2004,Bruun:2007} 
for studies of the dilute Fermi gas at unitarity. The main 
question is whether it is possible to extend a self-consistent 
kinetic theory into the domain $\eta/s<1$, and whether one can
describe not only flow properties, but also all other transport 
properties like diffusion and energy loss. Self-consistency requires 
that the life time of the quasi-particles is long compared to the 
characteristic thermal time, $\tau\sim 1/T$. A simultaneous
description of momentum diffusion, charge diffusion, and energy loss
is important because in kinetic theory there is a close connection
between these observables. In AdS/CFT-like fluids the relation 
between different transport processes is non-trivial. For example,
at infinite coupling shear viscosity  is finite while the heavy 
quark diffusion coefficient is zero. The current experimental 
situation in QCD is not entirely clear, with both AdS/CFT based 
\cite{Noronha:2009vz} and kinetic approaches \cite{Fochler:2008ts} 
demonstrating some success. 

\item Phenomenological studies address the quasi-particle structure
of nearly perfect fluids by studying fluctuations and correlations
of conserved charges other than energy and momentum. In liquid 
helium quasi-particles were studied using neutron scattering, and
in the dilute Fermi gas one can study the quasi-particle structure 
using radio-frequency spectra, see \cite{Baym:2007,Haussmann:2009}.
In the quark gluon plasma there is some evidence for quasi-particles
from charge fluctuations \cite{Koch:2005vg} and from the success
of the recombination model in reproducing the flow of identified 
particles at intermediate momenta \cite{Nonaka:2003hx}.

\end{itemize}

Acknowledgments: This work was supported in parts by the US 
Department of Energy grant DE-FG02-03ER41260 (T.S.) and
DE-FG02-08ER41540 (D.T.). D.T. is also supported by the 
Alfred P.~Sloan foundation. We would like to thank Gordon 
Baym for providing the impetus to write this review. We would 
like to acknowledge useful discussions with Dam Son, Edward 
Shuryak, and John Thomas. In preparing a revised version we
benefited from the remarks of an anonymous referee, and from
communications by G.~Aarts, C.~Greiner, A.~Sinha, and 
F.~Zwerger. 

\vspace*{1cm}


\end{document}